\documentclass[11pt,a4paper]{article}
\usepackage{graphicx}      
\usepackage{booktabs}       
\usepackage{natbib}        
\usepackage{hyperref}      
\usepackage[labelfont=bf]{caption}
\usepackage[utf8]{inputenc}
\usepackage[T1]{fontenc}
\usepackage{lmodern}
\usepackage{amsmath,amssymb,amsfonts}
\usepackage{graphicx}
\usepackage{booktabs}
\usepackage{multirow}
\usepackage{tabularx}
\usepackage{threeparttable} 
\usepackage{xcolor}
\usepackage{microtype}
\usepackage{caption}
\usepackage{subcaption}
\usepackage{enumitem}
\usepackage{titlesec}
\usepackage{appendix}
\usepackage{float}
\usepackage{array}
\usepackage{url}

\usepackage{tabularx}
\usepackage{array}

\newcolumntype{L}{>{\raggedright\arraybackslash}X}

\usepackage[margin=1in]{geometry}  
\usepackage{setspace}                 
\hypersetup{
    colorlinks=true,
    linkcolor=blue!80!black,
    citecolor=blue!80!black,
    urlcolor=blue!70!black,
    pdftitle={Mapping the AI Economy: Google's ATLAS v1.0},
    pdfauthor={Zanna Iscenko et al.}
}

\title{
    \textbf{
    \Large Google's AI \& Economy ATLAS v1.0: \\     Mapping Gemini Usage in the Economy\thanks{We extend our sincere thanks to Diane Coyle (University of Cambridge) and David Autor (MIT) for their contributions, guidance, and review. Other contributors who made this report possible include Rossi Abi-Rafeh, Yinkei De Maqua, Alon Halevy, Dave Hotchkiss, Preston McAfee,  Adriana Olmos, Yooki Park, Michael Pisa, Ruth Porat, Daniel Rock, Leanne Trujillo, Arthur Turrell, and Kent Walker.}
    }
\vspace{-0.5em} }

\author{
    \fontsize{10.5pt}{12pt}\selectfont 
    \begin{tabular}{c c c c}\
        Zanna Iscenko$^{1,}$\thanks{Corresponding authors. Emails: \href{mailto:zannai@google.com}{zannai@google.com}, \href{mailto:sstrand@google.com}{sstrand@google.com}} & Scott Strand$^{1, \dagger}$ & Yiyuan Chen$^1$ & Guillaume Aimard$^1$ \\
        Mihai Codreanu$^1$ & Vivek Sampathkumar$^1$ & Alex Imas$^2$ & Julian Jacobs$^2$ \\
        Evalyne Muiruri$^2$ & Juan Mateos-Garcia$^2$ & Jia Jen Ng$^2$ & Samirah Javed$^2$ \\
        Josh Martin$^1$ & Omar Ajmeri$^1$ & Denis Calin$^1$ & Andrew Kim$^1$ \\
        & Fabien Curto Millet$^1$ & James Manyika$^1$ & \\
    \end{tabular} \\[0.5em]
    \fontsize{9.5pt}{12pt}\selectfont  
    $^1$Google \quad $^2$Google DeepMind
    }

\vspace{-2em} 
\date{\normalsize July 23, 2026}

\begin{document}

\maketitle

\vspace{-2em}

\begin{abstract}
\noindent This paper introduces the AI \& Economy ATLAS (Activity, Task, Landscape, and Adoption Study), an ongoing economic research initiative using Google AI usage data. The first iteration of ATLAS is built on 15 million de-identified interactions across the Gemini App, Google AI Mode, and Gemini API. Using privacy-preserving algorithms as well as established and bespoke classification methods, we map AI usage to over 800 occupations, 4000 tasks, 300 household activities, 150 countries, and 140 languages. We then make a number of observations on what the data reveals about AI’s diffusion, and its usage at work and in day-to-day life. In the workplace, we show that while AI adoption spans occupations covering just above 88\% of US employment, penetration remains shallow and overwhelmingly collaborative in nature, with end-to-end task automation limited in scope. Outside of work, AI spans activities making up about 98\% of Americans' non-sleep time, with disproportionately high use in high-friction tasks such as engaging with government and professional service providers, likely delivering economic value that standard national accounts may miss. Globally, adoption scales with national wealth and has broad linguistic distribution, with English queries representing only around a third of volume. As we build upon ATLAS and expand its scope and capabilities, we will continue to provide large-scale empirical evidence to inform the public, policy and academic questions about the ongoing AI transformation.
\end{abstract}



\newpage

\section{Executive Summary}

In 1987, economist Robert Solow quipped that the computer age was visible everywhere except the productivity statistics. Today we see the signs of a new Solow paradox: AI appears to be everywhere, yet its impact remains hard to discern in many traditional measures of employment, productivity, and growth.

To help bridge this gap, we introduce Google's AI \& Economy ATLAS (Activity, Task, Landscape, and Adoption Study): an ongoing economic research initiative built on Google data. This initial version (``ATLAS v1.0'') encompasses 15 million de-identified interactions across the Gemini App, Google AI Mode, and the Gemini API. Operating under strict privacy-preserving standards, ATLAS clusters and maps these interactions to traditional labor frameworks like BLS occupational codes and O*NET tasks, as well as incorporating attributes such as intent and multimodality. We also introduce a range of novel lenses for analysis of AI use, including mapping non-work interactions to American Time Use Survey activities, characterizing work interactions by expertise and task characteristics, and investigating the languages of use. 

We are developing ATLAS to better understand how people are using AI at scale. Google AI Mode serves over 1 billion monthly active users (MAUs) and Gemini App over 900 million. In combination with Gemini API, these surfaces serve users across more than 800 occupations, 4000 work tasks, 300 household activities, 150 countries, and 140 languages.\footnote{As of May 2026, Google AI Mode serves over 1 billion MAUs, and Gemini App services over 900 million MAUs \citep{google2026a, google2026b}. Although work-related activity comprises a minority of conversational (non-API) AI use at 14\% of total usage, applied to these userbases this represents between 135 million and 260 million MAUs.} This first snapshot contributes new data and evidence to advance our understanding of AI's economic impact in several key areas. Ten early findings emerge from our analysis:

\begin{enumerate}
    \item \textbf{AI has diffused very broadly in both work and life.} 
    
    Adoption in the workplace spans all major sectors (e.g., from Professional and Business Services, to Construction, Leisure and Hospitality, and more). It also spans over 68\% of all occupations that collectively represent just above 88\% of total US employment, including both much-discussed occupations such as software developers and market researchers, and those less-discussed such as farmers, industrial engineers, and foresters. Outside of work, AI is touching all of the major ways in which humans spend their time, including leisure, household management, education, personal care, and travel. Overall, AI conversations span activities that make up 98\% of Americans' waking hours.

    \item \textbf{Though broadly used at work, the overall depth of AI's use is shallow.} 
    
    ATLAS data shows that AI is used for only 21\% of total tasks in the median occupation with any AI use. Only 3\% of occupations showed AI usage for over 75\% of their tasks; these occupations include software quality assurance analysts and testers, human resources specialists, and document management specialists.
    
    \item \textbf{While there is some task automation, the vast majority of use so far is collaborative and assistive to tasks and work.} 
    
    Non-routine cognitive tasks (e.g. hypothesis testing and creative design) make up only about 35\% of the professional tasks in the economy as a whole, yet they make up almost 65\% of work-related AI interactions in our data. In our initial attempt to taxonomize intent, only a small amount of this usage appears to be focused on automation based on our classification. Instead, usage of AI for non-routine cognitive work is centered on Partial Drafting and Generation, Review and Refinement, Ideation and Strategy, and Information Retrieval and Learning. Attempts to automate tasks end-to-end represent less than 10\% of AI conversations in non-routine cognitive work appearing in our data.

    \item \textbf{AI use is not only a white collar phenomenon; it is also assisting in physical and manual work.} 
    
    While nearly a third of heavily physical occupations show no observed AI usage, workers in many manual and technical trades are using AI on the job. In these roles, AI frequently acts as a hands-on collaborator for diagnostics, troubleshooting, and real-time learning. We also observe disproportionate multimodal use of AI (i.e., uses that involve images and video) in these contexts: for example, automotive technicians and industrial mechanics using AI to interpret complex test results, debug electrical wiring, and inspect machinery for wear, where the usage rate of multimodal AI is more than 2 times higher than the overall work baseline.

    \item \textbf{Work-related AI usage correlates strongly with higher wages and education.} 
    
    In the US workforce, a 1\% increase in an occupation's median earnings is associated with a more than 2.5\% increase in AI usage intensity. Weighted by Gemini conversations, the median salary across observed occupations is around \$83,000, roughly \$20,000 higher than the true employment-weighted national median. This relationship persists even after controlling for the occupation's educational attainment (which itself is positively correlated with AI usage).

    \item \textbf{While wages and expertise are typically correlated in the economy, ATLAS points to a complex relationship between AI usage and expertise.} 
    
    When classified according to expertise levels, tasks requiring lower-to-middle levels of expertise tend to see relatively higher AI usage than the highest expertise level tasks, despite ATLAS data also suggesting high-earning workers---and therefore those with more scarce skills and expertise---adopt AI at the highest rates.

    \item \textbf{AI is delivering real utility at home that standard economic metrics may miss.} 
    
    Over 86\% of interactions with conversational AI happen outside of work, and humans' daily time allocation strongly and positively predicts where people direct their AI questions. Many of these conversations are in ``productive household activities'', including household management, researching purchases, interior design and help with appliances and tools. If these household time-savings would average out to just 30 minutes per week, the unpaid productivity gains in the US alone could be worth approximately \$100 billion, using standard valuation methodologies.

    \item \textbf{AI usage outside of work appears particularly high for high-friction administrative tasks.} 
    
    ATLAS data covers a broad set of activities outside of work, and we find that a handful of high-friction professional and bureaucratic activities exhibit significant over-representation in the data relative to the actual time spent on them. These activities include interactions related to government services, legal topics, finance, and education. As an example, in the ATLAS data sample for this report, we find that government services and civic obligations are over-represented by a factor of almost twenty. We also observe that nearly half of all medical, legal, financial, and government AI consultations take place outside 9-to-5 business hours, bypassing temporal barriers to accessing these institutions.

    \item \textbf{Geographically, AI adoption tracks closely with national wealth, with notable exceptions.} 
    
    Globally, per capita AI usage closely mirrors national wealth, with a 1\% increase in GDP per capita associated with a 0.9\% increase in usage. This concentration means the lowest-adopting quintile of countries accounts for only 2\% of AI conversations, highlighting concerns about a persistent digital divide. However, multiple middle-income nations across Latin America and the Middle East buck this trend, adopting AI at rates comparable to Western Europe. The differences in adoption rates between countries are not easily explained by any single factor, such as internet access and interest in AI topics among the population.

    \item \textbf{There are surprising trends in AI use across languages and modalities.} 
    
    English accounts for only about a third of global conversations, and users do not show signs of systematically abandoning their native languages for complex professional tasks, as work and non-work activities show nearly identical language distributions. Global usage also varies across modalities; for instance, users in non-OECD countries generate images and videos for work at roughly twice the rate of those in advanced economies.
\end{enumerate}

Taking a step back, our initial ATLAS v1.0 data therefore sheds new light on several widespread claims about AI’s impact on the economy.  For instance, we do not find evidence in ATLAS to support the claims that AI is about to cause massive automation and displacement of white-collar work; that AI is irrelevant to blue-collar work; or that the purpose of AI is strictly to automate tasks.  Neither does the data support the claim that the global ``AI race” is strictly between the US and China from the perspective of leadership in adoption. However, appropriate prudence is required in interpreting these results: they constitute early observations and could evolve as the frontier of AI capabilities continues to advance, the ways in which businesses and individuals use these capabilities expand, and as the analytical methodologies economists and researchers use to measure their impact also progress. 

Methodologically, we are particularly cognizant that our initial classifiers on intent and expertise leave scope for greater sophistication over time. And there are many other limitations to ATLAS that we readily acknowledge. By its nature, ATLAS does not capture the ultimate productive output the user is working toward or how effective their interaction was; it only captures what people are directly doing with AI. Moreover, there is a much wider range of economically relevant AI usage that has yet to be reflected in ATLAS. These include AI-enabled products with billions of interactions like Google Workspace (used by over 3 billion, \citep{GoogleWorkspace2025}), Google Translate (used by over 1 billion people, \cite{Yao2026Translate}), and AI Overviews (used by over 2.5 billion, \cite{Loew2026Search}); enterprise platforms, such as Google Cloud Gemini Enterprise; frontier capabilities in several key areas, such as agentic coding and world models.

As a result of its limitations and the early state of the AI innings, there are also therefore many important claims in the public discourse that ATLAS v1.0 is unable to shed much light on - including the notion that AI might deepen the global digital divide or that AI could be playing a role in dampening entry-level hiring.  These areas merit deeper investigation over time, in combination with other data sources.

Looking forward, ATLAS itself also raises many additional questions requiring further investigation: Why do stark adoption disparities persist across countries even when normalizing for factors like internet access?  What factors explain usage spikes in specific emerging markets?  Does AI usage at home indeed help save time or lead to better outcomes for people?  How might we account for the value this generates in broader measures of economic welfare?  What else characterizes the usage of AI across different categories of work, including physical and manual work?  How might we better detect the emergence of entirely new, AI-enabled work categories beyond the early signals in our data?  

As we continue to build upon ATLAS and expand its scope and capabilities, we will strive to provide a sharper empirical map to inform these questions and serve as a navigational tool for the AI transformation. 

The rest of the report is structured as follows.  Section \ref{sec:data_methods} provides an overview of our data and methodology. Section \ref{sec:work_usage} discusses  observed patterns in work-related usage in ATLAS. Section \ref{sec:home_usage} describes AI adoption for non-work activities and puts it in context of the established statistical frameworks for measuring time use. Section \ref{sec:geo_diffusion} presents our findings on global diffusion and multilingual patterns in AI use. Section \ref{sec:discussion} discusses the results and summarizes directions for future research.

\section{Data and Methods Overview}\label{sec:data_methods}

The ATLAS v1.0 dataset relies on a fully automated privacy-preserving data processing pipeline that analyzes a sample of 14,653,926 de-identified user interactions across the Gemini App, Google AI Mode, and Gemini API between April 6 and April 19, 2026.

\subsection{Methodology Overview}\label{subsec:methodology_overview}

The data preparation and classification process involves several automated steps:

First, AI interactions that have been redacted of identifying information are classified as ``work'' and ``non-work'' related using an automated classifier, and then directed into unique pipelines. These conversations are then summarized, with each pipeline focusing strictly on the facets required for its downstream classification.

Second, using Observation Clustering and Taxonomy Organisation (OCTO), a bespoke clustering and hierarchical taxonomy assignment tool developed by Google DeepMind, these summaries are grouped into semantically similar clusters. The observations in the cluster are then summarized into a cluster label tailored to downstream classification needs.

Third, these clusters are mapped to official statistical taxonomies: the Bureau of Labor Statistics (BLS) 2024 American Time Use Survey (ATUS) Activity Lexicon for non-work interactions \citep{bls_atus_lexicon_2024}, and BLS 2018 Standard Occupational Classification (SOC) \citep{bls_soc_2018}, further broken down into occupational titles and tasks from Occupational Information Network (O*NET) Database v30.2 for work interactions \citep{onet_30_2}. We then augment the cluster dataset with additional bespoke classifiers unrelated to official statistical taxonomies. Detail about these specific annotations is presented in ATLAS report sections that rely on them.

Finally, the pipeline’s performance was evaluated using three distinct methods: measuring pipeline accuracy on synthetic data, inter-rater agreement evaluation with and without the AI ratings, as well as  human approval of AI labels. For the former, we generated a synthetic ground-truth dataset seeded with the most granular tiers of the O*NET-SOC and ATUS taxonomies and calculated the rate with which our pipeline recovered the ground truth categories.  

\subsection{Privacy and Data Governance}\label{subsec:privacy_data_governance}

Privacy preservation is a foundational component of the ATLAS methodology, with multiple lines of privacy protection implemented at every stage of the pipeline:

First, we redacted Personally Identifiable Information (PII). Prior to any processing, extensive Data Loss Prevention (DLP) filters automatically stripped from the data any PII including but not limited to contact details, financial records, government identifiers, and health information.

Second, we enhanced the de-identification of the data. All internal log identifiers (which are already de-identified) are replaced with mathematically unlinked Universally Unique Identifiers (UUIDs) to completely obfuscate user identities and prevent data in the pipeline being linked back to broader log context.

Third, we deployed data minimization through two layers of summaries. Processing of underlying text of conversations was minimized, as they were first summarized individually (with the full text discarded), and individual summaries then aggregated and re-summarized again as a cluster of related interactions. At both steps summaries focused only on facets relevant to taxonomy assignment (e.g., specialized professional tools and knowledge in a particular cluster of conversations).

Fourth, we implemented K-anonymization. Any cluster representing fewer than 10 unique users was automatically discarded and excluded from any further analysis.

The underlying data processing was fully automated. Original conversation texts or their individual summaries are not retained in the final ATLAS dataset, and access to the final data is tightly restricted to a small team of researchers under the oversight of internal Privacy Working Groups.

Gemini models play several essential roles throughout: summarizing conversations and clusters, extending the descriptions of taxonomy categories with additional detail, performing LLM classification for official taxonomy mappings and bespoke annotations, as well as creating synthetic datasets to validate performance.

To place our findings in context, it is helpful to note where ATLAS v1.0 differs in scope and design from the methodologies used by economic studies of AI use by Anthropic \citep{handa2025economictasksperformedai, anthropic2026aeiv6} and OpenAI \citep[e.g.,][]{NBERw34255}. In terms of data sources, our dataset pools a larger variety and quantity of user interactions than have been previously explored together, spanning a standalone conversational AI app, an AI search experience, and a developer-facing API entry point. In terms of technical methodology, although we rely on a clustering-based taxonomy mapping process similar to Anthropic, our implementation differs considerably---for instance, by improving scalability to support a large sample size (15M), introducing recursive traversal of multiple nested statistical taxonomies (combining the SOC and O*NET classification into a single pipeline), leveraging LLM-assisted taxonomy category annotation to improve classification context, randomizing classifier options to mitigate well documented biases in LLM classification arising from option order \citep[e.g.,][]{https://doi.org/10.48550/arxiv.2308.11483}, and validating classifier performance with synthetic data. Finally, in terms of the analytical scope of the dataset, we expand on the previously limited analysis of AI uses outside work by mapping AI interactions to an established statistical taxonomy for household time use (ATUS) and exploring multi-language adoption patterns. For work-related topics, we bring in new lenses from economic literature, such as the \citet{NBERw33941} task expertise and routine/non-routine and manual/cognitive task characteristics. Another innovation in ATLAS, to the best of our knowledge, is adjusting for Google app penetration differences between countries when we investigate cross-country AI adoption patterns.

These represent only a subset of the methodological differences between the respective reports; for a comprehensive description of our data sources and processing pipelines, please refer to the full methodology in Appendix~\ref{app:app1}. Appendix~\ref{app:classifier_validation} contains further detail on classifier validation methodologies and results.

\subsection{Limitations}\label{subsec:limitations}

The findings in ATLAS v1.0 have several important limitations.

First, taxonomized ATLAS v1.0 data does not currently contain paid usage of Gemini API, which includes enterprise usage via Google Cloud.\footnote{We observe the number of paid Gemini API requests by country, which are used the analysis of geographic dispersion. However, due to the content of these API requests not being available,  paid Gemini API does not contribute to any more granular analyses in the report.} As a result, despite the extensive volume of work activity on our three included surfaces, enterprise-level professional AI use cases may be under-represented. ATLAS also does not include AI interactions from several other Google experiences and products, including Google Workspace, AI Overviews, Google Translate, Google Maps, Google Flow, Google Antigravity, and Gemini Notebook. We plan to expand the scope of ATLAS data significantly in future iterations. 
    
Second, ATLAS v1.0 measures behavioral interactions, not definitive productivity outcomes. A completed conversation does not guarantee that the user accomplished their intended goal, saved time, or produced measurable economic value.
    
Third, ATLAS v1.0 measures how users are currently using AI, and not the full universe of tasks for which they \emph{could} bring. The data captures the current frontier of AI adoption, rather than the technology’s potential impact and reach. Users who have not yet adopted AI are not represented.
    
Fourth, ATLAS v1.0’s classifications are probabilistic. Deriving exact job titles or tasks from conversational text is inherently uncertain. While our classifiers perform at least at the human-expert levels for the validation exercises we performed, highly granular occupational or household activity findings carry more uncertainty than broader, major-group trends. 

This first iteration of ATLAS provides an early snapshot of a fast-changing landscape, drawn from a dataset we intend to refine and revisit frequently. It is the foundation for a long-term body of research from Google, and an invitation for economists, researchers, policymakers, and industry experts to collectively explore these massive economic questions with us. 

\section{Work Usage}
\label{sec:work_usage}

\subsection{The Broad Picture}

As AI models become increasingly capable, a key concern for policymakers and workers alike is understanding how this technology will impact labor markets, including the ways and types of work humans do. We intend to use ATLAS to understand which workers are using AI and how they are using it. We view this as a first step towards supplying society with information needed to make decisions about how to adapt to this technological change and make its potential gains broadly beneficial.

Efforts to understand the impacts of AI and attendant technological automation on labor market outcomes predate the advent of modern generative large language models \citep[e.g.,][]{Felten2019, Felten2021}. These efforts themselves are part of a broader literature attempting to model the labor market consequences of the already remarkable proliferation of information technologies over the past several decades. Labor markets during this period exhibited job and wage polarization, manifesting as an increase in the shares of both high skill, high wage occupations and low skill, low wage occupations. This phenomenon has been ascribed at least in part to the concurrent wave of digitalization \citep{Acemoglu2011}.

Economists have used task-based models of jobs---treating occupations as sets of tasks---to understand the consequences of technological shocks or automation and how the nature of work changes in response \citep[e.g.,][]{10.1162/003355303322552801, Acemoglu2011, 10.1257/jep.33.2.3}. Prior work, preceding the wide availability of large language models, necessarily could not anticipate the particular, novel consequences of this technology. For instance, \citet{Acemoglu2020} focused on the effects of industrial robots whereas generative AI technologies are raising particular concerns about the status and security of white-collar workers' occupations. However, \citet{10.1162/003355303322552801}, which applied the task-based framework to study computerization, provides testable predictions about which tasks will be impacted by analogous technological developments---in this case LLMs---and how those technologies would interact with the tasks performed by workers. In particular \citet{10.1162/003355303322552801} suggests traditional procedural (or `rules-based') computerization serves as a substitute for routine tasks while functioning as a complement for non-routine problem solving and complex communication tasks.

The adoption of generative AI appears to be occurring rapidly.\footnote{These numbers vary between different surveys and countries, but are high and showing increases over-time. For example, \cite{pew_2026_americans_ai} reports that 49\% of U.S. adults have used chatbots (up from 33\% in 2024), while \cite{ipsos_google_2026_ai_survey} finds in a study of 21 countries that 62\% have used generative AI in the last twelve months (up from 48\% in 2024).} The pace and scale of potential changes ushered in by this technology are raising understandable concerns about its effects \citep{ipsos_google_2026_ai_survey} though there is substantial variation in sentiment between countries. Amidst this rapid technological change, there has also been a flurry of attempts to understand which work is going to be impacted by AI and whether we might already be seeing labor market impacts.

\citet{Eloundou2024} made projections about the exposure of different tasks and occupations to LLM impacts, estimating that around 80\% of the U.S. workforce could have at least 10\% of their work tasks affected by the introduction of LLMs. As we show below, however, tasks being ``affected'' by LLMs is very different from those tasks being end-to-end automated.

\citet{richmond2026ai} categorized occupations based on their short-term automation risk, finding 18\% of jobs at relatively high short-term risk. \citet{handa2025economictasksperformedai} analyzed AI task usage, determining 57\% of usage suggests augmentation while 43\% of usage suggests automation of human capabilities. \citet{NBERw35271} argue that existing AI exposure metrics are ineffective at explaining observed workplace AI usage and propose an alternative approach to predicting AI diffusion based on comparative advantage.

A number of researchers have found no labor-market effects of AI or determined that it was too early to tell whether AI was having an impact or not \citep[e.g.,][]{AudolyGuerinTopa2026, gimbel2026what, NBERw33777, iscenko2026looking}. \citet{massenkoffmccrory2026labor} similarly find no systematic unemployment increase for workers with higher exposure but suggest these occupations are projected to have slower growth over the next decade based on BLS projections and that there may be evidence of slowed hiring of younger workers in exposed occupations. \citet{kolko2026ai} indicates that better data is needed to fully understand AI's labor market impacts. We agree that we can only better understand these questions through better data. We believe ATLAS in its current and future iterations can provide relevant insights on how AI is being used and inform these discussions.

To structure this analysis, we begin by examining which occupations are using AI, comparing AI usage shares to employment shares across major occupation groups. We then narrow our focus to the range of tasks within each occupation, measuring how broadly and deeply AI has penetrated individual roles. Next, we characterize the nature of those tasks, whether they are cognitive, routine, manual, or relational, and how AI usage patterns diverge from the predictions of prior technology frameworks. We then distinguish tasks by user intent, asking whether AI is being used to automate or to augment human work, and by the expertise level those tasks require. Finally, we examine the relationship between AI usage intensity and occupational earnings and education. Together, these lenses enable us to begin developing a detailed view of the implications of AI for the labor market that goes beyond broad exposure metrics.

\subsection{Occupation Data}

As described earlier, we mapped de-identified user conversation clusters to standard occupational and task classifications using the O*NET v30.2 data taxonomies.\footnote{Further detail on data construction and the mapping process can be found in Appendix \ref{app:app1}. The classification of LLM interactions into occupation or task taxonomies is a challenging endeavor and necessarily involves some imprecision. Additional detail on the validation of classifier assignments to tasks and occupations can be found in Appendix \ref{app:classifier_validation}. We intend to continue to refine these processes over time but are sufficiently reassured by our validation process that the existing results are reflective of the AI usage in the ATLAS sample.} Henceforth, we use \textit{detailed occupation} to refer to the most granular occupation available, which is the 8-digit O*NET SOC code or the associated 6-digit SOC aggregation\footnote{This is generally obtained by truncating the final 2 characters of the 8-digit O*NET SOC. There are a handful of cases where the OEWS and O*NET taxonomies are not perfectly compatible in this way. In those cases, the detailed O*NET occupation is mapped to the relevant detailed OEWS SOC code.} when OEWS data on occupation level statistics such as employment or earnings is used. We use \textit{major occupation} to refer to the highest level aggregation of occupation groups corresponding to the 2-digit OEWS SOC occupation codes.

The analysis in this section is based on AI interactions determined to be related to work tasks, aggregated across three Google sources (App, AI Mode, API).\footnote{For analysis related to occupations and work tasks, we think it is important to include API data in order to adequately capture key work use cases like coding. Our current approach is to treat Gemini interactions for these statistics as equivalent across the three sources (App, AI Mode, API). However, we recognize there may be important differences in the nature of the interactions on these different surfaces. In future work, we may attempt to incorporate any relevant nuances.} Throughout this section, we refer to these as ``Gemini'' interactions, to distinguish them from interaction with other AI products.

\subsection{Occupational AI Usage}

\begin{figure}[htbp]
    \centering
    \caption{White-Collar Professions Are Overrepresented in AI Usage Relative to US Employment}
    \includegraphics[width=\textwidth]{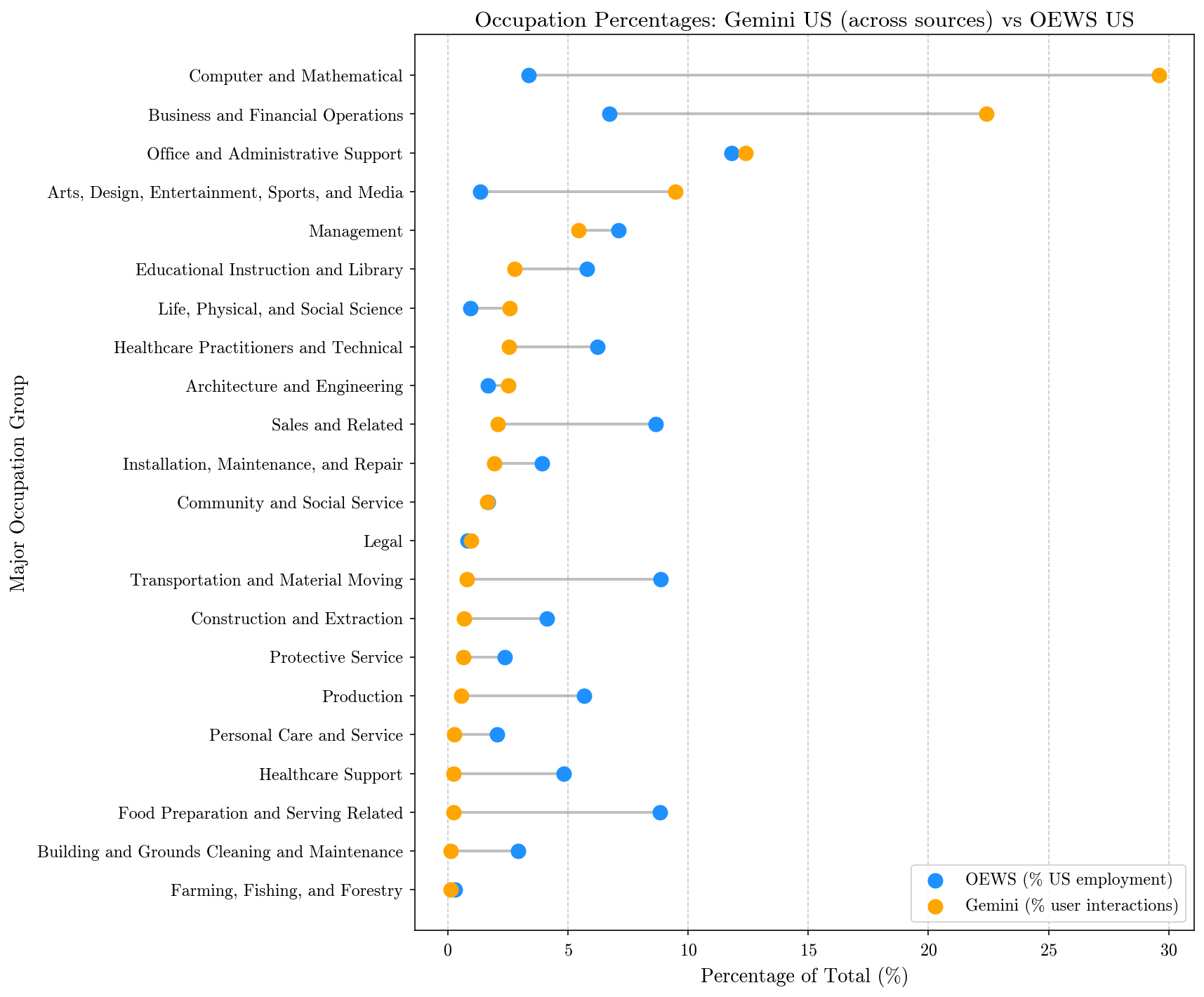}
    \label{fig:figure_2_1}
    \vspace{0.2cm}
    \parbox{\textwidth}{\footnotesize \textit{Notes:} This figure compares the shares of US work-related Gemini interactions within broad occupation groups to shares of civilian US employment in those groups. US employment shares at the Major Occupation Group (SOC) level sourced from Occupational Employment and Wage Statistics May 2024 data \citep{bls_oews_2024}. Gemini shares are derived from interactions aggregated across surfaces (App, AI Mode, API) classified into O*NET SOC occupations. OEWS data does not contain information on Military Specific Occupations so those are omitted.}
\end{figure}

Starting with the US, looking at Gemini usage relative to population employment shares (Figure~\ref{fig:figure_2_1}), ``Computer and Mathematical'', ``Business and Financial Operations'', and ``Arts, Design, Entertainment, Sports, and Media'' occupations are most over-represented in terms of AI use. Several occupations are under-represented relative to their population employment shares, including ``Sales and Related'', ``Transportation and Material Moving'', and ``Food Preparation and Serving Related'' occupations.

For detailed occupations (Table~\ref{tab:table_2_1}), we find that among the more over-represented occupations in our dataset are financial/market analysts, software developers, and systems administrators. This is consistent with our expectations about modern AI tools currently being more relevant for white-collar professions. Common tasks in these occupations where we observe high AI usage involve data collection and analysis, information monitoring, maintaining computing environments, developing software related to scientific analysis, and testing or validating software.\footnote{O*NET task ids 5433, 21596, 1318, 21667, 21670}

In a number of occupations, such as home health aides or fast food workers and cooks, we observe substantial under-representation in AI usage relative to the employed population; we also fail to observe AI usage in certain occupations at all (above our privacy thresholds). These are occupations which appear to be more focused on physical or interpersonal components and dissimilar from the knowledge work where AI usage is currently more prevalent. We explore these dynamics further later in this report.

\begin{table}[htbp]
    \centering
    \caption{Over and Under-Represented Detailed Occupations Relative to US Employment}
    \label{tab:table_2_1}
    \begin{tabularx}{\textwidth}{@{} L L L @{}}
\toprule
\textmd{Most Over-represented } & 
\textmd{Most Under-represented } & 
\textmd{Largest occupations not observed} \\
\midrule
Financial and Investment Analysts & 
Home Health and Personal Care Aides & 
Food Preparation Workers \\[0.6em]

Market Research Analysts & 
Fast Food and Counter Workers & 
Cooks, Fast Food \\[0.6em]

Software Developers & 
Cashiers & 
Dining Room and Cafeteria Attendants and Bartender Helpers \\[0.6em]

Network and Computer Systems Administrators & 
Retail Salespersons & 
Substitute Teachers, Short-Term \\
\bottomrule
\end{tabularx}

    \vspace{0.2cm}
    \parbox{\textwidth}{\footnotesize \textit{Notes:} Over/under-representation based on signed distance between Gemini and US employment shares. Gemini shares are derived from interactions aggregated across surfaces (App, AI Mode, API) classified into O*NET SOC occupations. O*NET SOC codes are mapped to corresponding SOC codes in the May 2024 OEWS data \citep{bls_oews_2024} to obtain US employment shares. OEWS data does not contain information on Military Specific Occupations so those are omitted. In addition, residual detailed occupation categories with titles including ``All Other'' are excluded due to uncertainty about the quality of classification into these labels and for consistency with other detailed occupation analysis in this section. Non-observation of occupations may come from a combination of the relevant workers not using AI and those groups being sufficiently small that they cannot be detected given the privacy protections we implement.}
\end{table}

Moving to global data, when comparing Gemini interactions from OECD\footnote{For all analyses referencing OECD groupings, note that Gemini data availability in Europe is more limited. In particular, API data is unavailable for classification. Therefore, the presence of coding applications (``Computer and Mathematical'' occupations) are likely understated for the OECD.} and non-OECD countries, we find that the usage from OECD countries is more disproportionately concentrated in ``Computer and Mathematical'' and ``Business and Financial Operations'' occupations. In contrast, interactions from non-OECD countries have a greater share of tasks associated with ``Office and Administrative Support'', ``Arts, Design, Entertainment, Sports, and Media'', and ``Educational Instruction and Library'' occupations. The occupation shares differ between country categories; the differences in the ``Arts, Design, Entertainment, Sports, and Media'' category appear particularly pronounced, with these occupations representing roughly 70\% higher share of Gemini interactions in non-OECD countries (15.9\%) vs OECD countries (9.4\%).

\begin{figure}[htbp]
    \centering
    \caption{Using Gemini for Work in Arts and Design Occupations is More Common in Non-OECD Countries While Computer and Math and Business and Finance Occupations Are More Common in the OECD}
    \label{fig:figure_2_2}
    \includegraphics[width=\textwidth]{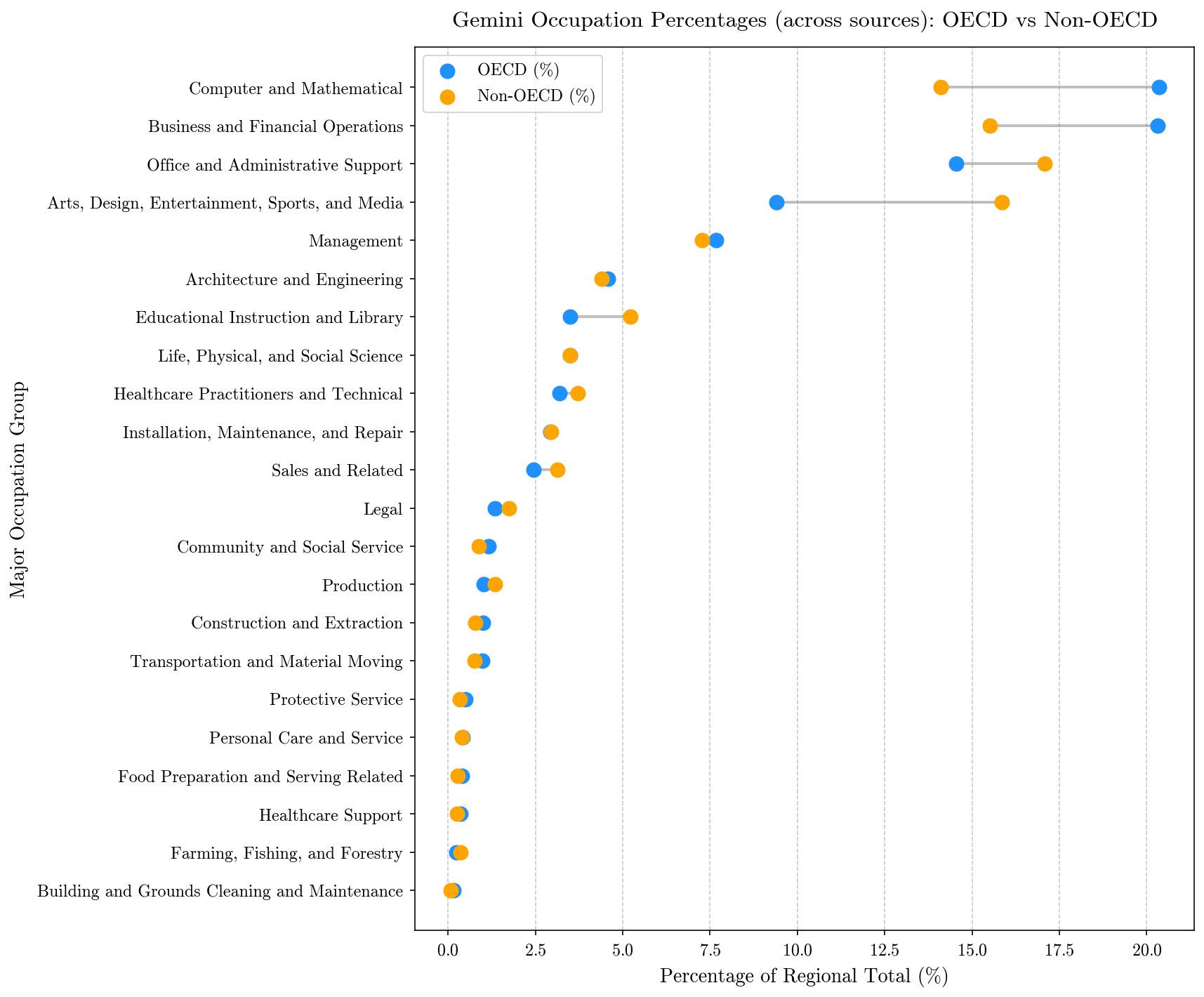}
    \vspace{0.2cm}
    \parbox{\textwidth}{\footnotesize \textit{Notes:} Gemini shares are derived from interactions aggregated across surfaces (App, AI Mode, API) based on the user's country/region and classified into O*NET SOC occupations. This chart compares the broad occupation group shares in the 38 OECD member countries to non-OECD countries.}
\end{figure}

Looking at the detailed occupations within these categories, the ones most disproportionately represented in the OECD countries' Gemini interactions include financial and investment analysts, network and computer systems administrators, credit analysts, human resource specialists, software developers, and accountants and auditors.\footnote{Note that the current ATLAS sample was drawn from Gemini interactions in early April 2026. So, tax-related queries may be influencing the relative prominence of associated occupations.} Meanwhile, the detailed occupations most disproportionately represented in non-OECD countries' Gemini interactions include graphic designers, data entry keyers, photographers, instructional coordinators, and word processors and typists.

The ``Office and Administrative Support'' occupations (data entry keyers, bookkeeping clerks, word processors and typists) which are more common in the non-OECD countries could point to the different nature of office work that may be exposed to AI usage in economies with different levels of development. Moving forward, it will be important to understand how AI usage evolves, where the most valuable knowledge work occurs, and whether AI tools enable workers in less developed economies to perform more complex tasks or these differences are driven more by occupational composition, surface/API coverage, or product penetration.

It is also worth noting the presence of graphic designers and photographers as occupations that are more prevalent in non-OECD countries' Gemini interactions. Gemini usage associated with film and video editors, producers and directors, and special effects artists and animators is also more common in non-OECD countries. We observe substantially more multimodal usage in non-OECD countries. Gemini work-related conversations\footnote{Image generation data is only available for Gemini App.} are roughly twice as likely to include a generated image or video in non-OECD countries relative to OECD countries. Moreover, multimodal use is more than twice as common in work conversations relative to non-work conversations. There is also variation in media generation across regions, with it being most common in Africa and least common in Europe.

\subsection{Occupation and Task Coverage}

Next, we turn our attention to considering the breadth of AI usage within the economy. Particularly, we consider the breadth of occupations where we observe non-negligible usage of AI tools and within those occupations, how comprehensively AI is being used to accomplish the constituent tasks.

\begin{figure}[htbp]
    \centering
    \caption{Gemini Usage is Observed Across 68\% of Detailed Occupations, Including the Vast Majority of White-Collar Professions}
    \includegraphics[width=\textwidth]{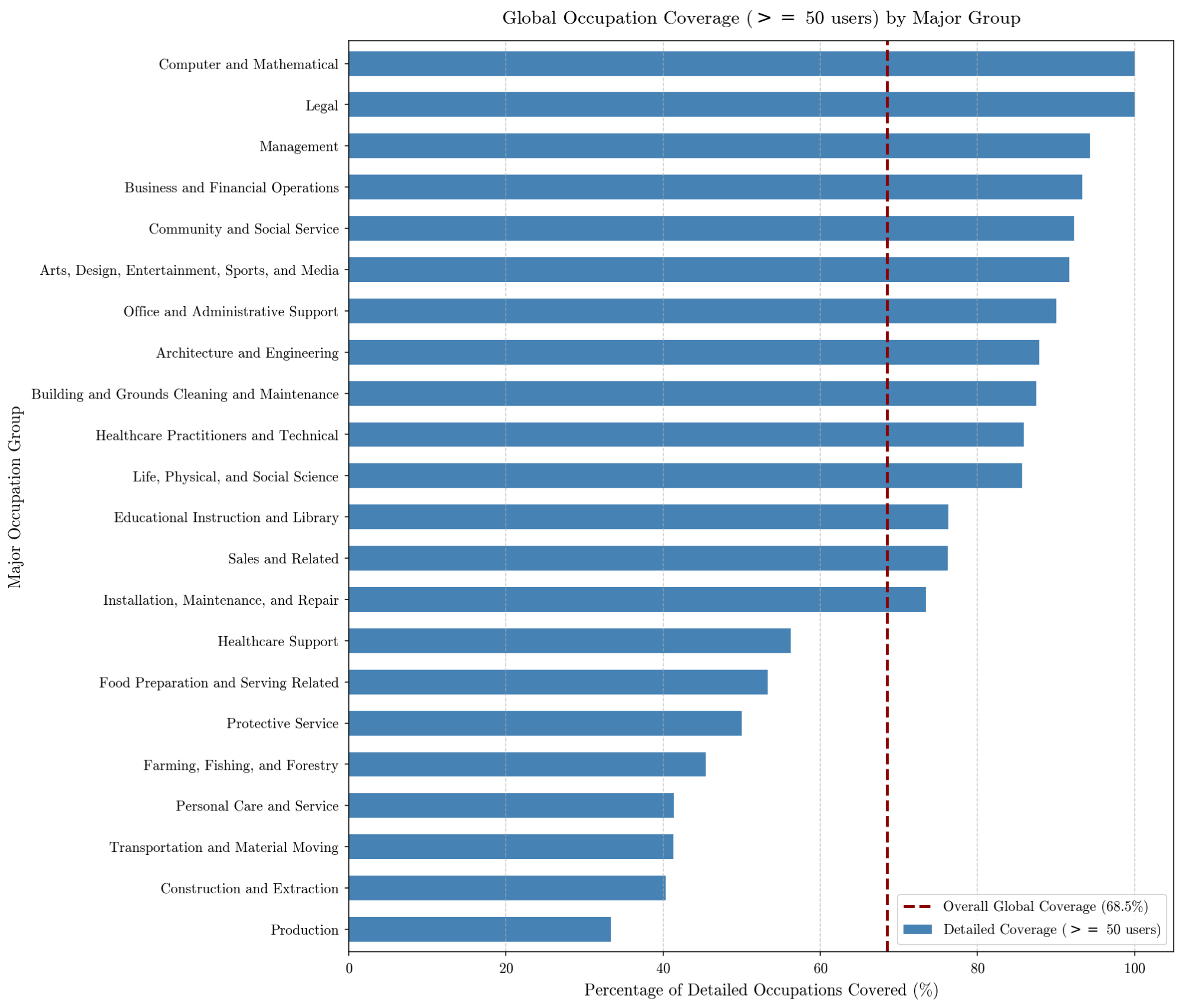}
    \label{fig:figure_2_3}
    \vspace{0.2cm}
    \parbox{\textwidth}{\footnotesize \textit{Notes:} This chart uses the 22 non-military SOC major occupation groups on the vertical axis and for each of those groups, depicts the share of component detailed occupations where we observe at least 50 Gemini users. The Gemini usage shares are derived from interactions aggregated across surfaces (App, AI Mode, API) classified into O*NET SOC occupations. O*NET SOC codes are mapped to corresponding SOC codes in the May 2024 OEWS data \citep{bls_oews_2024} so that US employment shares can be calculated. Military Specific Occupations are omitted and residual detailed occupation categories with titles including ``All Other'' are excluded for consistency with other detailed occupation analysis in this section. Data are also not present for code 45-3031 (Fishing and Hunting Workers). Therefore, the 761 remaining OEWS SOC occupations are the relevant universe of occupations used in this figure.}
\end{figure}

To measure occupation coverage, we consider the portion of the sample of almost 15 million Gemini interactions classified as work-related, which are assigned to O*NET tasks and occupations. Within each of the major occupation groups, we calculate the share of the associated detailed occupations where we observe tasks being completed by a minimum number of users (50 globally). These results are in Figure~\ref{fig:figure_2_3}. In aggregate, we observe usage above these thresholds in 68\% of detailed occupations globally but there is substantial variation across major occupation groups. We observe greater task coverage in detailed occupations in occupation groups, such as ``Computer and Mathematical'', ``Legal'', and ``Management'' occupations, which contain more office work.\footnote{Note that there are differing numbers of detailed occupations within each broad category. For instance, there are only 7 detailed Legal occupations while there are 50 Office and Administrative Support Occupations. However, the rankings of occupational coverage in Figure \ref{fig:figure_2_3} reveal an interesting divergence from the volume-based employment shares. For example, Legal occupations exhibit broad coverage across detailed roles even though their overall interaction volume remains modest relative to tech or finance professions.} The detailed occupations we observe above the stated minimum usage globally represent 88.4\% of employed civilian workers in the US.\footnote{In the absence of global employment data mapped to US occupation classifications (SOC), we currently rely on US employment statistics \citep{bls_oews_2024} to illustrate the materiality of the occupations observed in our data. We intend to continue exploring global AI usage related to work in future reports.}

Each detailed occupation in the O*NET data is associated with a set of constituent tasks, which cover the work done in that occupation. We measure the individual O*NET tasks where we observe Gemini usage globally across a minimum set of (25) users. Overall, we observe roughly 20\% of tasks meeting this usage threshold in our sample. For each detailed occupation, we calculate the share of tasks where we have achieved this level of saturation. For 29\% of detailed occupations, we observe zero task saturation. That is, for none of their tasks do we observe at least 25 users attempting that task with Gemini. This includes occupations such as stockers and order fillers, miscellaneous assemblers and fabricators, food preparation workers, fast food cooks, refuse and recyclable material collectors, and police and sheriff's patrol officers.

Around 30\% of occupations had at least a quarter of their tasks observed with Gemini usage above our threshold.\footnote{Comparing this to data from the Anthropic Economic Index, this is slightly lower than the 36\% from \cite{handa2025economictasksperformedai} and the reported ``combined across reports'' figure of 49\% from \cite{anthropic2026aeiv4}. Nonetheless, the different methodologies, both in terms of data sampling, as well as the stricter privacy preserving algorithm we enforce on our data does not make these figures not entirely comparable.} For 11\% of occupations, half of their tasks had observed Gemini usage. For 3\% of occupations, at least three-quarters of their tasks were saturated.\footnote{These occupation shares are based on the 923 detailed O*NET occupations. US employment data is only available for more aggregated 761 OEWS occupations used for Figure \ref{fig:figure_2_3}. A direct weighting of the detailed O*NET occupations within the OEWS taxonomy was unavailable. Therefore, estimating the employment covered at these thresholds requires some assumptions. We could 1) assume an OEWS occupation meets the relevant threshold if any of the constituent O*NET occupations did 2) require all of them to meet the threshold 3) estimate a share of the employment based on the share of constituent occupations meeting the threshold. (1) and (2) provide upper and lower bounds of coverage so we present those here. Occupations with at least a quarter of their tasks having AI usage cover 44-50\% of US employment. Occupations with at least half of their tasks saturated cover 26-31\% of US employment. Occupations with at least a three-quarters of their tasks having AI usage cover 9-10\% of US employment.} The latter group includes occupations like software quality assurance analysts and testers, human resources specialists, and document management specialists (See Table~\ref{tab:table_2_2}).

\begin{figure}[htbp]
    \centering
    \caption{For Occupations with Observed Gemini Usage, Workers Are Generally Using AI for Less Than One Quarter of Their Tasks}
    \label{fig:figure_2_4}
    \includegraphics[width=\textwidth]{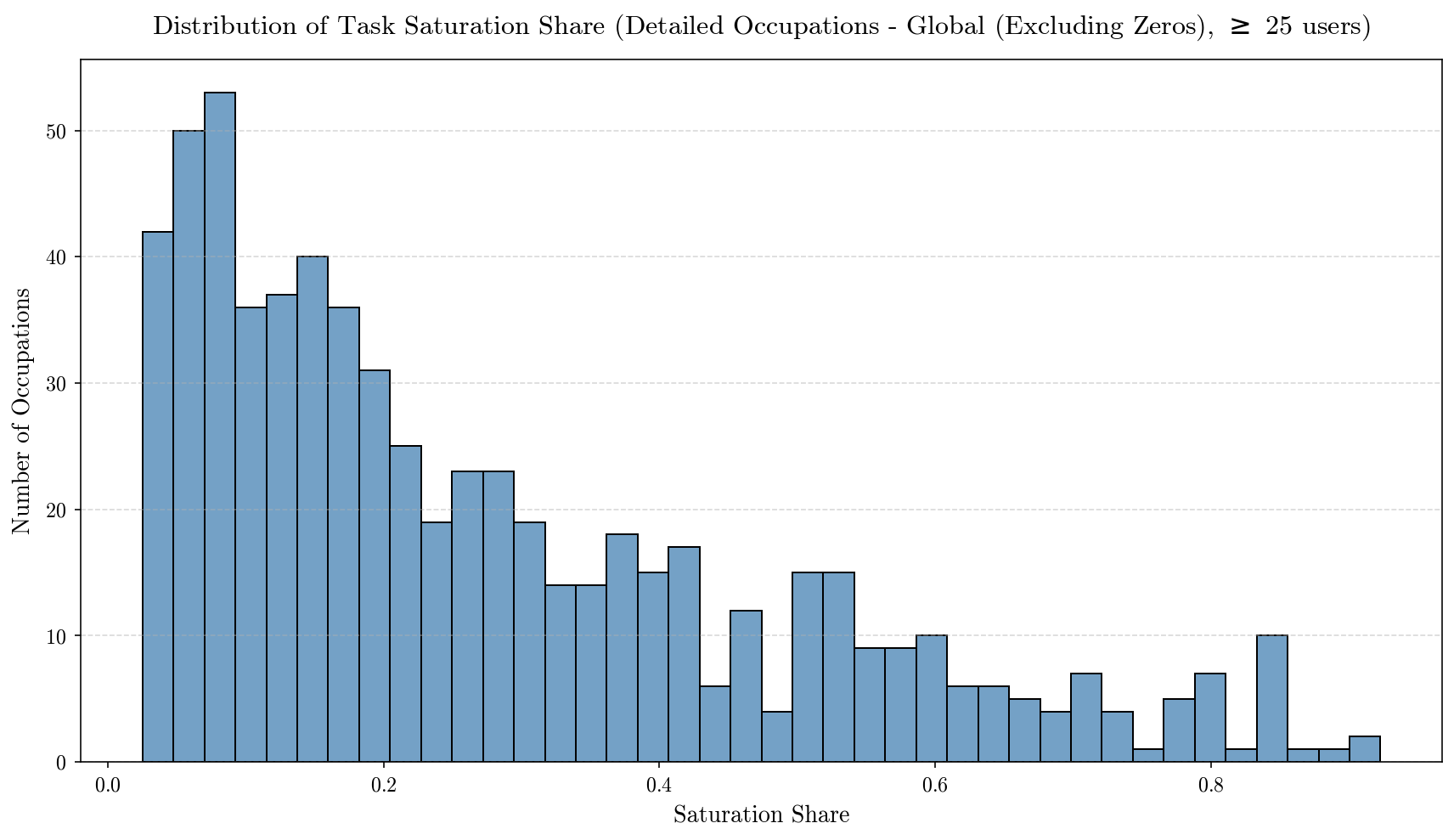}
    \vspace{0.2cm}
    \parbox{\textwidth}{\footnotesize \textit{Notes:} This chart displays the distribution of task saturation across detailed occupations with non-zero AI usage. ``Task Saturation'' is defined as the percentage of O*NET tasks within an occupation that see meaningful interaction volume (at least 25 unique users globally). Occupations where zero tasks meet this threshold are excluded. The vertical axis represents a count of occupations, while the horizontal axis states what percentage of tasks are saturated in those occupations. Gemini shares are derived from interactions aggregated across surfaces (App, AI Mode, API) classified into O*NET SOC occupations. Military Specific Occupations so those are omitted and residual detailed occupation categories with titles including ``All Other'' are excluded for consistency with other detailed occupation analysis in this section.}
\end{figure}

For the group of occupations where at least one task has above the threshold number of user interactions, we observe a median task saturation of 21\%. In other words, among occupations where we observe meaningful AI usage, workers are using AI for about a fifth of their occupation's tasks at the median. This conditional distribution is presented in Figure~\ref{fig:figure_2_4}.\footnote{Previous work on measuring AI exposure using O*NET task data have aggregated to the occupation level using O*NET's task importance ratings but consistent with \cite{Eloundou2024}, we do not observe substantial changes to our findings. Applying this weighting does not meaningfully impact the distribution of task saturation we observe in our data. (See Appendix \ref{app:app_3})}

Taken together, these results offer possible support for the notion that AI is currently serving primarily as a complement to existing work.  In line with classic determinants of technology-driven productivity dispersion across workers and organizations \citep{10.1257/jel.49.2.326}, AI appears useful for a subset of tasks performed within occupations, but they do not currently appear to be comprehensively used for performing the work currently done by humans. It is possible that levels of saturation may grow as new AI breakthroughs emerge. It is also possible that novel workflows and tasks will develop around evolving AI capabilities that will maintain a degree of complementarity between workers and AI systems.

\begin{table}[htbp]
    \centering
    \caption{Occupations With the Most/Least Task Saturation Among Those With Gemini Usage}
    \label{tab:table_2_2}
    \begin{tabularx}{\textwidth}{@{} L L @{}}
\toprule
\textmd{Least Saturated} & 
\textmd{Most Saturated} \\
\midrule
Special Education Teachers, Middle School & 
Market Research Analysts and Marketing Specialists \\[0.6em]

Kindergarten Teachers, Except Special Education & 
Human Resources Specialists \\[0.6em]

Special Education Teachers, Preschool & 
Software Quality Assurance Analysts and Testers \\[0.6em]

Midwives & 
Document Management Specialists \\[0.6em]

English Language and Literature Teachers, Postsecondary & 
Network and Computer Systems Administrators \\
\bottomrule
\end{tabularx}
    \vspace{0.2cm}
    \parbox{\textwidth}{\footnotesize \textit{Notes:} Gemini shares are derived from interactions aggregated across surfaces (App, AI Mode, API) classified into O*NET SOC occupations. Military Specific Occupations so those are omitted and residual detailed occupation categories with titles including ``All Other'' are excluded for consistency with other detailed occupation analysis in this section. Saturation is defined as the share of all O*NET tasks associated with a detailed occupation where we observe at least 25 Gemini users. Occupations where zero tasks meet this threshold are excluded.}
\end{table}

\subsection{Task characteristics}

We have seen that there is substantial variation in the extent to which an occupation's tasks are being performed with AI. Next, we explore the attributes of tasks which affect how much AI usage we observe in our data. We also attempt to detail the relationship between that AI usage and human work. It is critical to acknowledge that AI usage is not synonymous with automation (the substitution of human effort with machines for specific tasks). Instead, workers can also use AI tools for collaboration (the complementary enhancement of human productivity). Moreover, the type of tasks workers perform with AI will shape both the relevant labor market impacts and societal consequences of continued AI diffusion. If, for instance, AI enabled workers to offload mundane or routine tasks, their time may be freed up to pursue more creative or enjoyable tasks.

In order to better understand the distribution of AI task usage in our data, we use the prompt provided in \citet{NBERw33941} to map ONET tasks to five categories\footnote{The prompt used was taken from A.2 of the June 18 version of the working paper. Classification was performed with Gemini 3.1 Flash Lite.}: Routine Cognitive (RC), Routine Manual (RM), Non-Routine Cognitive Analytic (NC), Non-Routine Manual (NM), and Non-Routine Interpersonal (NI). Under this categorization, routine tasks are ones that are codifiable, meaning they can be fully specified through a set of structured instructions or rules. Manual tasks involve physical labor. Cognitive tasks are those that involve cognitive processes. And, interpersonal tasks are ones that require communication, human interaction, and social skills. This taxonomy is grounded in how the labor market literature has conceptualized digital technology shocks in the past. Later on, we'll discuss how what we observe with Gemini usage differs from the models put forward in that literature.

\begin{figure}[htbp]
    \centering
    \caption{Occupations Without Substantial Gemini Usage Have Greater Shares of Manual Tasks}
    \label{fig:figure_2_5}
    \includegraphics[width=\textwidth]{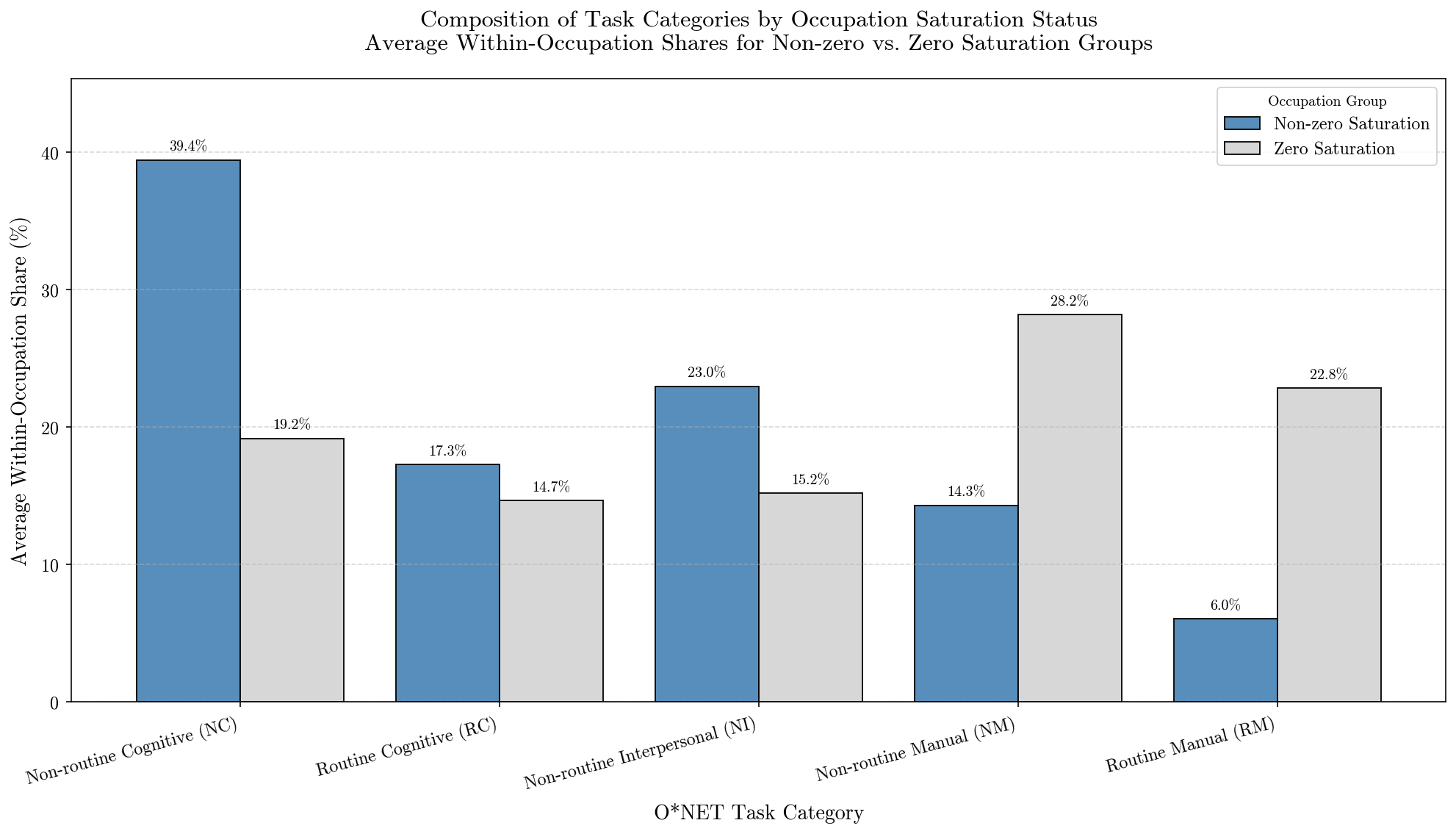}
    \vspace{0.2cm}
    \parbox{\textwidth}{\footnotesize \textit{Notes:} This figure uses the categories from \citet{NBERw33941}, assigning each O*NET task, covering occupations across the economy, to one of the five categories on the horizontal axis. It then compares the group of detailed O*NET occupations with zero task saturation (omitted from Figure \ref{fig:figure_2_4}) to the remaining occupations. We define a task as being saturated if we observe at least 25 unique users globally performing that task. An occupation falls in the zero saturation group if none of its component tasks meet this threshold. The shares here are the average of the within-occupation shares for the zero / non-zero task saturation groups.}
\end{figure}

In Figure~\ref{fig:figure_2_5}, we compare the group of detailed O*NET occupations (29\%) where we observed no tasks with sufficient usage across at least 25 users to the remaining occupations with non-zero saturation. What we observe is that the most salient feature on the extensive margin appears to be the physicality of the tasks in an occupation. Manual tasks represent higher shares of the constituent tasks in the occupations with zero task saturation. Cognitive and interpersonal tasks, especially non-routine cognitive tasks are more common in the occupations where we observe non-negligible AI usage across at least one of the component tasks.

As was described earlier, the literature has previously conceptualised digital technology shocks through the lens of skill-biased technological change (SBTC) and routine-biased technological change. In \citet{10.1162/003355303322552801} and \citet{Acemoglu2011}, traditional digitalisation waves diffused by automating routine cognitive and manual labour. This dynamic also meant enhanced productivity for a subset of skilled workers such that automation and augmentation were simultaneous phenomena. As digital technologies proliferated, they generally produced \textit{job polarization}, with rising demand for high-wage and low-wage labour, but a decline in the relative share of middle-wage work.

Previously non-routine cognitive work, which required abstract reasoning, problem-solving, and interpersonal communication, tended to experience digital technology complementarity and corresponding wage gains. The decline in the middle skill jobs, meanwhile, was driven by rising demand for highly skilled labor but also a shift to low skilled work. LLMs such as Gemini could present a potential discontinuity with this understanding of technological automation and augmentation dynamics. Current AI systems exhibit advanced capabilities in reasoning, synthesis, textual generation, and multimodal execution, allowing them to perform broadening swathes of non-routine cognitive tasks.

\begin{figure}[htbp]
    \centering
    \caption{Non-Routine Cognitive Tasks Are Over-Represented in Gemini Usage Relative to the O*NET Task Universe While Manual and Interpersonal Tasks Are Under-Represented}
    \includegraphics[width=\textwidth]{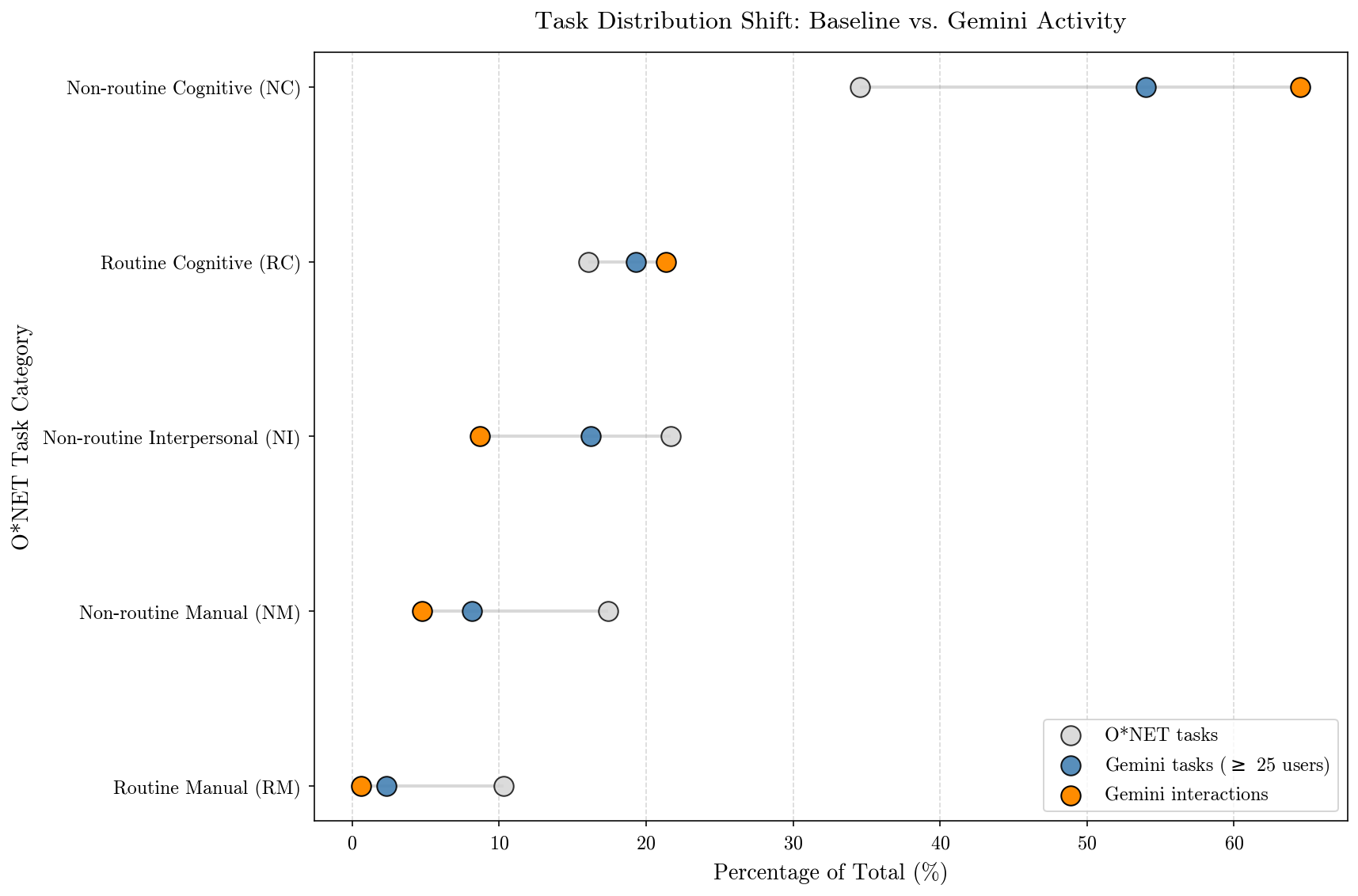}
    \label{fig:figure_2_6}
    \vspace{0.2cm}
    \parbox{\textwidth}{\footnotesize \textit{Notes:} This figure uses the categories from \citet{NBERw33941}, assigning each O*NET task, covering occupations across the economy, to one of the five categories on the vertical axis. The horizontal axis represents the shares of tasks in each of those categories for the relevant distribution, starting with the O*NET task universe. It then compares the O*NET taxonomy distribution to the representation of work-related tasks in the ATLAS data. Gemini shares are derived from interactions aggregated across surfaces (App, AI Mode, API) classified into O*NET task ids. The two Gemini data points for each category represent shares based on a) whether an O*NET task appears in the Gemini data with at least 25 users b) the number of interactions associated with each task.}
\end{figure}

On the intensive margin\footnote{As we observed earlier, there is a large majority of O*NET tasks where we do not observe substantial usage. For the remaining analyses, we focus on the intensive margin within this set of tasks with AI usage. Furthermore, note that this AI usage distribution across tasks is heavily skewed (with a skewness of greater than 16). The mean Gemini interaction volume at the mean is more than five times greater than the interaction volume at the median. At the 90th percentile, interaction volume is more than nine times greater than the median.}, we again rely on the \citet{NBERw33941} task categories applied to O*NET tasks to uncover the types of tasks where we observe more Gemini usage in our data. Our analysis unsurprisingly suggests a heavy concentration of Gemini interactions for cognitive domains. Cognitive tasks make up roughly half of the O*NET task taxonomy. Within the set of tasks where we observe non-negligible\footnote{At least 25 users.} Gemini usage, cognitive tasks---encompassing both routine and non-routine---account for 73\% of the total, a figure that increases to 86\% when considering Gemini interaction volume. In contrast, manual and interpersonal tasks are less represented in Gemini usage than in the O*NET taxonomy. Manual and interpersonal tasks make up 28\% and 22\% of the O*NET taxonomy, respectively, but comprise roughly 5\% and 9\% of the Gemini interaction volume.

We note particular over-representation of Gemini usage for non-routine cognitive analytic tasks. The \citet{NBERw33941} classifier identifies these tasks as those requiring analytical skills---such as judgment, strategic thinking, problem-solving, and creativity. These tasks are non-codifiable, meaning the rules for accomplishing them are not sufficiently well understood to be specified explicitly in computer code and executed by traditional machines. These are the tasks where AI unlocks potential that was challenging to realize with past paradigms of deterministic, rules-based computing. Non-routine cognitive analytic tasks comprise 35\% of all O*NET tasks but 65\% of Gemini interactions in our data.

While we noted earlier that occupations where we don't see tasks with substantial AI usage tend to have greater shares of manual tasks, we do observe some AI usage even within occupations with high manual task shares.\footnote{The median occupation's tasks are 12.5\% manual} This usage is often in tasks that are complementary to manual work. For example, roughly 44\% of Industrial Machinery Mechanics' tasks are manual. However, we observe thousands of AI uses for non-routine cognitive tasks such as analyzing test results and machine error messages.\footnote{O*NET task id 11821}

Other occupations that have above median shares of manual tasks and high AI usage for various tasks include electrical and electronics repairers (25\% manual), computer user support specialists (19\% manual), photographers (18\% manual), and shipping, receiving, and inventory clerks (18\% manual).

We do also observe some AI usage for manual tasks, especially the non-routine variety, in certain occupations. For example, over ten thousand conversations in our sample relate to automotive service technicians and mechanics (83\% manual) using AI for testing vehicle components and systems, rewiring systems, and inspecting parts for wear.\footnote{O*NET task ids 23522, 23545, and 23541} These occupations often also have higher multimodal usage. For example, automotive service technicians and mechanics have multimodal conversation shares more than 2 times higher than the average in work AI interactions. As we will see in the next section, the ways in which people use AI for manual tasks differs from their use for cognitive tasks.

\subsubsection*{Discussion: The Adoption Margin and Implications for Inequality}

The finding that AI usage in ATLAS is prevalent across both routine and non-routine work raises important questions on how the technology will impact wage inequality and the skill distribution. As illustrated in \citet{imas2026who}, while ``micro'' lab and field studies have largely found that those lower in the expertise distribution were able to catch up to experts, shrinking the expert premium in performance, studies looking at real-world ``macro'' evidence have largely found that AI has exacerbated differences between experts and non-experts.

A key friction explaining this ``micro-macro'' gap is the endogenous adoption margin, i.e. the fact that in practice, adoption and effective usage of AI are not randomly assigned but self-selected by workers and organizations \citep{imas2026who}. This margin leads to two potential scenarios for skill-biased impacts and wage inequality:

\begin{enumerate}
    \item \textbf{The ``Catch-up'' Scenario (Equalization)}: Under a scenario of broad, uniform adoption---subsidized by falling compute costs, expanded free tier, intuitive user interfaces, and widespread organizational training---AI could act as a skill equalizer. Controlled task-level experiments indeed suggest that when access and usage are held constant, less experienced or lower-performing workers benefit disproportionately from AI assistance, effectively compressing the performance and wage distribution \citep[e.g.,][]{Brynjolfsson2025, Noy2023, NBERw34851}.
    
    \item \textbf{The ``Run-away'' Scenario (Exacerbation)}: Alternatively, if adoption remains endogenous and highly concentrated, the benefits of AI may accrue primarily to those who are already advantaged. Observational data---including ATLAS data---reveals that real-world AI usage is strongly skewed toward higher-income regions (as we will see in Section \ref{sec:geo_diffusion}, and also echoed in previous studies like \citet{appelmccrorytamkin2025geoapi}), higher-skilled occupations, and often with senior or managerial workers \citep{anthropic2026aeiv5}. In this scenario, existing advantages are locked in and potentially amplified through several channels:
    \begin{itemize}
        \item \textit{Task Selection}: When task choice is left to the worker, lower-skilled workers may apply AI inappropriately, resulting in performance drops, whereas higher-skilled workers select tasks where the tool provides high utility \citep[e.g.,][]{Otis2026}.
        \item \textit{Complementary Judgment}: Effective utilization of AI in non-routine cognitive work requires high-level human capital inputs such as architectural judgment, error-detection, and code-reading ability to verify and integrate AI outputs \citep{beane2025human, paradis2024doesaiimpactdevelopment}.
        \item \textit{Seniority-Biased Demand}: If senior workers leverage AI more effectively, firms may substitute away from entry-level hiring while keeping senior employment stable, leading to a seniority-biased technological change that shifts career ladders and exacerbates wage premiums \citep{HosseiniMaasoum2025}.
    \end{itemize}
\end{enumerate}

Ultimately, whether AI functions as a force for decreased or increased inequality will determine whether we see wage compression or divergence as a result. And this will depend at least partly on institutional scaffolding, including the degree to which organizations make AI tooling available to both low- and high-skill workers, develop in-house innovations to make AI useful for their specific business needs, and invest in formal training programs, management support, and protected learning time to broaden effective adoption across the entire workforce.

\subsection{Task Intent}

We have observed that cognitive tasks, both routine and non-routine, are the most prominent uses of AI for work in our data and we observe smaller shares of interpersonal and manual task usage. However, the usage of Gemini for these tasks does not necessarily imply automation of those tasks. AI can be used to complement human capabilities, rather than wholly replacing them. The nature of the interaction between human work and AI usage can have significantly different labor market implications. Automation, with AI substituting for human work, may lead to reductions in wages or employment. If, on the other hand, human work and AI usage are complements and it enhances worker productivity, it may lead to wage gains. Historically, these complementing and automating impacts have occurred simultaneously. Moreover, the types of work that are automated or augmented also carry implications for worker satisfaction.

We cannot observe activity outside of Gemini so we cannot definitively say what portion of larger work, involving human effort, AI is being used to complete. However, we can attempt to discern the intent of users when interacting with Gemini. We developed a classifier to assign groups of conversations to the following categories: \textit{Task Automation}, \textit{Partial Drafting and Generation}, \textit{Review and Refinement}, \textit{Ideation and Strategy}, and \textit{Information Retrieval and Learning}.\footnote{\textit{Task Automation} is defined as the AI being asked to execute the core task or a major sub-task end-to-end. \textit{Partial Drafting and Generation} covers requests to generate a substantial draft, template, or component for performing the task but the final output would need to be substantially edited, populated with data, or integrated into the bigger whole for the assigned task to be considered completed (e.g. verifying specific transactions when the assigned task is to prepare tax returns). \textit{Review \& Refinement} covers AI being tasked \textit{primarily} with editing or checking human-produced documents (e.g. checking for consistency in a translation). \textit{Ideation and Strategy} is defined as AI acting as a thought partner (e.g. synthesizing multiple sources and discussing how it applies to a business task). Finally, \textit{Information Retrieval and Learning} involves seeking factual answers or tutorials related to performing a task.} As described earlier, de-identified conversations are summarized and then grouped into anonymized clusters. We then pass these aggregate summaries of these anonymized conversation clusters, accompanied by the occupation title and O*NET task statements assigned to each cluster, and instruct Gemini 3.1 Flash Lite to select a label that best fits the role the AI is asked to perform in the cluster summary in relation to the work task the user is performing. This is a challenging task, especially as the de-identified conversation clusters we currently use for this classification could nest multiple user intents, and this approach is a preliminary attempt to deduce the relationship between AI usage and human work within a task. We hope to refine our approach over time but we feel it is important to provide some insight on the extent to which different types of human work may be subject to full automation, compared to a broader range of more collaborative uses, in our snapshot of the data.

\begin{figure}[htbp]
    \centering
    \caption{Using Gemini for End-to-End Task Automation Is Rare Except for Routine Cognitive Tasks}
    \label{fig:figure_2_7}
    \includegraphics[width=\textwidth]{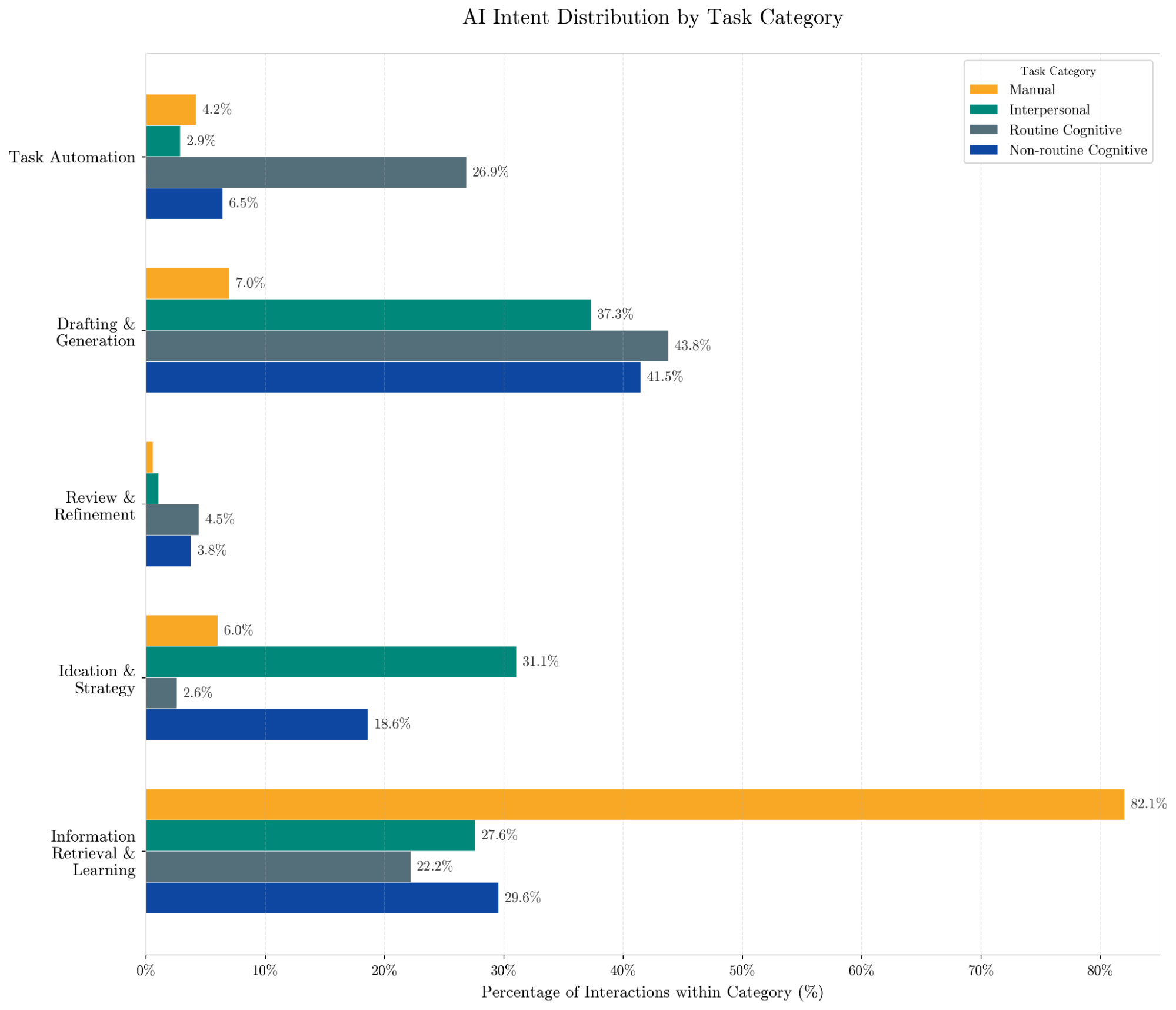}
    \vspace{0.2cm}
    \parbox{\textwidth}{\footnotesize \textit{Notes:} For de-identified conversation cluster summaries, Gemini is used to label the user intent to determine whether they are seeking to automate the task or complement human work occurring outside of Gemini. This chart depicts the shares of conversations assigned to the intent categories on the vertical axis for each of the task categories with different colored bars. The task categories are based on \citet{NBERw33941}. Routine and Non-routine Manual tasks have been combined into a single category.}
\end{figure}

Focusing on cognitive tasks, we find that while non-routine cognitive work is the most prevalent form of AI usage in our data, much of this work appears to be less focused on automation based on our classification (see Figure~\ref{fig:figure_2_7}). Instead, usage of Gemini for non-routine cognitive work is centered on Partial Drafting and Generation, Review and Refinement, Ideation and Strategy, and Information Retrieval and Learning. Automation represents the intent of less than 10\% of AI conversations in non-routine cognitive work appearing in our data. There is a pronounced contrast in comparison to the distribution observed for routine cognitive work. In this subset, more than a quarter of AI conversations in our data are targeting task automation. We reiterate that these findings should not be considered definitive, but we consider the differences between routine and non-routine cognitive work to be potentially meaningful.\footnote{While Anthropic uses only a broad categorization between ``augmentation'' and ``automation'', ``augmentation'' still represents the majority of interactions, accounting for 52\% of conversations, compared to 45\% for pure ``automation'' and has been failing over time \citep{anthropic2026aeiv4}.}

Interpersonal and manual tasks exhibit even less focus on automation based on our classification. Interpersonal tasks are spread across drafting and generation, ideation and strategy, and information retrieval and learning. Among the more common interpersonal tasks for which we observe Gemini usage are addressing employee relations issues, locating vendors of materials and interviewing them, and planning and conducting activities for a balanced program of instruction. Manual tasks for which we observe Gemini usage are overwhelmingly concentrated on information retrieval and learning. Common examples include installing and performing minor (computer) repairs, testing vehicles and their components, and setting up employee (computing) equipment.\footnote{We note that even in the case of observing higher rates of end-to-end automation in tasks, these do not necessarily equate job automation, as coordination costs, complementary tasks and organizational frictions are highly prevalent \citep{NBERw30177, NBERw34639}.}

\subsection{Expertise}

Another attribute of the tasks which affects how certain kinds of AI usage might affect workers' prospects is the tasks' expertise. \citet{NBERw33941} argue that the value of labor exposed to automation is dependent on whether the tasks automated are, loosely, supporting tasks for the occupation or core expert tasks. The expertise level of the occupation changes as more or less expert tasks are automated away, changing the value of the labor in that occupation. Specifically, \citet{NBERw33941} model occupations as bundles of distinct tasks and establish that technology has distinct market effects depending on whether it displaces an occupation's expert tasks or its inexpert supporting tasks. Automating more expert tasks erodes barriers to entry and depresses occupational wages, whereas automating inexpert tasks raises the scarcity of remaining human expertise, increasing wages but lowering employment.

Our measures of task expertise\footnote{See Appendix \ref{app:app1} and \ref{app:app_3} for the distribution of expertise scores} are based on an exact replication of the methodology in \citet{NBERw33941}. Expertise for each task is defined as 100 - the average Standard Frequency Index (SFI) of the lemmatized terms in its O*NET task statement, where SFI is an index combining word frequency and word entropy provided for 140,000 English language words in \citet{Zeno1995-sf}.\footnote{See Section 3.1 of \citet{NBERw33941} for a more detailed explanation of the steps in the calculation and adjustments performed to discipline outlier SFI scores.} The intuition for this methodology lies in the Efficient Coding Hypothesis (ECH)---the idea that speakers face a trade-off between the technical precision of their language and how commonly understood the words they use are. In this context, \textit{expert} words---i.e. those associated with specialized knowledge or skills---should be relatively rare in common usage but low entropy, meaning they predictably belong to a particular domain when observed.

\begin{figure}[htbp]
    \centering
    \caption{Gemini Usage Is Concentrated in Cognitive Tasks Across the Expertise Spectrum With Usage for Low-Expertise Non-Routine Cognitive Tasks Being Particularly Overrepresented}
    \label{fig:figure_2_8}
    \includegraphics[width=\textwidth]{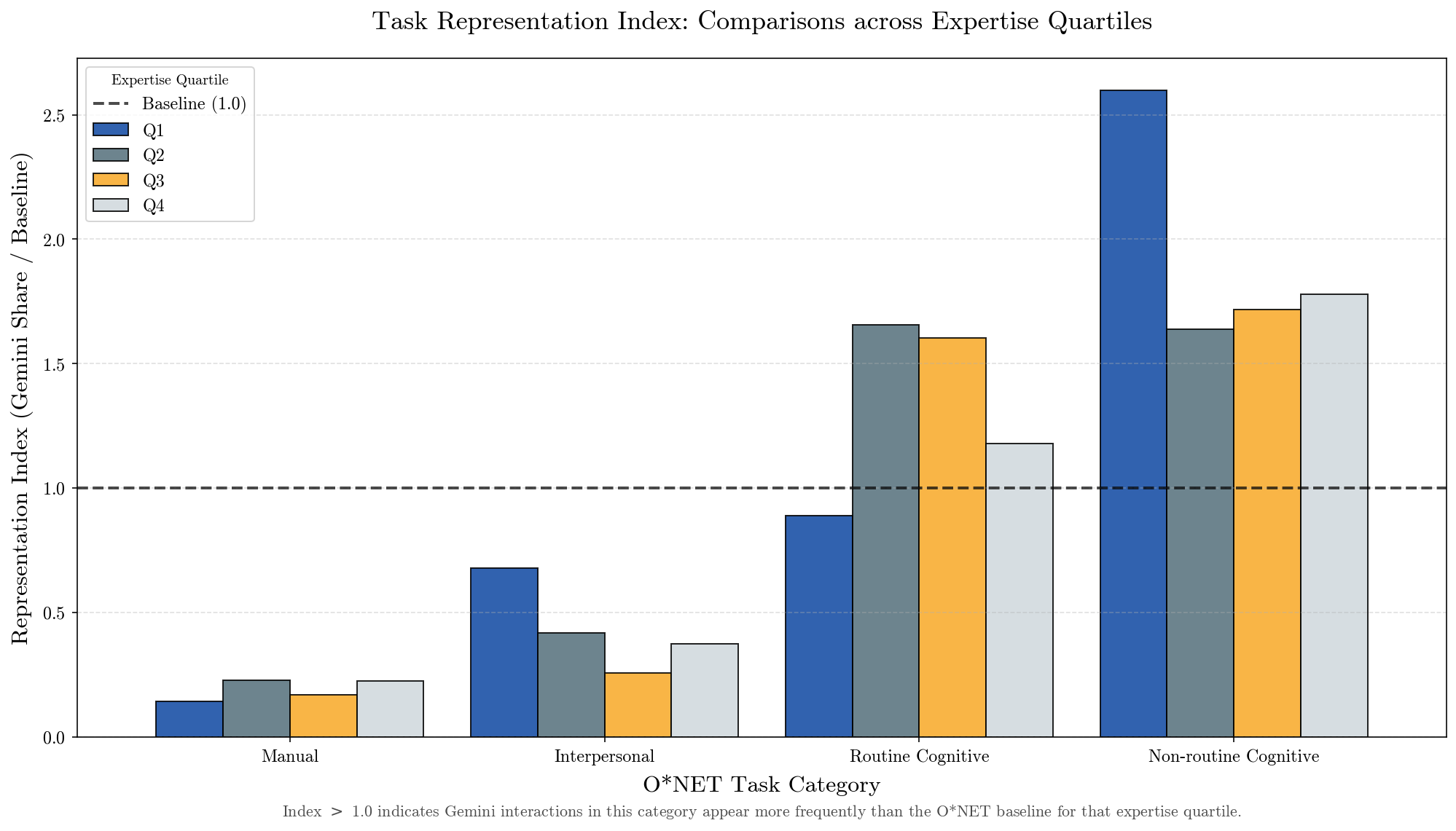}
    \vspace{0.2cm}
    \parbox{\textwidth}{\footnotesize \textit{Notes:} The baseline O*NET taxonomy is used to categorize tasks into expertise quartiles, where expertise scores are calculated based on the method in Section 3.1 of \citet{NBERw33941}. The task categories on the horizontal axis are also based on \citet{NBERw33941}. Routine and non-routine manual tasks have been combined into a single category. The bars in the figure represent the share of Gemini interactions assigned to the categories on the horizontal axis divided by the corresponding category share in the baseline O*NET taxonomy. This calculation is done separately for each expertise quartile.}
\end{figure}

Next, using the expertise measure based on expert word content, we identify quartiles of task expertise in the universe of nearly nineteen thousand O*NET tasks. Then, within each expertise quartile, we compute the share of Gemini work activity assigned to each task category (routine/non-routine cognitive, interpersonal, manual). We normalize these shares by dividing by the relevant task category's share in that expertise quartile of the O*NET task universe (see Figure~\ref{fig:figure_2_8}). As we observed earlier, AI usage is currently more prevalent in cognitive tasks and especially non-routine cognitive tasks. However, in our data, Gemini use is most over-represented relative to the O*NET taxonomy baseline task distribution, for the lowest expertise cognitive tasks, especially the non-routine variety. Common tasks of this type that we observe in our data include rewriting material such as news reports into specified languages and writing and reviewing product specifications for goods to be purchased.\footnote{O*NET tasks 9331 and 1157.}

More research is needed to determine the reasons for this pattern and its consequences for the labor market. These may be the tasks where the current marginal returns to AI usage are the highest given the current nature and performance of generative AI models. If model performance improves further, usage could increase for higher expertise tasks. If we observe greater technological progress in applications in physical domains (robotics, autonomous vehicles, etc.), manual tasks may also see greater AI usage. What is clear is that AI-related impacts will not be cabined to routine tasks. Further work is also needed to understand the entanglements of AI usage and skills, as well as their downstream labor market implications.

\subsection{Earnings and education}

We find that workers in occupations with higher median earnings and education levels are more intensive users of Gemini. We measure Gemini usage intensity as the total number of US conversations in our sample classified into a detailed occupation category, divided by the population employment of that occupation in OEWS data \citep{bls_oews_2024}. Again, we restrict our current analysis to the tasks and occupations we observe in our current sample of Gemini usage. In future work, we hope to explore these relationships in greater detail.

\begin{figure}[htbp]
    \centering
    \caption{Gemini Usage Is Higher in Occupations With Higher Median Earnings and Educational Attainment}
    \label{fig:figure_2_9}
    \includegraphics[width=\textwidth]{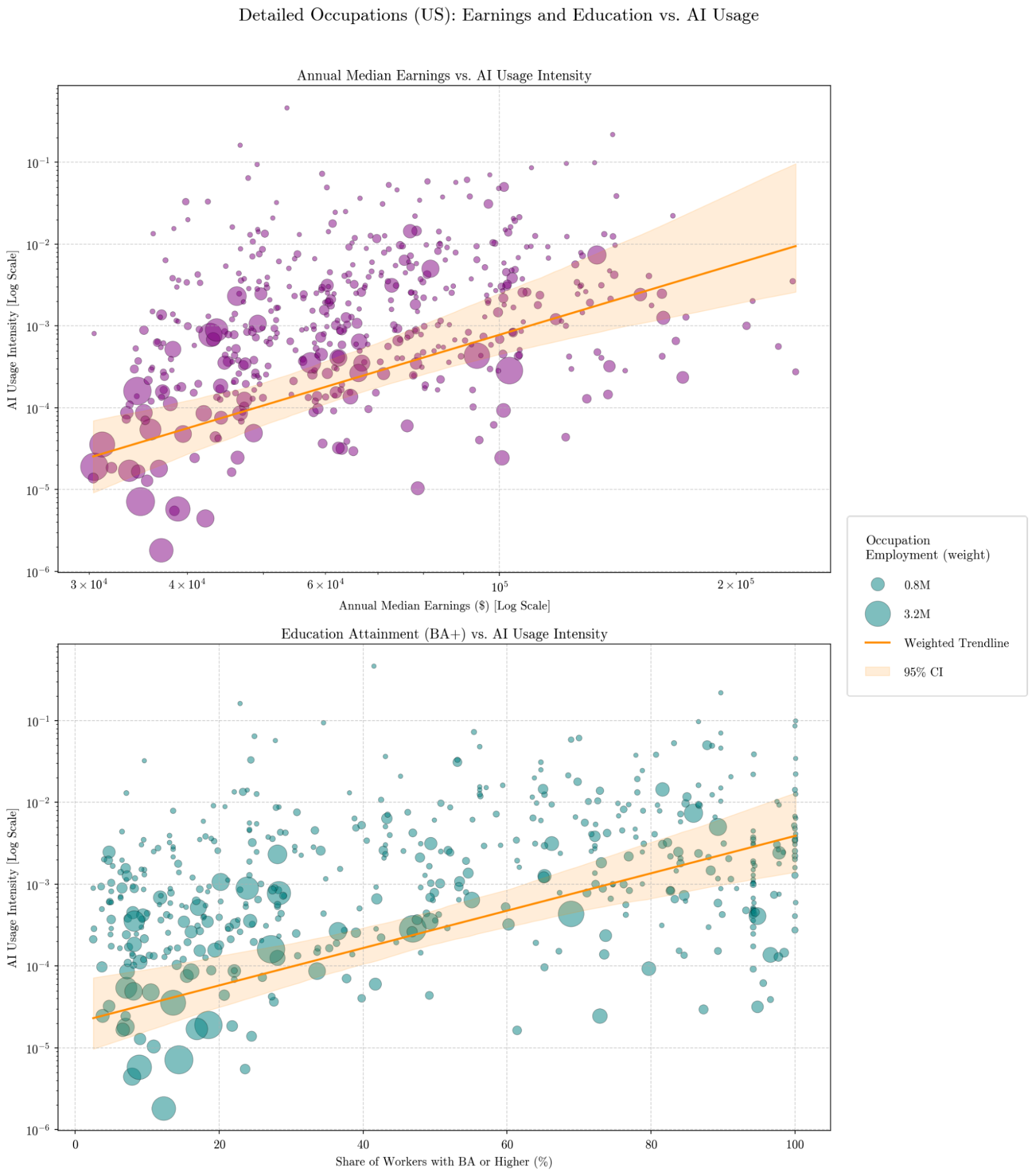}
    \vspace{0.2cm}
    \parbox{\textwidth}{\footnotesize \textit{Notes:} Gemini usage intensity is a measure of our observation of AI usage adjusted for the employment population associated with a given detailed occupation. Gemini usage intensity equals the total AI conversations for a detailed occupation in our sample divided by the total employment in that occupation in the OEWS 2024 data. That is also the source for occupation-level annual median earnings. OEWS data does not contain information on Military Specific Occupations so those are omitted. In addition, residual detailed occupation categories with titles including ``All Other'' are excluded due to uncertainty about the quality of classification into these labels and for consistency with other detailed occupation analysis in this section. The source for education is Table 5.3 from the BLS employment projections, which has the Educational attainment for workers 25 years and older by detailed occupation. \citep{bls_educ_attainment}}
\end{figure}

Figure~\ref{fig:figure_2_9} presents the relationship between AI usage intensity and two occupation level characteristics (where available): (1) the occupation level (log) median earnings (2) the share of an occupation's workers with a bachelor's degree or higher.\footnote{The source for earnings is the same 2024 OEWS data used for employment totals earlier. The source for education is Table 5.3 from the BLS employment projections, which has the Educational attainment for workers 25 years and older by detailed occupation. \citep{bls_educ_attainment}} Occupation level data is weighted by population employment. There are two main findings from this figure. First, we see AI usage intensity in our data being positively correlated with both occupational earnings and educational attainment. These trends survive in a regression including both covariates.\footnote{The coefficients (robust standard errors) from this regression are 1.86 (0.44) for log median earnings and 0.0076 (0.003) for the bachelor's plus share, indicating that a one percent increase in median earnings is associated with 1.86 percent higher AI usage intensity and that a one percentage point increase in occupational bachelor's plus share is associated with 0.76 percent higher AI usage intensity. The R2 value is 0.337. The corresponding coefficients on univariate regressions are 2.68 (0.38) for log median earnings and 0.017 (0.003) for the bachelor's plus share.} Second, we observe that the positive relationship is not uniform in either the median earnings or education figure. There is substantial dispersion in AI usage within occupations having the same median earnings or education levels highlighting the need for a nuanced understanding of these relationships.

Another way to conceptualize the skew in AI usage towards higher-income workers is to consider the following exercise. If one calculates the median annual earnings of US civilian workers\footnote{This is using the same sample of occupations from the regressions and charts related to earnings. It excludes the residual "All Other" detailed occupations. It is also restricted to the sample of occupations with non-zero Gemini usage to be able to compare the resulting numbers with different weightings. These numbers are all calculated by taking a weighted average with the referenced weights in each case.}, weighted by employment, it is \$62,252 (\$68,354 at the mean). If instead, one weighted this by Gemini conversations, the result would be \$82,919 (\$93,142 at the mean). Weighing by token usage extends this effect further, with the median weighted earnings then at \$86,157 (\$95,842 at the mean).\footnote{Relatedly, \citet{Bick2026Management} shows that firm management quality---specifically performance-focused personnel practices like merit-based rewards and active promotions (measured via the World Management Survey) are a powerful predictor of AI adoption. Interestingly, firm encouragement is the single biggest predictor of AI adoption. Nonetheless, this relationship may suffer from reverse causality, as AI adopting firms may outperform peers \citep{NBERw31222}.}

The distributional consequences of these findings will need to be understood in relation to the preliminary analyses presented earlier about the tasks people are performing with AI, and the nature of the relationship between that AI use and human work. Similar analyses are often read differently by different audiences. One audience might interpret the high concentration of AI use among highly educated, higher-wage white-collar workers as evidence that these professionals are actively automating themselves out of existence. Our preliminary evidence points to a different dynamic. If these workers are instead augmenting their work by automating routine cognitive tasks and simultaneously collaborating with AI in their performance of non-routine cognitive work, this points to greater returns to human skill in these non-routine dimensions. Ultimately, this could lead to a deepening of wage inequality between this augmented cohort and those less able to use AI in their work. As we carry out further research, we aim to develop a clearer picture of how these observations interact.

\section{Home Usage}
\label{sec:home_usage}
\subsection{The Broad Picture}

AI is diffusing fast, even when compared to other transformative technologies like personal computers, smartphones, or the internet  \citep{Bick2026, KalyaniQJE, sajadieh2026, curtomillet2026reflections}. One reason for this speed of diffusion is the high adoption in the household. In fact, generative AI may have more users in the household than in the workplace \citep{NBERw32966}.\footnote{This may be due to having lower initial set-up (fixed) costs than other technologies like the computer. By the same time in the life cycle, computers were up to 6--7 times more likely to be used in work than in non-work settings \citep{NBERw32966}.} Despite these trends, there's surprisingly little known about both how AI is used in the household as well as its potential impact on intra-household productivity.\footnote{While household activities are harder to quantify in economic and national account data, they are often economically (as well as socially) very valuable. Moreover, due to limited time budgets there is often a margin of substitution between household and market activities \citep{becker1965, cb816db9-bac3-3b58-9c7d-1751b51fc3f2, Coyle2018}. These margins of substitution may become even more important in the future, due to advances in e.g. domestic robotics products.}

In this section, we analyze AI usage patterns in the household. We start with about 10 million AI usage logs, coming from what we name ``conversational AI''. In practice, we restrict our sample to the Gemini App and Google AI Mode (AIM), excluding API logs, which are much more likely to contain sophisticated, disjoint work tasks that are generally not representative of home usage.

As described in Section \ref{sec:data_methods}, we map summarized clusters of user conversations to American Time Use Survey (ATUS) \citep{bls_atus_lexicon_2024} to enable analysis of AI use in a structured framework. The ATUS framework is uniquely suited for this task as it provides a comprehensive and standardized classification of the full spectrum of daily activities, encompassing both formal market work and non-work personal life. A randomly selected, representative sample of US respondents provides a sequential, chronological diary of a full 24-hour day. For each reported activity, the survey logs what specific activity was performed, when it occurred, where it took place, and who was present. These entries are then coded into a standardized, three-tiered hierarchical taxonomy. For instance, a respondent cleaning their kitchen is systematically categorized from broad to granular under Tier 1: Household Activities (Code 02), Tier 2: Housework (Code 0201), and Tier 3: Interior Cleaning (Code 020101). We sequentially match AI usage data to these three tiers of ATUS, creating, to our knowledge, the most comprehensive overview of how AI is used across daily non-paid work activities.\footnote{We exclude a very small share of the sample which is related to sleeping, as this was considered a task that is not AI enhanceable (despite a few ATLAS conversations on this topic). We also note that the first figure will include results on `work' category that includes conversations mapped directly from the upstream level of the work/non-work classification.}

Our data spans 74\% of all classified statistical categories in the ATUS taxonomy, covering activities that represent about 98\% of Americans' average daily non-sleep time usage.

\begin{figure}[htbp]
    \centering
    \caption{Over 86\% of Conversational AI Usage Occurs Outside of Formal Work, Dominated by Leisure, Education, and Household Activities}
    \label{fig:figure_3_1}
    \includegraphics[width=\textwidth]{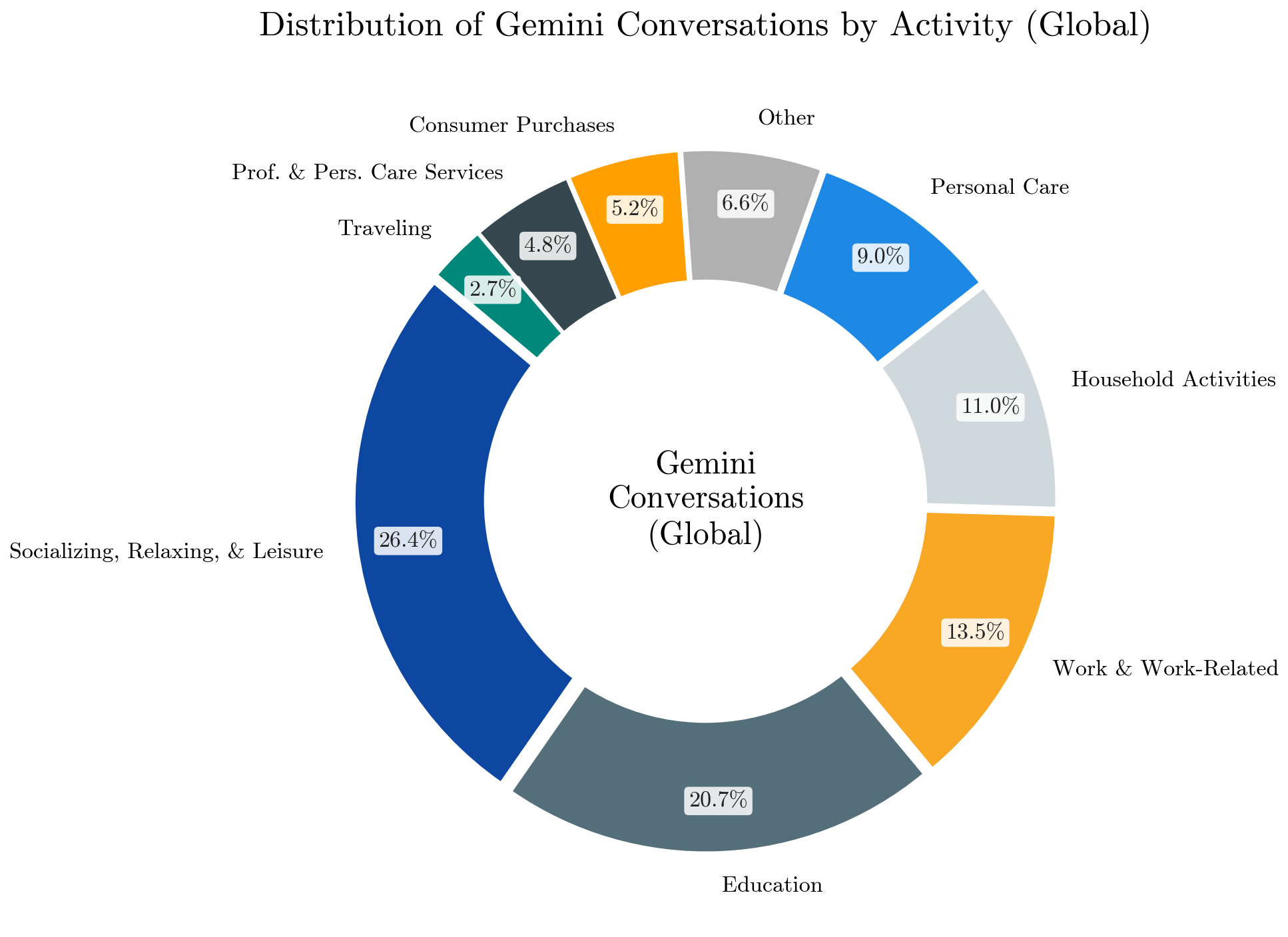}
    \vspace{0.2cm}
    \parbox{\textwidth}{\footnotesize \textit{Notes:} This figure shows the global share of conversational AI requests (AI Mode and Gemini App) across major life activities, classified according to the American Time Use Survey (ATUS) Tier 1 major activity categories (excluding sleeping). Work conversations shares is inclusive of the share that was considered ``work'' in the first step of our pipeline. Non-work interactions account for 86.5\% of all AI conversations, with nearly half concentrated in two domains: Socializing, Relaxing, \& Leisure (26.4\%) and Education (20.7\%). Household Activities (11\%), Personal Care (9\%), Consumer Purchases (5.2\%), and Professional \& Personal Care Services (4.8\%) account for approximately 30\% of total usage. Categories accounting for less than 2\% of total conversations are grouped under ``Other''.}
\end{figure}

Figure~\ref{fig:figure_3_1} shows the global distribution of activities by the most aggregate ATUS classification (tier 1). Non-work conversations overwhelmingly dominate, with ``Work \& Work-Related'' tasks having a share of 13.5\% of total conversations, meaning that over 86\% of AI conversations in our data occur in unpaid personal life, household production, and leisure activities.\footnote{While we only focus on conversational AI in this subsample, the striking domestic concentration also appears in other AI lab data. For example, in OpenAI's Signals data \citep{NBERw34255}, approximately 69\% of ChatGPT interactions occur in non-work, consumer contexts. Similarly, \cite{ipsos_google_2026_ai_survey} survey data shows that a lot of the top usages are personal in nature, including related to understanding or explaining new topics or to save time on tasks or projects.}

Nearly half of all global conversations are concentrated in two domains: ``Socializing, Relaxing, \& Leisure'' (26.4\%), that encompasses activities like playing computer and video games, watching TV and movies, reading, and social gatherings and ``Education'' (20.7\%), that includes activities like taking classes, learning and skill building. The next most common tasks are practical household management tasks comprising ``Household Activities'' (11\%), ``Personal Care'' (9\%), ``Consumer Purchases'' (5.2\%), and ``Professional \& Personal Care Services'' (4.8\%), making up about 30\% of usage. Conversely, highly physical or location-bound activities like ``Traveling'' (2.7\%), Leisure and Sports, or Telephone Calls (included in the Other category) represent a smaller portion of interactions. Nonetheless, while the smaller percentages reflect the comparative advantage of large language models in assisting with cognitive and bureaucratic tasks rather than direct physical execution, their presence does highlight that AI can serve as a complement or source of information for even predominantly physical tasks or location-bound tasks.

\subsection{Comparisons with Human Time Usage and Heterogeneity}

In which tasks do humans find AI most useful? An indication of the answer to this question lies in which tasks humans disproportionately delegate or ask AI to help. To make this comparison, in this section, we contrast the volume of AI conversations against the actual hours humans spend working in different task categories.\footnote{Importantly, we recognize that information retrieval about a task is not a substitute for physically executing it. Comparing these time-shares is not to make a claim about AI substituting task, but rather a method to control for baseline human attention.}

\begin{figure}[htbp]
    \centering
    \caption{AI Assistance Over-Indexes in Cognitive and Bureaucratic Tasks, and Under-Indexes in Physical and Location-Bound Activities}
    \label{fig:figure_3_2}
    \includegraphics[width=\textwidth]{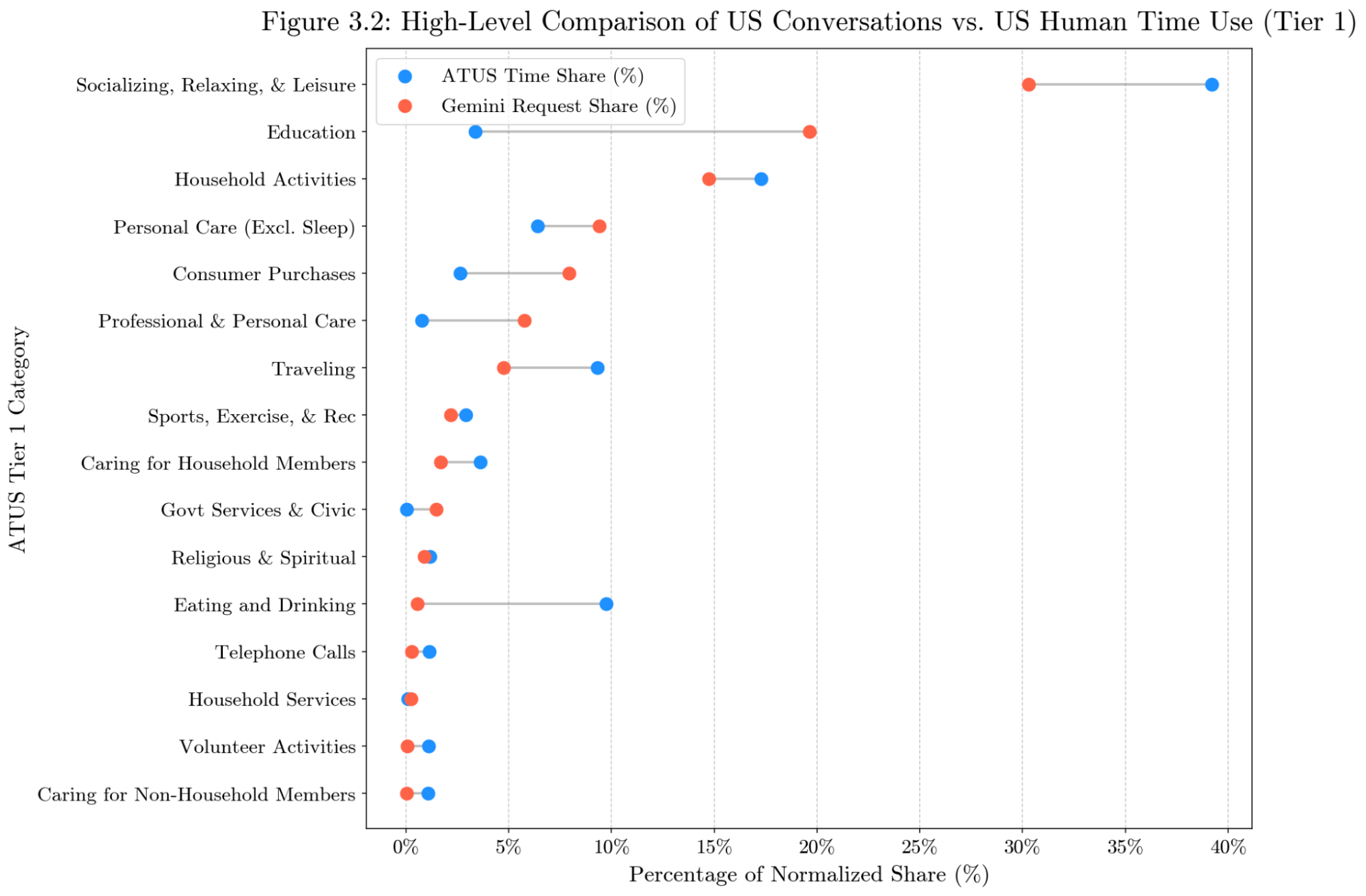}
    \vspace{0.2cm}
    \parbox{\textwidth}{\footnotesize \textit{Notes:} This figure compares the share of U.S. non-work AI conversations against actual human time allocation across ATUS Tier 1 major categories using dumbbell plots. Both shares are calculated out of total non-work active time (excluding work and sleep) to ensure direct comparability. Cognitive, informational, and administrative tasks such as Education, Consumer Purchases, and Professional \& Personal Care Services  are heavily over-represented in AI conversations relative to human time spent. Conversely, inherently physical, experiential, or location-bound activities like Eating and Drinking Traveling, Telephone Calls, Sports, Exercise, \& Recreation and Caring for Household and Non-Household Members are substantially under-represented.}
\end{figure}

Figure~\ref{fig:figure_3_2} compares the share of non-work ATLAS AI conversations in the US against the time Americans report spend on the same tasks in the American Time Use Survey (ATUS)\footnote{We are excluding sleeping and work activities, and are normalizing across ATUS Tier 1 categories. Details on data construction and the mapping process can be found in Appendix \ref{app:app1} and \ref{app:classifier_validation}.}. For a balanced comparison, we focus on the US AI usage, where the ATUS provides a highly granular, regularly updated standardized comparison set. We observe that ``Education'' is significantly overrepresented, with its AI conversation shares about 5.8 times greater than the actual time Americans allocate to it (despite ``Education'' usage being lower in the US than outside of the US). Similarly, ``Professional \& Personal Care Services'' over-indexes by more than seven-fold, and ``Consumer Purchases'' over-indexes by nearly three-fold. The biggest relative outlier is ``Government Services \& Civic Obligations'', which has almost a twenty-fold higher representation in AI usage than human activities.

Conversely, activities that are inherently physical, experiential, or location-bound are under-represented.\footnote{We test this more formally in Appendix \ref{app:household_section}. We find that the raw mean of the over-representation ratio (measuring the ratio between AI conversation shares and actual time usage) is about twice as high for non-physical than physical tasks, while the median is just above two thirds bigger.} The starkest negative gap is found in ``Eating and Drinking,'' which is about 18 times under-represented. Similarly, ``Traveling'', ``Sports, Exercise, \& Recreation'', and ``Caring for Household Members'' show negative gaps.\footnote{We note that a custom classifier found that roughly 82\% of ATUS Tier 3 activities could be classified as non-automatable given current capabilities. Our findings reveal a substantial volume of AI engagement within these domains.} Although non-physical tasks exhibit higher AI representation ratios on average, it is remarkable that physical activities still capture a massive aggregate share of total chatbot consultations. This indicates that user adoption is driven not solely by the demand for total task substitution, but by the demand for complementary decision and information support during physical execution. Categories like ``Socializing, Relaxing, \& Leisure'' (the leading category in terms of AI usage) and ``Household Activities'' are relatively close to parity when compared to actual time spent.

\begin{figure}[htbp]
    \centering
    \caption{Human Time Allocation Strongly Predicts AI Conversation Volume Across Daily Activities}
    \label{fig:figure_3_3}
    \includegraphics[width=\textwidth]{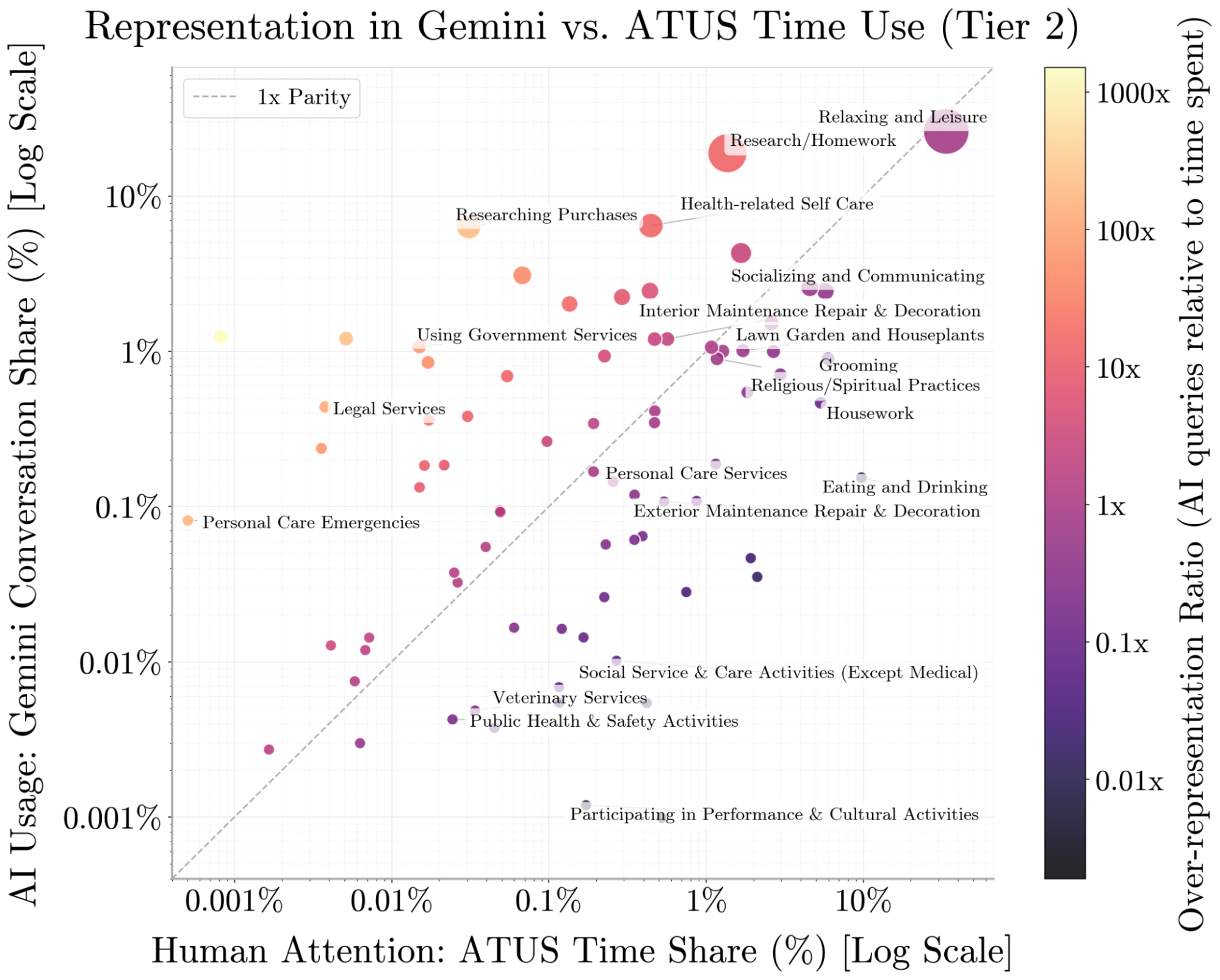}
    \vspace{0.2cm}
    \parbox{\textwidth}{\footnotesize \textit{Notes:} This scatter plot illustrates the relationship between the share of human time allocation (x-axis) and the share of U.S. AI conversations (y-axis) across ATUS Tier 2 (intermediate-level) activity categories, plotted on logarithmic scales to accommodate broad variations in category size. Each point represents an individual Tier 2 task category (excluding sleeping, working, and uncoded activities). The solid 45-degree line represents exact parity, where an activity captures the same share of AI requests as human time. Points above the 45-degree line indicate activities that are over-represented in AI usage relative to human time allocation (e.g., education, financial/legal/civic services), while points below the line indicate under-represented activities (e.g., physical grooming, dining). Overall, the positive upward slope demonstrates that human daily time use is a primary predictor of AI conversations.}
\end{figure}

In Figure~\ref{fig:figure_3_3} we go one step down on the level of aggregation, and provide a visualization of the relationship between US AI conversations shares and ATUS time usage at the tier 2 (intermediate) level. We plot this relationship in log-transformed shares, to enable a visual representation that is not perturbed by outliers. We observe a general positive correlation between human time use and AI conversation shares. For a more disaggregated view, in Appendix Figure \ref{fig:tier3-correlations} we present the results at the tier 3 (minor) level, where we similarly observe a positive correlation between time usage and AI conversations.

In Table~\ref{tab:table_3_1}, we see a few (more granular) categories that are over-represented and under-represented in our data. The most over-represented tasks are in education, comparison shopping, financial services, and legal and social services. More specifically, Research/Homework, Researching Purchases, Health-related Self Care, Financial Services, hobbies such as writing and conversations about Computer Use for Leisure are the most over-represented, while the most under-represented (but still present in our data) are Television and Movies, Eating and Drinking, as well as Washing, Grooming and Cleaning.\footnote{More details can be found in Appendix Figure \ref{fig:abs-differences-representation} that has the most over- and under- represented categories by the absolute percentage change differences.}

\begin{figure}[htbp]
    \centering
    \caption{Regardless of the Activities Granularity Level, Time Allocation Explains a High Percentage of the Variance in AI Conversations}
    \label{fig:figure_3_4}
    \includegraphics[width=\textwidth]{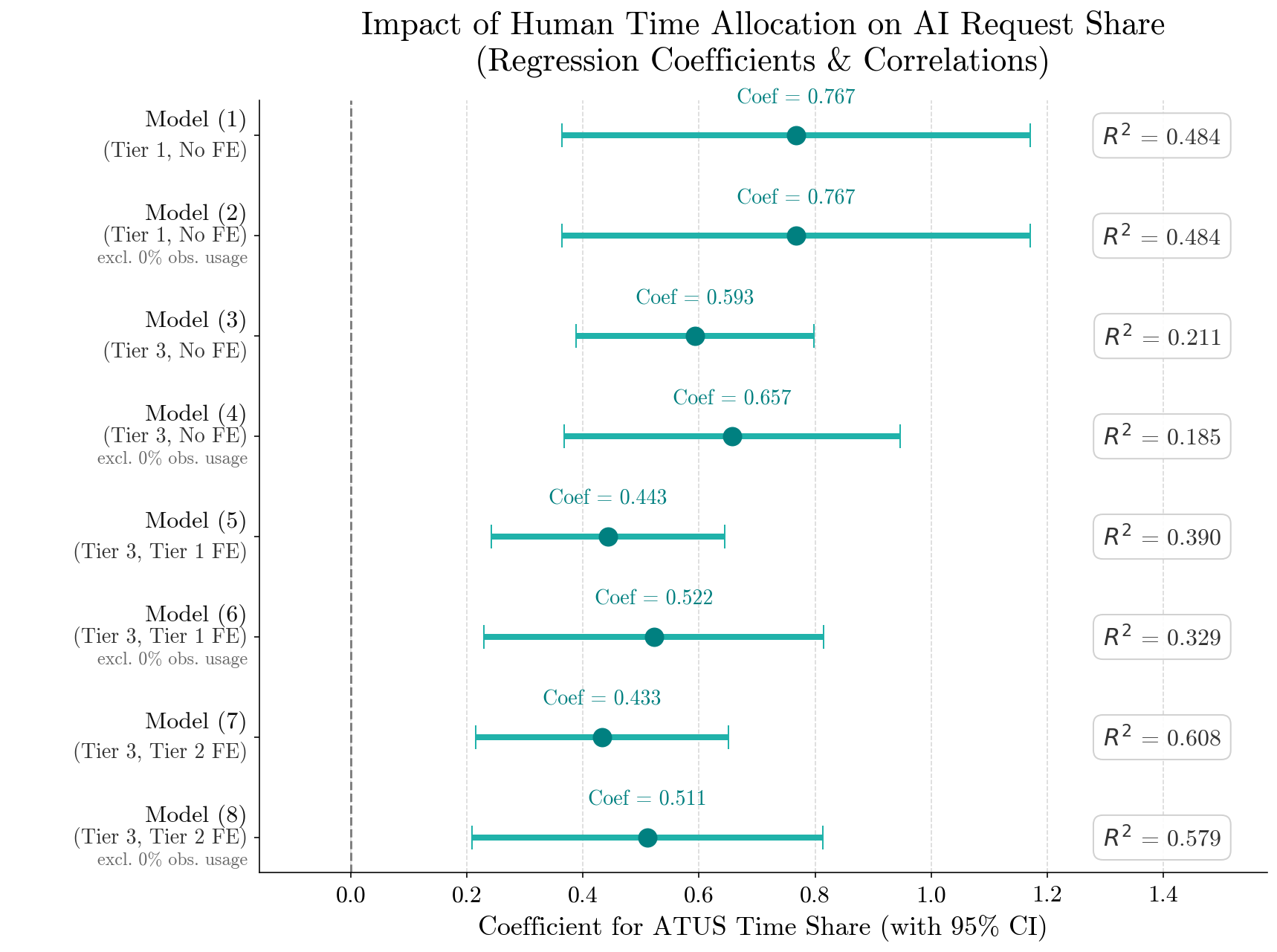}
    \vspace{0.2cm}
    \parbox{\textwidth}{\footnotesize \textit{Notes:} This figure displays linear regression estimates (with robust standard errors) quantifying the positive relationship between human time use shares and AI conversation shares across ATUS task categories (excluding work and sleeping). Across both broad (Tier 1/2) and granular (Tier 3) activity levels, human time allocation strongly and statistically significantly predicts where AI is utilized. At the most aggregate level, human time use explains nearly 50\% of the variance in AI conversations shares (slope = 0.77), and remains highly predictive even at the most granular task level. These across-the-board correlations seem to suggest that conversational AI operates as a general-purpose household technology embedded across the full spectrum of daily life rather than a niche tool confined to specific workflows.}
\end{figure}

For a more technical view, Figure~\ref{fig:figure_3_4} presents the formal regression estimates of the relationship between the US share of AI conversations with actual US time allocation across ATUS categories. Across specifications, human time allocation strongly and positively predicts AI conversation shares. At the most aggregated level, AI conversation shares explains nearly 50\% of the variance in human time use, yielding a slope of 0.77. At the most granular level, this relationship remains robust, with slopes higher than 0.4, and explaining about 20\% or more of the variance in American time usage. This relationship suggests that AI can be considered a general purpose household technology, with use cases that transcend just a few time task categories.

\begin{table}[htbp]
    \centering
    \caption{Specific Tasks Most Over-Represented in AI Usage vs. Most Under-Represented Relative to Human Time Use}
    \label{tab:table_3_1}
    \begin{tabularx}{\textwidth}{@{} L L @{}}
\toprule
\textmd{Most Over-represented} & 
\textmd{Most Under-represented} \\
\midrule
Research/homework for class for personal interest & 
Television and movies (not religious) \\[0.6em]

Research/homework for class for degree, certification, or licensure & 
Eating and drinking \\[0.6em]

Health-related self care & 
Washing, dressing and grooming oneself \\[0.6em]

Hobbies, except arts \& crafts and collecting & 
Interior cleaning \\[0.6em]

Comparison shopping & 
Relaxing, thinking \\[0.6em]

Using other financial services & 
Socializing and communicating with others \\[0.6em]

Writing for personal interest & 
Food and drink preparation \\[0.6em]

Computer use for leisure (exc. Games) & 
Taking class for degree, certification, or licensure \\[0.6em]

Appliance, tool, and toy set-up, repair, \& maintenance (by self) & 
Laundry \\[0.6em]

Vehicle repair and maintenance (by self) & 
Reading for personal interest \\
\bottomrule
\end{tabularx}
    \vspace{0.2cm}
    \parbox{\textwidth}{\footnotesize \textit{Notes:} This table lists the top 10 most over-represented and top 10 most under-represented detailed daily activities (ATUS Tier 3 minor categories) in the United States, ranked by absolute percentage point difference (Gemini conversation share minus ATUS human time share). Over-represented tasks are predominantly cognitive, research-intensive, or administrative (e.g., academic homework, comparison shopping, health/financial research). Under-represented tasks consist of routine physical, or location-specific chores (e.g., watching television, eating/drinking, personal hygiene, interior cleaning, laundry). Residual catch-all categories ("not elsewhere classified" / n.e.c.), travel codes, and unobserved niche tasks are excluded.}
\end{table}

\subsection{Non-Market Welfare: Valuing Household Time Savings}

But do humans find value in these AI conversations? And if they do, can we roughly quantify it? Accounting for the economic benefits of tasks done in the household is a hard task that economists have long thought about, dating at least to \citet{kuznets1934national}. Non-market activities are not part of the GDP \citep{d30fe982-b2ca-36d9-afa7-fe11e723435a, https://doi.org/10.17863/cam.84770, Ramey_2009}, and if such activities replace activities done outside of the household they can even decrease what GDP measures \citep{becker1965}.

To tackle this question, we narrow down our analysis to particular types of daily activities that may be more likely to produce economic value.

\begin{figure}[htbp]
    \centering
    \caption{Conversational AI is Used Across a Wide Set of Productive Household Tasks and Is Particularly Common in Selected Professional, Government Services and Civic Obligations}
    \label{fig:figure_3_5}
    \includegraphics[width=\textwidth]{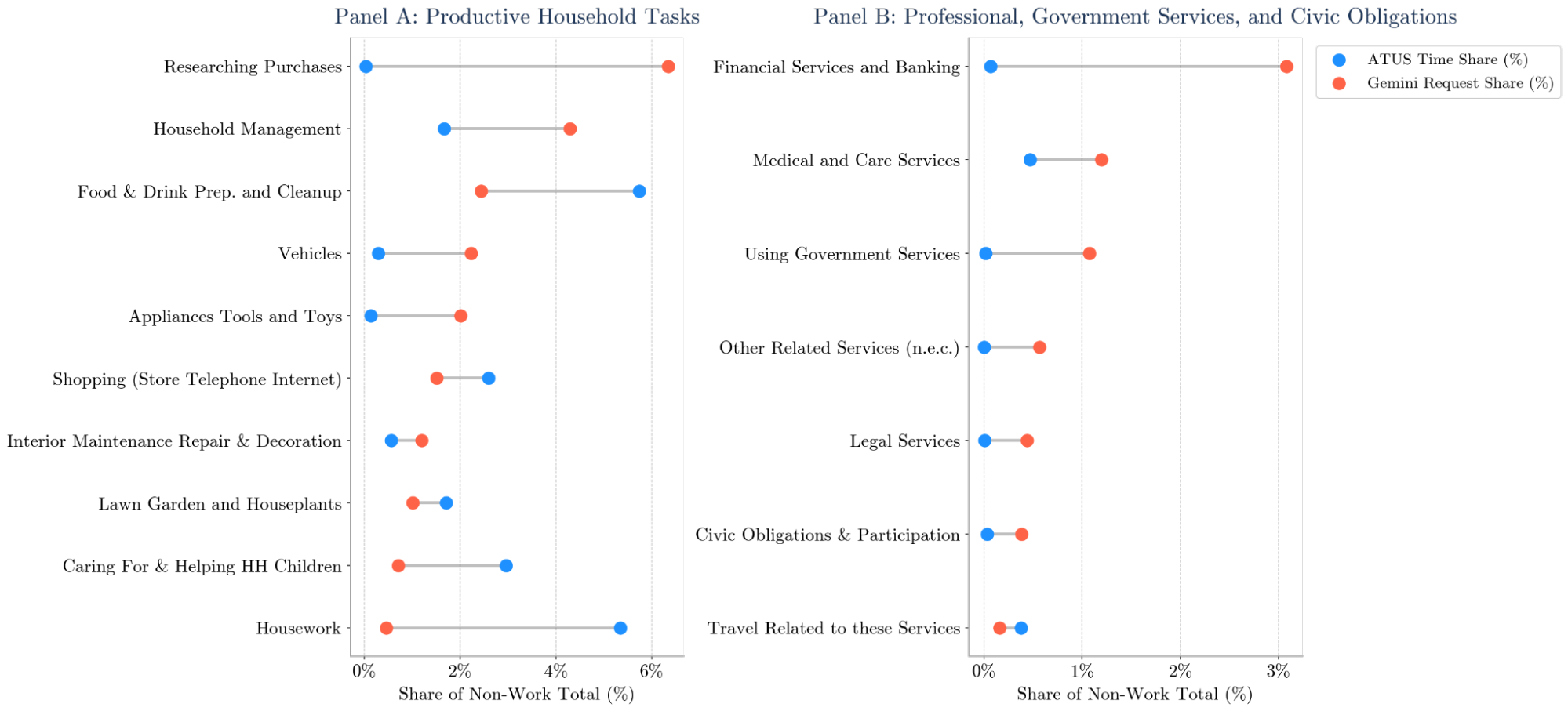}
    \vspace{0.2cm}
    \parbox{\textwidth}{\footnotesize \textit{Notes:} This figure compares U.S. AI conversation shares with ATUS human time allocation across two sets of economic activities using dumbbell plots. All shares are calculated out of total non-work active time (excluding work and sleep). Panel A (Productive Household Tasks) displays top activities meeting the economic "third-person criterion", tasks you could theoretically pay a third party to perform, such as cooking, cleaning, home maintenance, shopping, vehicles services and childcare. While humans spend more total hours on physical execution (cooking, housework), AI usage is heavily concentrated in the cognitive organizational and planning stages, led by researching purchases and household management (budgeting, bill paying, grocery lists). Panel B (Professional, Government Services and Civic Obligations) examines a set of domains pertaining to Medical Care, Financial Services, Legal Services, and Government Services \& Civic Obligations. Across these services, AI usage vastly exceeds human time shares, highlighting how households leverage AI in these types of important, but rarer high-friction services.}
\end{figure}

In Figure~\ref{fig:figure_3_5}, we present a two-panel comparison of AI conversation shares and ATUS time allocation at the Tier 2 level, using dumbbell plots to highlight the gap between where humans spend their time and where they direct their AI assistance. Panel A focuses on ``productive household tasks''---activities that satisfy the ``third-person criterion'' \citep{reid1934economics}. This economic concept defines a household activity as ``productive'' if you could theoretically pay a third person to do it for you (such as cooking, cleaning, home maintenance, shopping, or childcare), as opposed to personal care or leisure (like sleeping or watching a movie) which cannot be delegated. Panel A shows that the top two categories in terms of Gemini conversation shares are researching purchases and household management, which includes activities from household budget planning and paying bills to making grocery lists. While the physical execution of chores like food preparation and house cleaning is higher in ATUS than in AI conversations, AI usage is still significant in many of these categories. Looking systematically into conversation clusters, AI conversations in these categories seem to be highly concentrated in the cognitive planning and organizational phases. This suggests that AI may help users optimize and streamline key activities in household production.\footnote{Relatedly, concurrent work by \cite{NBERw34255} finds that ChatGPT provides economic value through 'decision support,' particularly in knowledge-intensive contexts, with 'Practical Guidance' and 'Seeking Information' ranking among the most common conversation topics.}

Panel B zooms in on a selected set of intermediate ATUS categories, in selected Professional, Government Services and Civic Obligations. These are cognitive or administrative tasks that represent significant transaction costs for households, such as Medical Care, Financial Services, Legal Services, Government Services, and Civic Obligations. Across all these domains, we see substantial AI usage. AI conversations shares in these domains are generally much higher than the actual time humans spend on them. Historically, accessing these services required individuals to often substitute away from market labor during standard business hours.

To illustrate this point further, we provide an even more detailed Tier 3 (most granular) breakdown of Government and Civic Tasks in the Appendix Figure \ref{fig:gvt-tasks-decomp}, which shows that the largest share of usage involves obtaining licenses and paying taxes, fines, or fees. This is followed by civic obligations (like voting, jury duty, or local council meetings) and questions regarding social services. Appendix Figure \ref{fig:gvt-tasks-word-map} shows that the most prevalent terms in summaries of clusters of these conversations, often have terms such as ``requirements'', ``compliance'', and ``procedures''. Just as expected from the granular ATUS classification, overarching themes center on around law \& police, social services, immigration \& visas, and taxes, indicating that users are often turning to AI to help with complex bureaucratic processes. In Appendix Figure \ref{fig:tasks-working-hours} we try to dig deeper into one possible reason for this over-representation, mainly that AI is metaphorically keeping government or professional services offices open around the clock. More specifically, we analyze the temporal distribution of AI queries across four domains: Medical Care \& Services, Financial Services \& Banking, Legal Services, and Government \& Civic Duties. Historically, accessing medical, legal, financial, or government services required an individual to substitute away from formal market labor during standard business hours, potentially creating a shadow cost of lost wages or lost workplace productivity. We find that around half of these bureaucracy or government-related conversations happen outside of standard working hours, allowing users to get tailored advice on nights, early mornings and weekends. The shares of conversations outside of these working hours are similar for the other categories analyzed displayed in Figure \ref{fig:figure_3_5}.\footnote{This is in line with new evidence from \cite{anthropic2026aeiv6}.}

In the final part of this subsection, we attempt to provide a range of the possible aggregate economic benefits of AI use to households. We start with the observation that the U.S. Bureau of Economic Analysis (BEA) spends considerable time putting together estimates of non-market activities for an accurate representation of historical economic development \citep{bridgman2022accounting, 10.1257/pandp.20231105}. In line with previous research and work from the BEA, we focus on several categories of `quantifiable productive household work': cooking, shopping, housework, childcare, gardening, odd jobs, and maintenance. This distinction is what allows economists to assign a market wage replacement value to non-market time, thereby omitting the potential welfare gains provided by AI in other activities like leisure. According to ATUS responses in 2024, an average American spends about 17.8 hours per week on ``productive'' household tasks as we defined them above.

Because there is no definitive historical or causal data on exactly how much time AI saves in the home, the analysis relies on a range of scenarios of time savings. While there is a fast growing literature on the effects of AI on productivity and time savings,\footnote{A growing (mostly experimental) body of evidence shows that generative AI significantly reduces task completion time; for instance, \cite{Noy2023} find a 40\% decrease in time taken for professional writing tasks, while \cite{DellAcqua2026} find that management consultants finish complex tasks about 25\% faster when they have AI access. Our own previous analysis \citep{strand_ai_at_work} showed that users with access to an AI form-creation tool spent 27\% less time editing their forms and for a text-heavy tasks sample, AI assistance reduced completion time by up to 32\%. \cite{Brynjolfsson2025} find that customer support agents using AI resolved issues about 15\% faster on average, while \cite{ChenStratton2026} find that software engineers code almost 9\% faster due to AI. There are few estimates for household tasks. \cite{blank2026householdimpactgenerativeai} use household internet browsing data from US home devices to find that household AI adoption may translate to an effective time savings of roughly 43\%--64\%, but these estimates may differ for non-browsing activities. Nonetheless, the continued advancement of generative and multimodal AI capabilities suggests that these efficiency gains will likely increase over time.} there are very few papers looking specifically at the household economy. \citet{blank2026householdimpactgenerativeai} estimate about a 43-64\% time saving for some productive household activities like browsing for shopping, while \citet{HERTOG2023122443} estimate in the near future that ``50--60\% of the total time spent on unpaid domestic work could be saved through automation'', but their measure also includes potential time savings from robotics, unrelated to current LLM capabilities. Clearly, these activities are not necessarily representative sets of tasks for the broader set of activities we measure in ATLAS, so we use a much more conservative range of averages between 0.5-5\% of time savings from using AI.\footnote{We are also noting that we do not assume that AI is physically performing tasks (e.g., maintenance, cooking, or housework); rather, the underlying assumption is that using AI to plan, research, or manage these tasks yields efficiency gains that translate into direct or indirect time savings.} Before presenting any calculations, we also note that assigning time savings is a flawed (but we hope useful) metric, as many improvements AI may bring at home will likely be manifested in the quality and breadth of household activities, rather than less time to achieve the same outputs.

Valuing time at the BEA’s conservative \$12.02/hour replacement wage—generates considerable economic value.\footnote{We note that, to be conservative, we did not adjust this figure for the last 5 years of inflation, and also did not consider a higher rate, despite the fact that AI seems to be helping with a set of tasks that are more specialized, such as legal and tax help, which would command higher wages.} An average 0.5\% time saving (approx. 5.4 minutes/week) yields \$56 annually per person, scaling to \$14.9 billion across the U.S. adult population. An average 2\% saving creates just below \$60 billion in US economic value, while the more optimistic average 5\% upper bound would create about \$149 billion. Further research is needed to confirm the extent to which individuals derive financial or time benefits from AI usage inside of the household, and measure potential broader welfare gains, arising, for instance, from better decisions about health, schooling or government services. However, we note that our initial estimates are not completely out of line compared with survey measures of AI benefits for consumers. For example, \citet{Nguyen2026} estimates about \$172 billion consumer surplus from AI usage in the beginning of 2026.

Because these substantial productivity gains occur entirely on the unpaid side of the production boundary, they are completely omitted from traditional GDP metrics \citep{10.1257/mac.20210319}. Nonetheless, they represent a meaningful expansion of household welfare and a reduction in the ``shadow time tax'' of daily life management. We also note that time savings from household AI use are unlikely to be evenly distributed given both the historical \citep{England2002} and current \citep{Milkie2025} gender composition in housework activities. According to ATUS data, men spent in recent years as low as 30\% less time per day doing what we defined as ``productive household activities'' compared to women, suggesting that the potential upside for women is considerably higher. Similarly, previous work found that technology-driven household savings may accrue more to women than men \citep{HERTOG2023122443}. However, this distributional effect could be reduced or even reversed unless the gender gaps in AI adoption \citep[e.g.,][]{cranney2026global, ALDASORO2024111814, educsci14121363} shrink.

\begin{figure}[htbp]
    \centering
    \caption{Across Countries, Higher Income Correlates With More Productive Household AI Use, While Developing Nations Concentrate Heavily on Education}
    \label{fig:figure_3_6}
    \includegraphics[width=\textwidth]{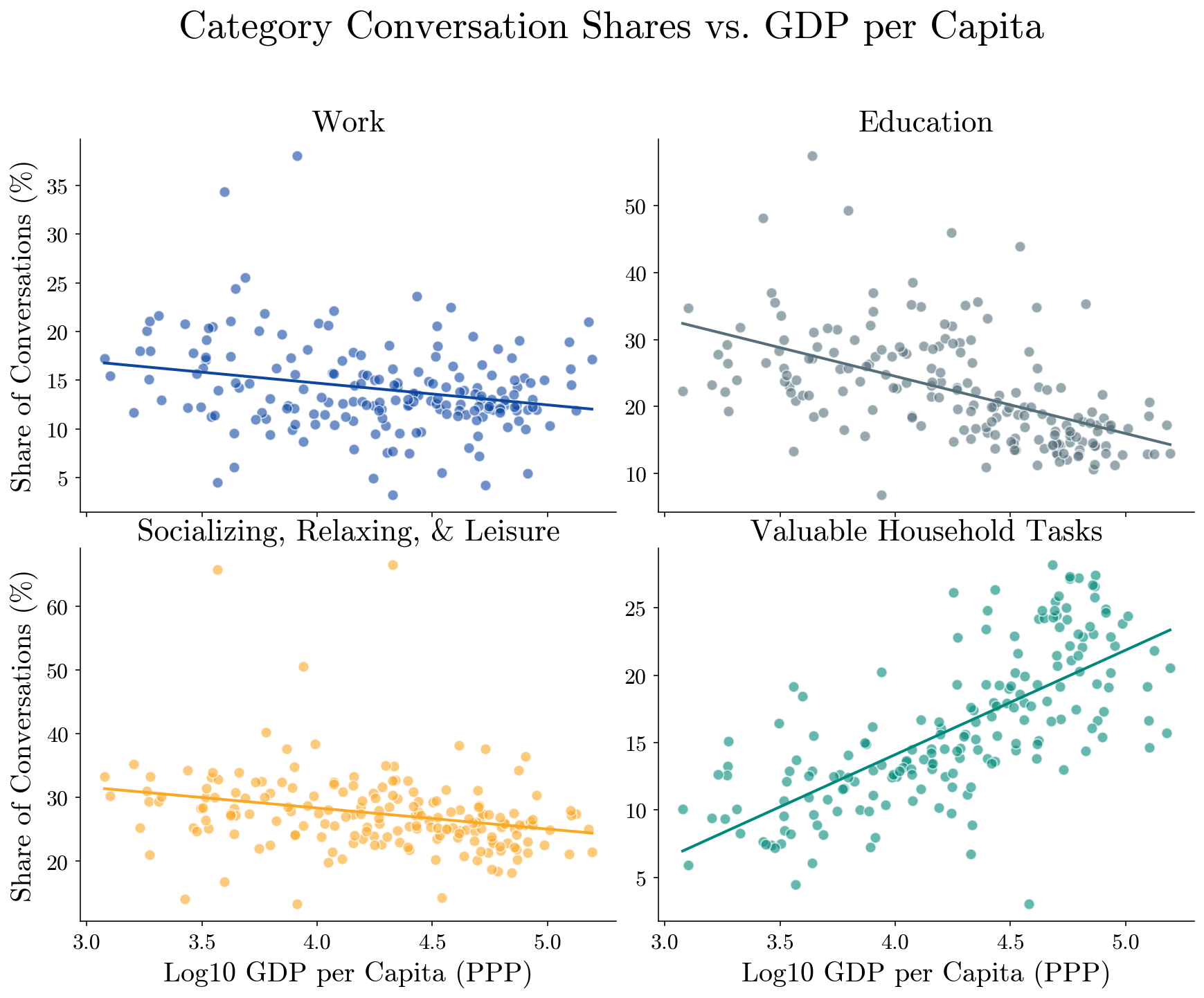}
    \vspace{0.2cm}
    \parbox{\textwidth}{\footnotesize \textit{Notes:} This figure presents country-level scatter plots and linear macro trend lines exploring how national income (Purchasing Power Parity GDP per capita, log-transformed) relates to the distribution of conversational AI usage across four core life domains: Work \& Work-Related, Valuable Household Tasks, Education, and Socializing, Relaxing, \& Leisure. Each data point represents an individual country or territory. Higher national income is statistically associated with a lower share of work-related AI conversations and a significantly higher share of productive household task assistance. Notably, while Education represents a large share of AI use across all economies, it captures an even more dominant share of AI conversations in lower- and middle-income developing countries, highlighting AI's role in expanding educational access and learning support globally.}
\end{figure}

Finally, we examine whether these benefits are likely to be evenly distributed in terms of geographic regions. To help inform this question, in Figure~\ref{fig:figure_3_6}, we show there is considerable geographic variation in the observed shares of different household activities. More specifically, we analyze the correlation between a country's income level and the composition of its AI usage in ATLAS data. We observe a somewhat negative (but statistically significant) correlation between national income and the share of AI conversations dedicated to Work \& Work-Related activities\footnote{This seems to be corroborated by survey data. In \cite{ipsos_google_2026_ai_survey} in emerging markets, the proportion of AI users who employ the technology specifically to assist with their work is exceptionally high, led by Nigeria (91\%), UAE (83\%), India (80\%), and Brazil (75\%).}, as well as a much stronger positive relationship between income and productive household tasks. However, the most striking finding is that although education was already over-represented in the US, it is an even larger share of AI conversations in the developing world.\footnote{The same qualitative results are found if instead of income, we use internet usage.} This could reflect cultural/demographic differences, but may also be a sign of the possibility that AI tools are being used to increase access to or shift the quality of schooling.\footnote{At a technical level, generative AI can be a highly cost-effective informational delivery mechanism. Previous research in Sierra Leone \citep{bjorkegren2025aileapfrogwebevidence} shows that average web search results can consume up to three thousand times more data than a text-based AI response. Because of this data-efficiency gap, bandwidth and processing costs for AI-driven information can be roughly 87\% less expensive than traditional web search. This allows AI to act as infrastructure in low-connectivity environments, providing relevant resources to schoolteachers, who frequently rate AI outputs as helpful. Nonetheless, we acknowledge that the literature on the impact on education is more mixed (with some negative results, such as \citet{Stromberg2026}) potentially due to increased likelihood of cheating or decreased time spent thinking for schoolwork.} A deeper dive into the reasons for these correlations and the divergence of usage between different geographic areas could be a promising area of future research.

\subsection*{Guest Comment: Time, Productivity, and the Household Production Boundary}
\noindent\textit{By Diane Coyle, Bennett Professor of Public Policy, University of Cambridge}

\vspace{0.3cm}

Economists have appreciated the importance of the allocation of time to different activities at least since Gary Becker highlighted (in a classic 1965 paper, \cite{becker1965}) the choices people make among paid work, household work and volunteering, and leisure. He pointed out that people combine time and market goods to consume, and thus started to integrate production and consumption in the household. Yet although it is clear that decisions are jointly made about paid, market activities and unpaid household activities, the household side of the boundary is often ignored in much economic analysis. The main reason for this seems to be that statistical agencies collect far less data on unpaid than on market activities, even though the available estimates suggest its scale (in terms of time spent valued at an equivalent wage) can be a massive fraction of the nation’s GDP \citep[e.g.,][]{bridgman2012accounting}. There are some exceptions. For example, an influential paper \citep{Jones2016} indicated that the higher level of GDP per capita in the US compared to some European countries overstates the difference in living standards when the greater leisure enjoyed in France and Germany is taken into account. Becker’s original insight was extended in Ian Steedman’s 2001 book ``Consumption Takes Time'' \citep{Steedman2001}, which explores the impact of the dual time and money budget constraints on consumer choice. \citet{https://doi.org/10.17863/cam.84770} point out that in a service-oriented economy like the US, time is a useful lens on productivity as well as consumer choice, and that measured productivity needs to be interpreted with care when technology is shifting activities across the production boundary between the market and household sectors. Innovations may seem to reduce productivity in economic statistics when they shift activity into the household because only the market sector is measured \citep{Coyle2018}.

New insights into how people are spending their time in the household are timely for two reasons. One is that AI, like earlier digital services, is expanding the possibilities open to them. The other is that in all advanced economies a growing share of marketed activities consists of services that are substitutable with household production. Official BEA data for the US show that eight headline categories of such work (including childcare, home healthcare and food services) have expanded as a share of total non-farm payrolls from about 11\% in 1990 to 15.7\% in 2025. Other market services such as coaching, interior design, or filing government forms are small in number but expanding. So both the scope of potential substitutions from market to household is greater than in the past and the technological possibilities for substitution have expanded.

\begin{figure}[htbp]
    \centering
    \caption{Two Decades of Household Time Reallocation Reveal Some Shifts Toward Digital Leisure and Home Management and Away from Socializing}
    \label{fig:figure_3_7}
    \includegraphics[width=\textwidth]{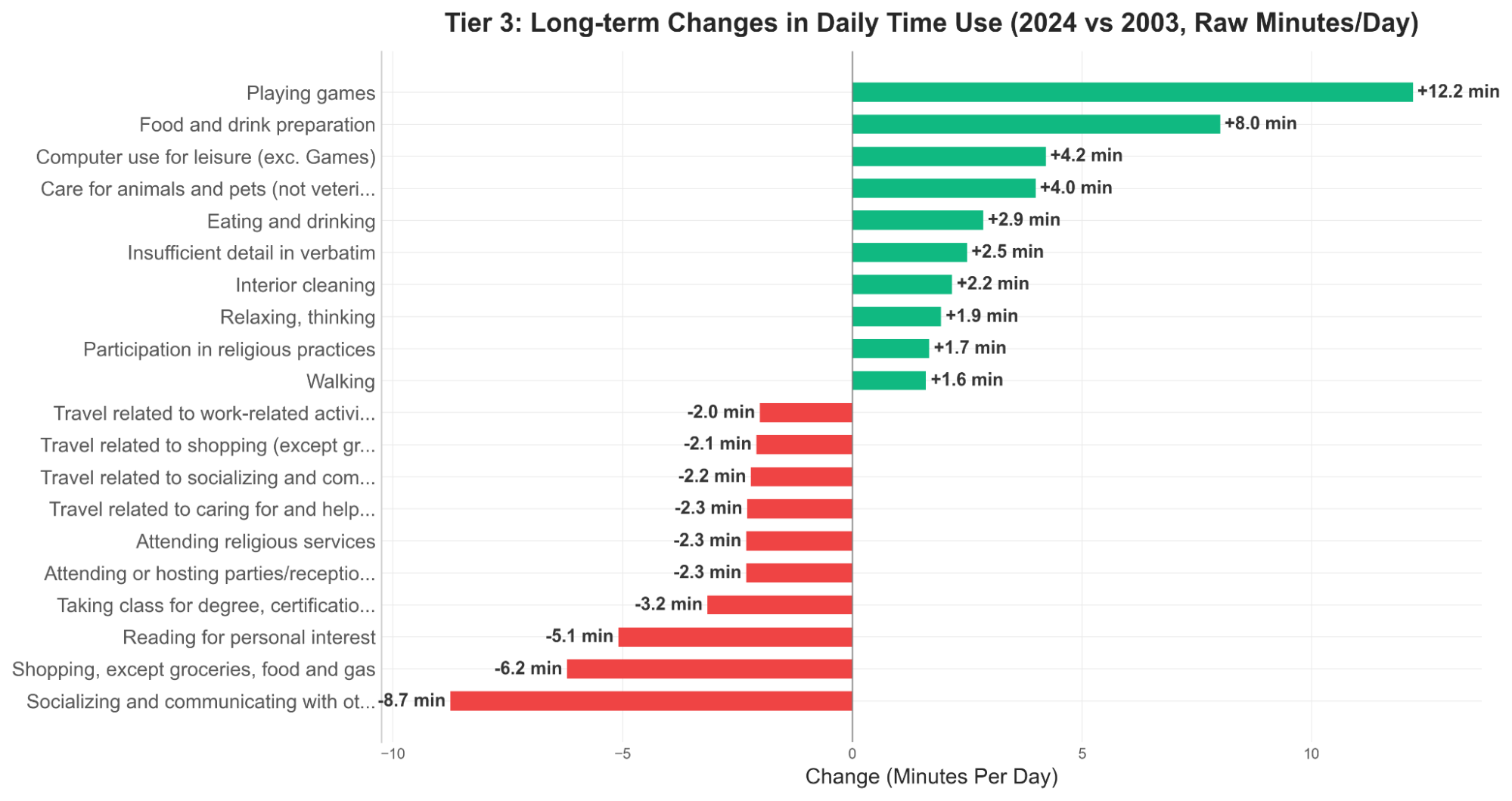}
    \vspace{0.2cm}
    \parbox{\textwidth}{\footnotesize \textit{Notes:} This figure displays the top 10 largest increases (green bars) and top 10 largest decreases (red bars) in average daily human time allocation (raw minutes per day) across granular ATUS Tier 3 activities between 2003 and 2024 (excluding sleeping and market work). The historical data illustrate some shifts in domestic life prior to and during the digital era, such as rising time spent on computer use for leisure, pet care, and physical fitness, alongside declines in traditional social communication, reading and routine household chores.}
\end{figure}

Productivity improvements can take two forms: either speeding up an activity or improving its quality. And consumer wellbeing similarly can be increased either by spending more time on more enjoyable activities or on time spent on a given activity becoming more enjoyable (although the fact this is hard to measure has given rise to a large literature). The American Time Use Survey (ATUS) provides the most regular and detailed statistics on how household time is allocated. The statistics shift slowly over time; Figure~\ref{fig:figure_3_7} shows the biggest gainers and losers recorded in the survey over a 20-year period to 2024, reflecting some (mainly) unsurprising social and technological changes. However, the longer-term trends can be significant. Think of the role of household durables (such as the washing machine, vacuum cleaner, and refrigerator) enabled by the general purpose technology of electricity. These enabled (mainly) women to enjoy more leisure in the home and to work outside the home. The impact on the labour market was consequential; and indeed the trend for more women to work in paid jobs (sometimes also paying other people to do their household work) helped boost measured productivity growth in the 1970s and 1980s \citep[e.g.,][]{10.1257/mac.20210138, NBERc12826}.

Technology-use timelines are much shorter with digital and AI, compared to the decades-long shifts previously. There is every reason to expect that AI will lead to significant changes in time use in both paid work and in household activities. Most recent economic analysis has focused on tasks in paid jobs. But the picture is incomplete without analysing unpaid household activities alongside this. In terms of household productivity, using AI could either enable people to carry out some activities faster and/or better (for example, it can draw up your weekly grocery order; and it can AI plan the week’s menus with healthier ingredients and recipes provided). What’s more, the use of AI within households is likely to affect both the formal labour market and economy (as it might lead to substitutions from market to household activities) as well as the composition of household activities. For example, demand for market services such as therapy or tax preparation might decline and demand for other market goods and services increase. To the extent AI use helps people cut costs---say by making it easier to shop around---it will also alter their combined choice of how to spend time and money.

In sum, as AI use in the household continues to increase, it will lead to changes in activity on both sides of the production boundary between the `official' economy and the domestic economy. As households are simultaneously both consumers and producers, these changes will depend on the extent to which AI lifts their combined time and money budget constraint (by saving time in absolute terms), and on the substitutability of different activities for each other in the household’s demand function. Neither a standard production function applied to the household nor models of household consumption demand alone can capture the decision problem fully; Becker’s classic \citep{becker1965} model is a partial integration of the household production and consumption decisions but omits the time needed to consume. What’s more, the act of production is often also identical with the act of consumption, especially in areas of online services where AI seems to be increasingly used. One approach to understanding the shifts at the level of the macroeconomy is to consider how the technology might alter the productivity of providing `basic goods' \citep{RePEc:ucp:jpolec:v:74:y:1966:p:132}; \citet{Hulten2022} take this approach to market production but it would also extend to household production. \citet{Fellner2015} formally model aggregated household choices over time and the quantity and quality of products demanded, and show there are multiple equilibria depending on how the time and money constraints bite.

\begin{figure}[htbp]
    \centering
    \caption{AI Assistance Concentrates in Education, Services, and Household Management Rather Than Physical Domestic Chores}
    \label{fig:figure_3_8}
    \includegraphics[width=\textwidth]{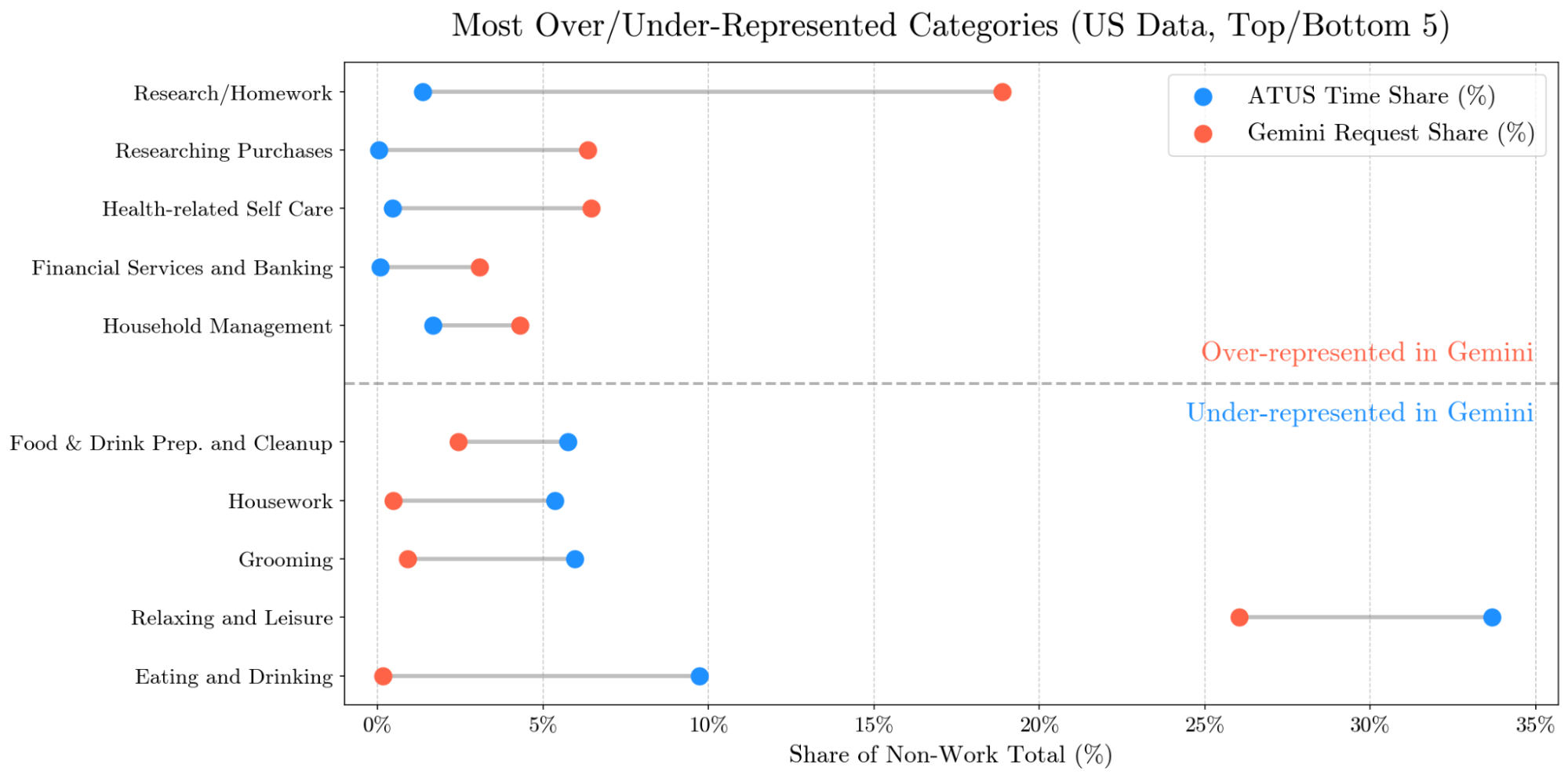}
    \vspace{0.2cm}
    \parbox{\textwidth}{\footnotesize \textit{Notes:} This figure compares AI conversation shares against ATUS human time allocation across the 10 most over-represented and 10 most under-represented intermediate (Tier 2) activity domains (excluding work and sleep). The dumbbell plots highlight the gap between human time allocation and AI consultation. Relative to human time, AI is heavily sought out for knowledge-intensive and organizational activities such as education, financial and legal services, and practical household management. Conversely, AI is least utilized for direct physical and location-bound domestic activities, including food preparation, housework, and personal grooming. These patterns demonstrate where AI currently acts as a high-leverage complement across the household production boundary.}
\end{figure}

ATLAS promises to offer a timely and detailed way to monitor such changes in activities and demand. The results already offer useful insights. For example, as can be seen in Figure~\ref{fig:figure_3_8}, compared to the ATUS, Gemini is being used more for activities such as education, accessing services (such as banking or public services) and household management; and less for physical tasks such as preparing food and housework. Such insights are not only important in themselves but also for helping track the impact of AI on the economy. Making sense of trends in the labour market and measured productivity requires taking account of both sides of the production boundary. These are separated in most economic analysis, but not in the choices individuals are making; and AI is affecting household time as much as it is measured working time.

\newpage
\section{Global Diffusion}

\label{sec:geo_diffusion}

\subsection{Methodologies for cross-country comparisons of AI adoption}

While previous transformative technologies had a general slow and unequal diffusion within a country \citep{KalyaniQJE}, these differences tend to be even bigger at an international level \citep{10.1257/aer.100.5.2031}. This is important, since the speed of international diffusion of major technological breakthroughs has served as a key determinant of cross-country productivity and economic growth \citep{1929ccba-ba3a-3051-b4bc-b696bde37fcf, repec:eee:grochp:2-565}. At the same time, the diffusion itself has historically been a factor of investment and industrial composition of different countries \citep{10.1257/aer.91.2.328, 10.1257/0022051042177685}.

In line with a historical speed up in the pace of technological adoption globally \citep[e.g.,][]{Hobijn2003}, AI survey data shows an unprecedented pace of global adoption \citep{Bick2026, sajadieh2026}. Despite this, there is still concern over the immediate impact AI may have on countries’ development trajectory \citep[e.g.,][]{NBERw28453, Cerutti2025Global}. In this section, we use ATLAS to track the global expansion of AI adoption, across different geographies.

Measuring economically meaningful diffusion using data from individual AI providers presents inherent challenges, a complexity acknowledged in Microsoft's global AI diffusion studies \citep[e.g.,][]{misra2025measuringaidiffusionpopulationnormalized, misra2026aidiffusion}, Anthropic's international index tracking adoption \citep[e.g.,][]{appelmccrorytamkin2025geoapi}, and by OpenAI \citep[e.g.,][]{NBERw34255}. Any observed patterns of usage across countries are driven by a complex cocktail of factors, including:

\begin{enumerate}
    \item \textbf{Access}: what proportion of population has access to internet connectivity and devices capable of supporting use of the provider’s AI product such as smartphones or PCs,
    \item \textbf{Interest}: understanding of and interest in using AI technologies in general,
    \item \textbf{Relative product reach and availability}: what products are available and how deeply the product being measured has penetrated the given country’s market of AI users relative to available alternatives,
    \item \textbf{Product and country multilinguality}: the quality of the underlying model’s output in the local language(s) and the country residents’ knowledge of English,
    \item \textbf{Contextual use cases}: countries have different underlying industry and occupational structures, as well as different institutional and user contextual needs.
\end{enumerate}

Access and interest are clearly relevant to understanding the macroeconomic trajectory of AI adoption in a given country, although ideally they need to be measured separately as they have different policy implications. The capabilities and reach of individual products, however, complicate this measurement. For example, if an AI product were the only alternative in a country but its neighbours had a proliferation of AI product choice, naive use of our usage data would show the country as a regional leader. If an AI product does not support a majority language in a country (but alternative AI models do), this could result in usage being observed only by the residents capable of speaking supported languages, and in a reduced intensity if users’ bilingual skills limit the topics they can discuss, resulting in under-estimation of the intensity of AI adoption in the country as a whole.

The analysis in this section is based on AI interactions over three Google sources (App, AI Mode, API). Throughout this section, we refer to interactions in App and AI Mode aggregated together as ``Conversational AI'', to distinguish them from interactions with API and other AI products.

Given the importance of the topic of the global community’s interest in genuine patterns of relative AI diffusion rather than Google’s regional footprint, throughout the next subsection, we adjust raw conversational AI usage data to account for variations in platform penetration and usage intensity using country-by-country web traffic referral shares.\footnote{For this data, we are using StatCounter’s \citep{statcounter2026aichatbot} GenAI Chatbot Tracker, a global web analytics service that aggregates traffic data via tracking scripts embedded on millions of websites. This data has two major limitations. First, it only covers Gemini Apps—excluding sources like AI Mode (AIM)—which forces us to assume that wherever Gemini Apps' penetration is below its global average, AIM's popularity drops proportionally. Second, it measures web referrals rather than actual conversations, which could skew results if different platforms have varying referral rates. Because of these limitations, the adjusted absolute usage values should be read as an estimated proxy. However, assuming Gemini Apps referrals positively correlate with the combined usage of both tools, this proxy still offers a much more accurate relative measure of global diffusion than using raw figures.} We also present the unadjusted figures of global conversational AI usage in Appendix \ref{app:geography} for transparency. As there is no global equivalent of referral traffic for API usage (and no grounds for extrapolating from conversational AI), we present raw usage rankings for Gemini API.

\subsection{International AI Adoption}

This section explores how conversational AI is spreading globally, drawing from the geographically classified data available in ATLAS. This data shows that AI adoption is far from uniform. By looking at these global patterns, adjusting for internet infrastructure, and separating casual use from professional work, we can see where AI users are located and how usage varies across geographies.

\begin{figure}[htbp]
    \centering
    \caption{Conversational AI Usage Per Capita Reveals a Strong Concentration of Demand Within Traditional Advanced Economies Alongside Several Standout Regional Leaders}
    \label{fig:figure_4_1}
    \includegraphics[width=\textwidth]{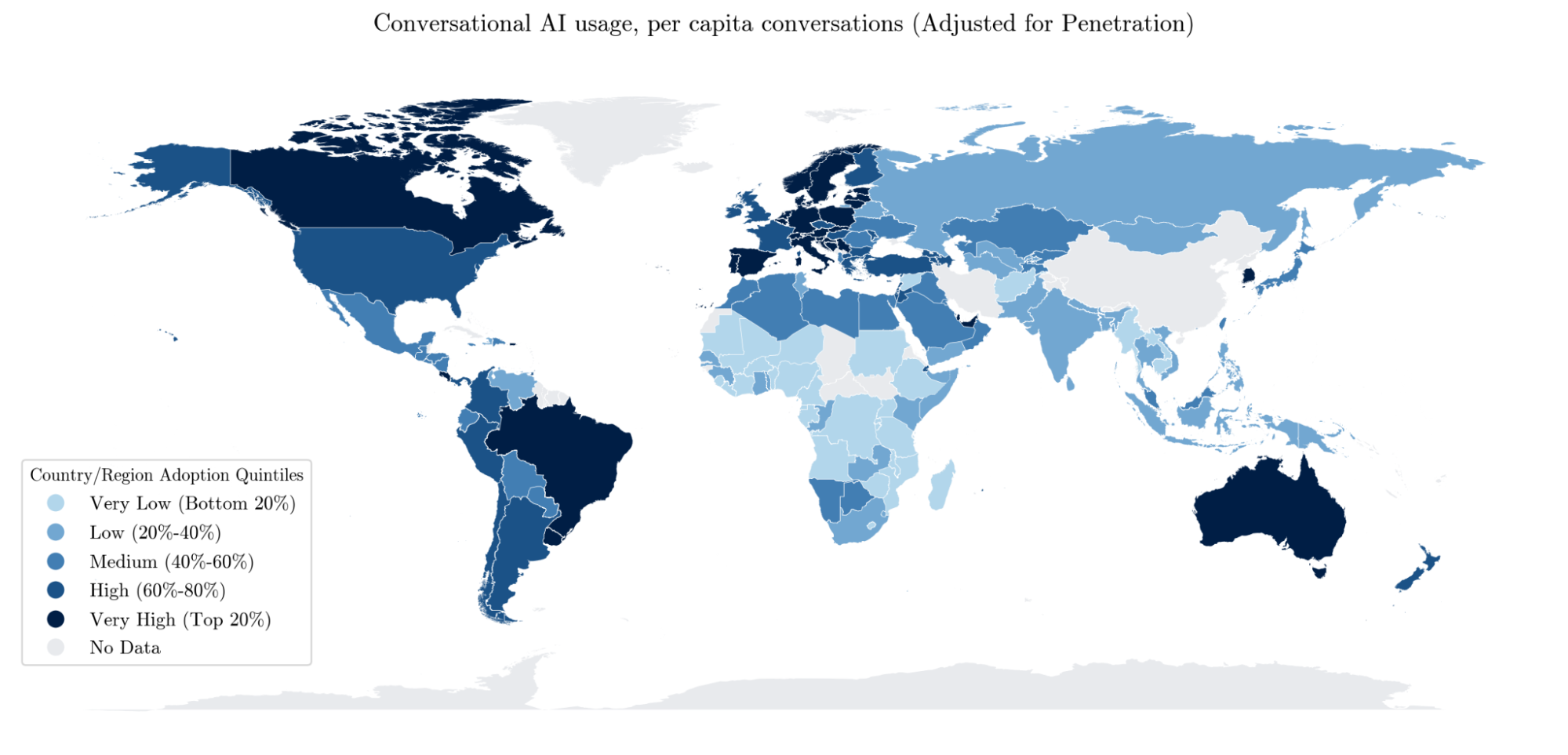}
    \vspace{0.2cm}
    \parbox{\textwidth}{\footnotesize \textit{Notes:} This map displays the global distribution of conversational AI (Gemini Apps and AI Mode) usage intensity per capita by country/region quintiles. Calculations are based on weighted conversation counts across both platforms. Several countries/regions are excluded due to consumer-level access being unavailable, or queries primarily occurring through enterprise Workspace accounts. Per-capita rates are calculated using \cite{worldbank_population_2024} population estimates from the World Development Indicators (WDI) database. For countries/regions with missing data in WDI, we use  values from the \cite{imf_weo_oct2024} WEO. The quintiles are based on country-level conversations per capita and are penetration-adjusted. Only countries/regions with more than 1,000,000 inhabitants are included.}
\end{figure}

Figure~\ref{fig:figure_4_1} shows the landscape of cross-platform, per capita AI conversational usage across the world, broken down by country/region usage quintiles. It reveals a strong concentration of demand within traditional advanced economies, alongside several standout regional leaders. North America, Europe, and Australia form a prominent high-intensity block, anchored firmly within the ``Very High'' (Top 20\%) adoption quintile.\footnote{Firm-level survey data across major high-income economies similarly demonstrates rapid adoption of AI capabilities by businesses, though with notable variations by firm size, age, and initial productivity \citep{NBERw34836}.}

Outside of these established tech hubs, the Middle East shows a strong Conversational AI adoption, T\"urkiye, the United Arab Emirates, Qatar and Israel sitting in the highest tiers, with a generally high level of usage of the remainder of the region. Latin America also shows high AI usage. A few South American countries/territories, including Chile, Peru, Brazil, Argentina, and Colombia, are also into the ``High'' and ``Very High'' (60\%--100\%) quintiles, alongside wealthier nations, which may be partially driven by extensive digital device usage \citep{2025}.

In contrast, several major economic and technology powers sit further down the adoption curve. Despite their large populations and large contributions to the global STEM graduate population \citep{oliss2023global}, India and Russia fall into the ``Low'' tier. Sub-Saharan Africa, along with parts of Central and South Asia, are heavily concentrated in the ``Low'' and ``Very Low'' quintiles, where infrastructure bottlenecks and physical connectivity limitations likely restrict baseline access.

\begin{figure}[htbp]
    \centering
    \caption{Relative Search Interest for AI-Related Topics Concentrates Heavily in Emerging Economies}
    \label{fig:figure_4_2}
    \includegraphics[width=\textwidth]{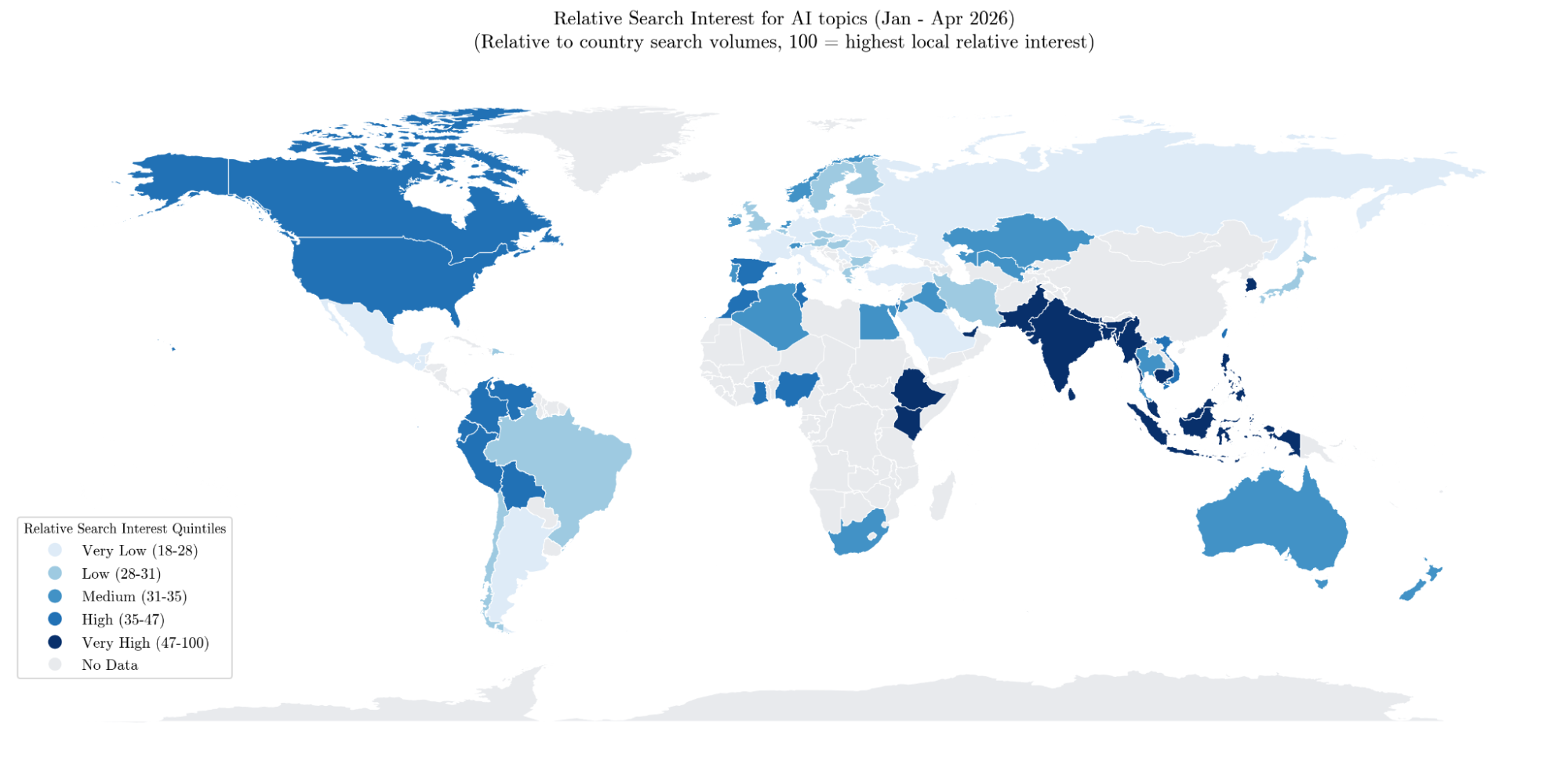}
    \vspace{0.2cm}
    \parbox{\textwidth}{\footnotesize \textit{Notes:} This map displays the global distribution of relative search interest for AI-related topics across country/region quintiles on Google Trends between January and April 2026. Values are normalized relative to total local search volume, with 100 representing the highest local relative interest. Areas with low search query volumes for AI topics are marked as 'No Data' (shown in light gray) and excluded for quintiles definition.}
\end{figure}

While per capita conversational usage is heavily concentrated in traditional advanced economies, a different picture emerges when examining global search interest. Figure~\ref{fig:figure_4_2} displays relative search interest for AI-related topics during early 2026, mapping where AI represents the highest share of local search queries rather than absolute volume. Crucially, the ``Very High'' search interest quintile is heavily dominated by emerging economies in South Asia (including India, Bangladesh, and Nepal) and Southeast Asia (such as the Philippines, Indonesia, Thailand, and Vietnam), alongside regional standouts in East Africa (Kenya and Ethiopia). In these countries/regions, public curiosity and active research into AI topics are exceptionally high relative to overall search activity, signaling strong grassroots interest in the technology's capabilities.

This divergence highlights a notable ``interest-adoption gap'' across different development tiers. In highly mature tech markets like Western Europe and Japan, relative search interest sits in the ``Low'' or ``Very Low'' quintiles, despite these areas driving the highest per capita conversational usage (as shown in Figure~\ref{fig:figure_4_1}). This pattern suggests that in advanced economies, conversational AI may have already transitioned from a novel subject of active search query exploration into an embedded, background utility in daily workflows. Conversely, in many emerging markets, massive interest and digital curiosity have yet to fully translate into high levels of per capita conversational usage.

\begin{figure}[htbp]
    \centering
    \caption{High-Usage Regions Drive Conversational AI Volume Disproportionately to their Population}
    \label{fig:figure_4_3}
    \includegraphics[width=\textwidth]{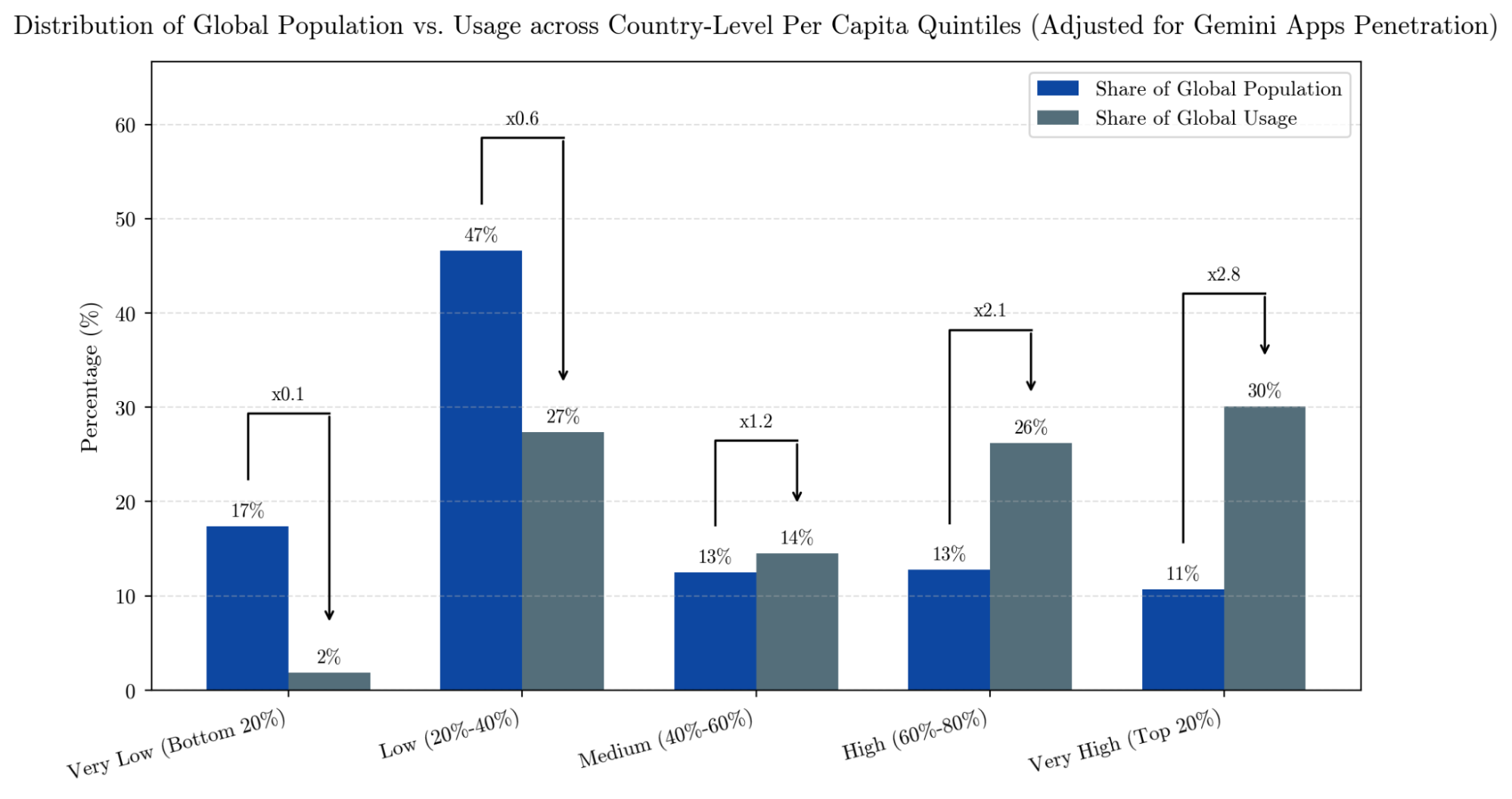}
    \vspace{0.2cm}
    \parbox{\textwidth}{\footnotesize \textit{Notes:} This figure compares the share of global population against the share of global conversational AI usage across the five per-capita usage quintiles at the country/region level. Multipliers denote the ratio of usage share to population share for each quintile (for instance, countries and regions in the ``Medium'' quintiles represent 1.2x their population in terms of usage). The quintiles are based on country-level conversation volume per capita, adjusted for Gemini Apps penetration. China and countries/regions with populations under 1,000,000 are excluded. Note that the analysis aggregates country-level data, which may mask within-country inequality in adoption and usage.}
\end{figure}

Figure~\ref{fig:figure_4_3} confirms the significant imbalance in global conversational AI adoption. We sort the countries/regions into quintiles of per capita conversational AI use from Figure~\ref{fig:figure_4_1}, and for each quintile calculate the share of total AI usage and share of the global population by countries/territories in our data. The contrast between the two extremes is particularly sharp. While the countries/regions in the lowest usage quintile in our sample account for roughly 17\% of the world’s population, they generate just 2\% of global usage. In contrast, the top quintile in our sample represents only 11\% of the population but drives  30\% of total AI conversations, amounting to 2.8 times its proportional share.\footnote{These results are robust to using free services only, suggesting that the relationship between use and gdp per capita is fully not explained by differences in ability to pay for AI services.}

\begin{figure}[htbp]
    \centering
    \caption{AI Conversational Usage (Gemini Apps and AI Mode) Shows a Clear, Positive Correlation With GDP Per Capita (PPP)}
    \label{fig:figure_4_4}
    \includegraphics[width=\textwidth]{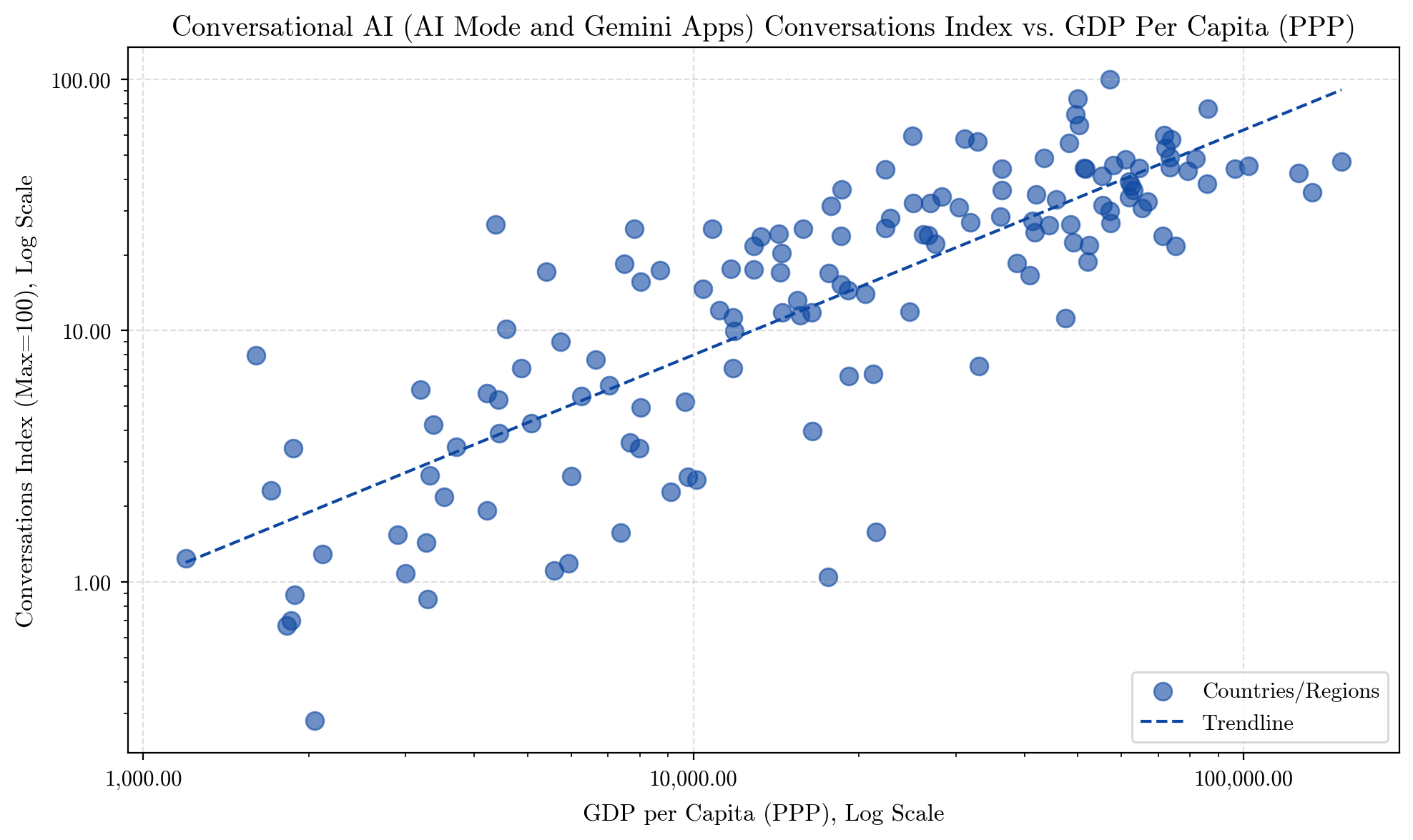}
    \vspace{0.2cm}
    \parbox{\textwidth}{\footnotesize \textit{Notes:} This figure displays the relationship between penetration-adjusted conversations per capita and GDP per capita (PPP) on a log-log scale. Per-capita conversations are adjusted for Gemini's penetration using StatCounter data and are indexed to a maximum value of 100. GDP per capita data is sourced from the \cite{worldbank_gdppc_2024} World Development Indicators (WDI) database. For countries/regions with missing data in WDI, we use values from the \cite{imf_weo_oct2024} World Economic Outlook (WEO) where available. Only Countries/Regions with more than 1,000,000 inhabitants with official availability of Gemini Apps are featured; Mainland China is excluded given that Gemini Apps access is through Workspace only.}
\end{figure}

Figure~\ref{fig:figure_4_4} shows the clear correlation between adjusted-AI conversational adoption intensity and national wealth. While this analysis does not establish direct causation---as macroeconomic relationships are highly complex and driven by many different variables---it confirms a strong, consistent association between log GDP per capita and AI usage (here proxied by per-capita, penetration-adjusted conversational AI counts). The trendline features a slope of approximately 0.9, meaning that a 1\% increase in GDP per capita translates on average to a 0.9\% increase in the number of AI conversations.\footnote{These findings are similar to \cite{appelmccrorytamkin2025geoapi} that found that a 1\% increase in GDP per capita being associated with a 0.7\% increase in Claude usage per capita.} The outlier countries, both in terms of over- and under- adoption of AI relative to national wealth trendline, do not appear to follow an obvious overarching geographic or economic pattern.\footnote{While per capita conversational adoption is heavily concentrated in richer contexts, public sentiment is anxious. \cite{pew_2026_americans_ai} indicates that almost two thirds of Americans feel AI is advancing too rapidly, and \cite{ipsos_google_2026_ai_survey} finds that U.S. concern is roughly double its excitement. In contrast, emerging markets exhibit a powerful ``optimism surplus,'' with excitement exceeding half in India, Brazil, and the UAE, which likely helps overcome baseline access constraints.} Further comprehensive research of the complex factors driving differences in cross-country AI adoption is clearly needed even though it is outside the scope of the current ATLAS report.

\begin{figure}[htbp]
    \centering
    \caption{Access to the Internet Acts as a Bottleneck on Per Capita Conversational AI Usage, but Does Not Explain the Full Disparity}
    \label{fig:figure_4_5}
    \includegraphics[width=\textwidth]{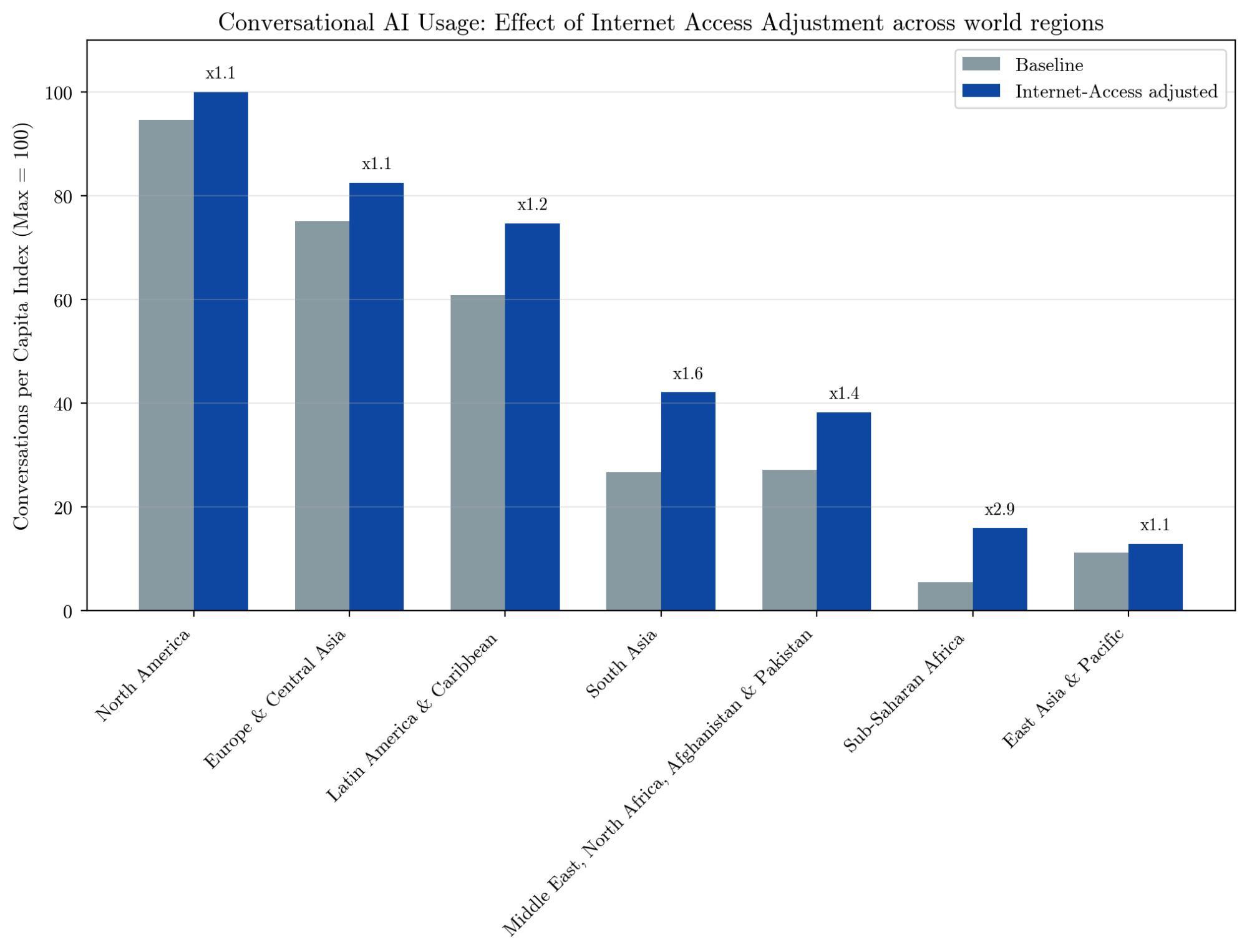}
    \vspace{0.2cm}
    \parbox{\textwidth}{\footnotesize \textit{Notes:} This figure illustrates the regional impact of adjusting adoption intensity for internet access. The baseline metric is conversations per capita normalized by Gemini App's usage share, while the internet-access adjusted metric utilizes internet users (calculated using \cite{worldbank_internet_2024} internet access percentages from the World Development Indicators (WDI) database) in the denominator instead of the total population. Both metrics are scaled to a maximum regional value of 100. The comparison is aggregated at the regional level, which masks country-specific variation and digital divides within regions.}
\end{figure}

Figure~\ref{fig:figure_4_5} examines the impact of adjusting regional AI usage for actual internet access levels, comparing conversations per capita against conversations per internet user. The results confirm that physical connectivity is an important bottleneck; for instance, adjusting for online populations nearly triples the adoption metric for Sub-Saharan Africa. However, this infrastructure adjustment alone does not fully close the global usage gap, indicating that other underlying drivers beyond basic access are also at play.\footnote{When we analyzed if disparities in international adoption can be systematically explained by national differences in AI sentiment, we found little systematic correlation with surveyed public enthusiasm or concern.}

\begin{figure}[htbp]
    \centering
    \caption{Global Gemini API Usage Per Capita Shows a Pronounced Concentration in Traditional, High Income Economies}
    \label{fig:figure_4_6}
    \includegraphics[width=\textwidth]{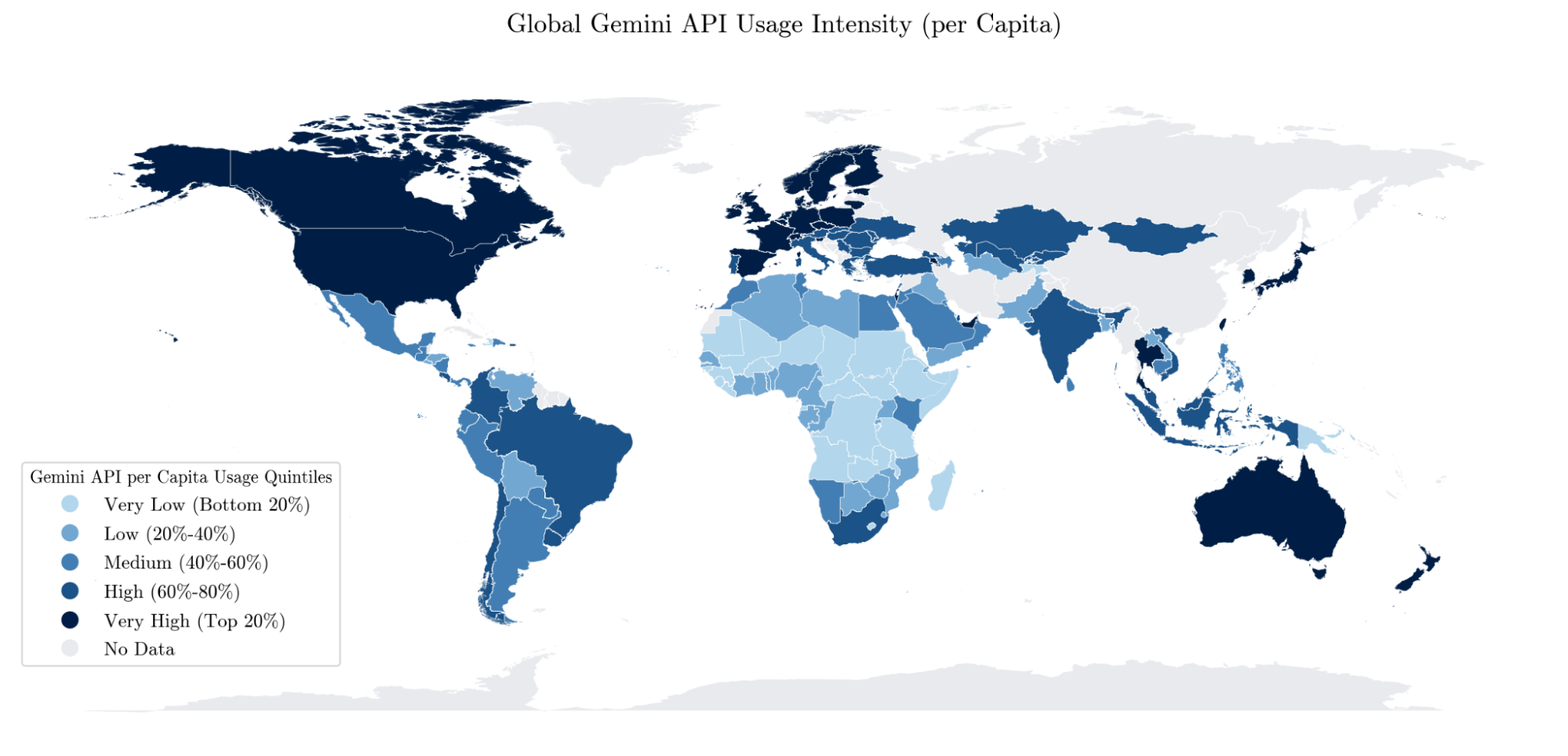}
    \vspace{0.2cm}
    \parbox{\textwidth}{\footnotesize \textit{Notes:} This figure displays the per-capita intensity of Gemini API requests by country/region quintiles. Unlike other metrics in this report, conversation counts here include enterprise/educational and paid usage. Usage locations are determined by the IP address originating the API request, meaning the map reflects the server's location rather than the end-user's (e.g., requests handled by a US-based server for a chatbot in Mexico will be mapped to the US). Per-capita rates are calculated using \cite{worldbank_population_2024} population estimates, and the figure does not adjust for Gemini API penetration. For countries/regions with missing data in WDI, we use values from the IMF World Economic Outlook (WEO) where available. The analysis is restricted to countries with official Gemini API availability and more than 1,000,000 inhabitants. The quintiles are based on total request counts per capita and are not adjusted for penetration.}
\end{figure}

Figure~\ref{fig:figure_4_6} maps global Gemini API usage per capita as reflected in ATLAS by country and region quintiles. While the distribution is broadly similar to the conversational AI landscape shown in Figure~\ref{fig:figure_4_1}, the concentration within traditional, high-income economies is even more pronounced, with the US, Canada, the UK, Japan, Australia, and the majority of the EEA all falling into the highest usage quintile.\footnote{Interestingly, the US appears more pronounced in API (work use) than in conversational usage. This is corroborated by \cite{Bick2026Management}, whose representative worker surveys document a persistent adoption gap between the U.S. and Europe, both at the worker and firm level. Like other previous technologies, good management seems to be corroborated with technology adoption \citep{10.1093/oxrep/grab009}. } This shift suggests that API usage---which is inherently more technically demanding and work-oriented---remains heavily concentrated within these established digital hubs.

This distribution likely reflects the different practical demands of API integration compared to consumer application usage.\footnote{The geographic patterns in the API data are a best-effort approximation, as the originating server of an API request may not always accurately represent the location of the end user. We currently have no reliable way to measure the nature and direction of any systematic distortion arising from this geolocation method.} While interacting with a standalone chatbot requires minimal technical expertise, using an API---especially in high volumes---depends on skilled software engineers to maintain the integration, reliable cloud infrastructure to manage data pipelines, and the capital necessary to scale the software. This financial requirement is further compounded for global deployments because, unlike consumer AI that relies on free or subscription tiers, APIs are billed strictly by token consumption. Due to structural tokenisation biases, non-Western text requires more tokens per word, inadvertently driving up both API expenses and latency for non-European languages \citep{Lundin2026}. Consequently, the development and commercialization of AI tools appear to remain concentrated within established technology hubs where the economic and structural incentives are far more favorable.

Due to conversation logging limitations, we are unable to explicitly classify API calls from European countries and any paid API usage as work/non-work with the methodology used in the rest of ATLAS data. In calculating the proportion of work usage, we assume that all paid usage is work-related usage, and omit European countries from the analysis due to lack of data. Among the 106 countries/regions for which we hold data, only 16\% show less than 90\% work conversations, and less than 2\% show less than 80\% work conversations. Given this relatively consistent, high work concentration of work conversations, we don’t provide additional breakdowns in relation to non-work patterns for Gemini API.

To understand where AI is actively being integrated into the global labor market, we must investigate beyond general adoption and isolate professional, work-related engagement. We filter the data used for general Conversational AI usage figures to include only work-related to conversations.\footnote{Doing so implicitly assumes for the purpose of this analysis that users don’t use a different Conversational AI platform for their work and non-AI usage, and that the share of work/non-work conversations is broadly consistent across platforms.}

\begin{figure}[htbp]
    \centering
    \caption{Highly Connected, High-Income Countries and Regions Lead Work-Related Conversational AI Adoption on a Per-Capita Basis}
    \label{fig:figure_4_7}
    \includegraphics[width=\textwidth]{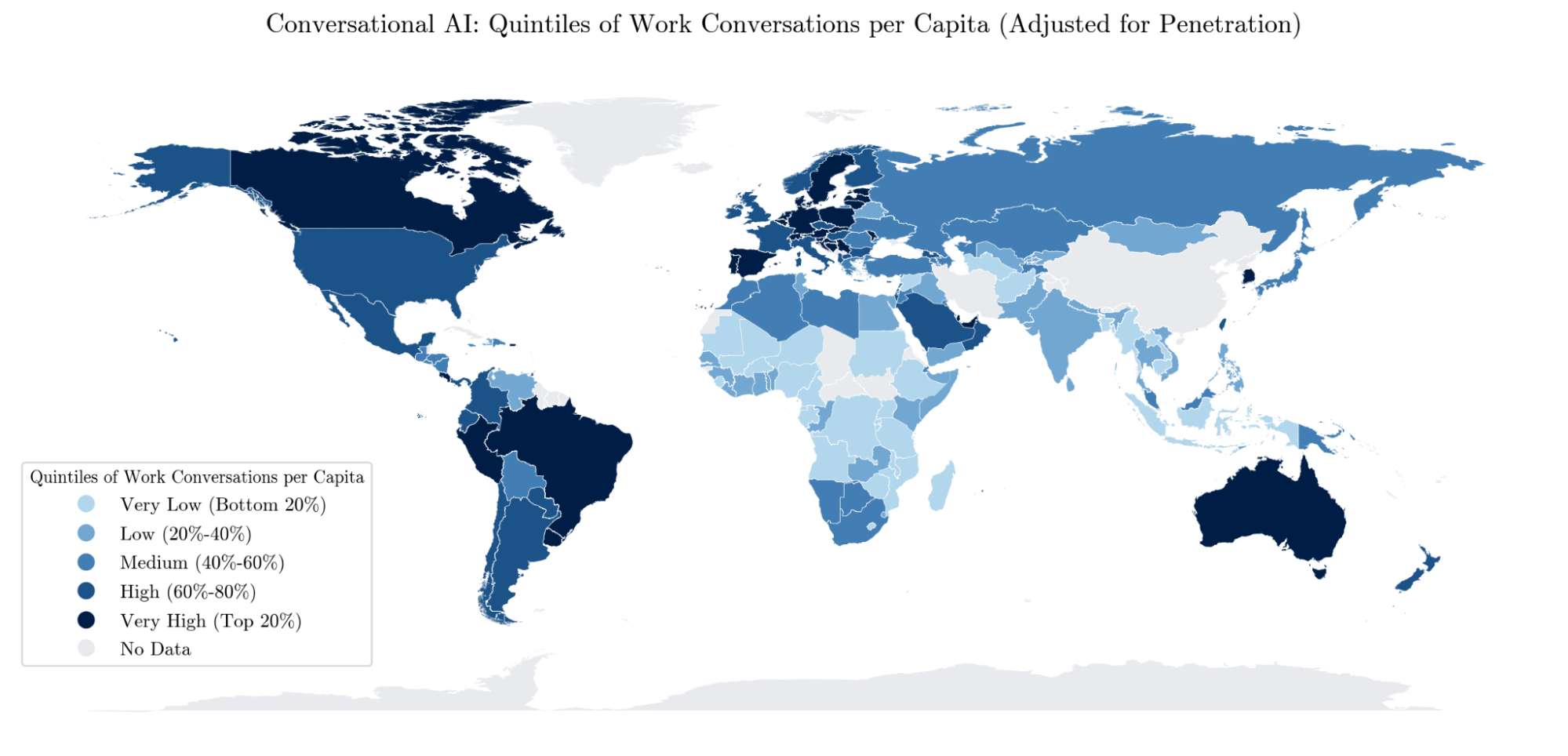}
    \vspace{0.2cm}
    \parbox{\textwidth}{\footnotesize \textit{Notes:} This figure displays the per-capita intensity of work-related conversations, adjusted for Gemini Apps overall penetration, by country/region quintiles. The metric divides work conversations per capita by the StatCounter Gemini usage proxy. China and countries with populations under 1,000,000 are excluded. To prevent extreme outliers from small sample sizes, countries with a Gemini usage proxy share of 0.1\% or lower are excluded. The intensity rates depend on the accuracy of both the automated work classifier and the web traffic share proxy.}
\end{figure}

Figure~\ref{fig:figure_4_7} maps the per-capita intensity of work-related conversations across conversational AI surfaces. The resulting geographic distribution is broadly consistent with overall usage patterns, showing very few variations. Highly connected, high-income countries continue to lead work-related AI adoption on a per-capita basis.

Latin America also shows a strong usage of Conversational AI for work-related usage, with most of the region falling the upper per-capita usage tiers.

\begin{figure}[htbp]
    \centering
    \caption{Developing Nations Lead in the Share of Work-Related AI Use}
    \label{fig:figure_4_8}
    \includegraphics[width=\textwidth]{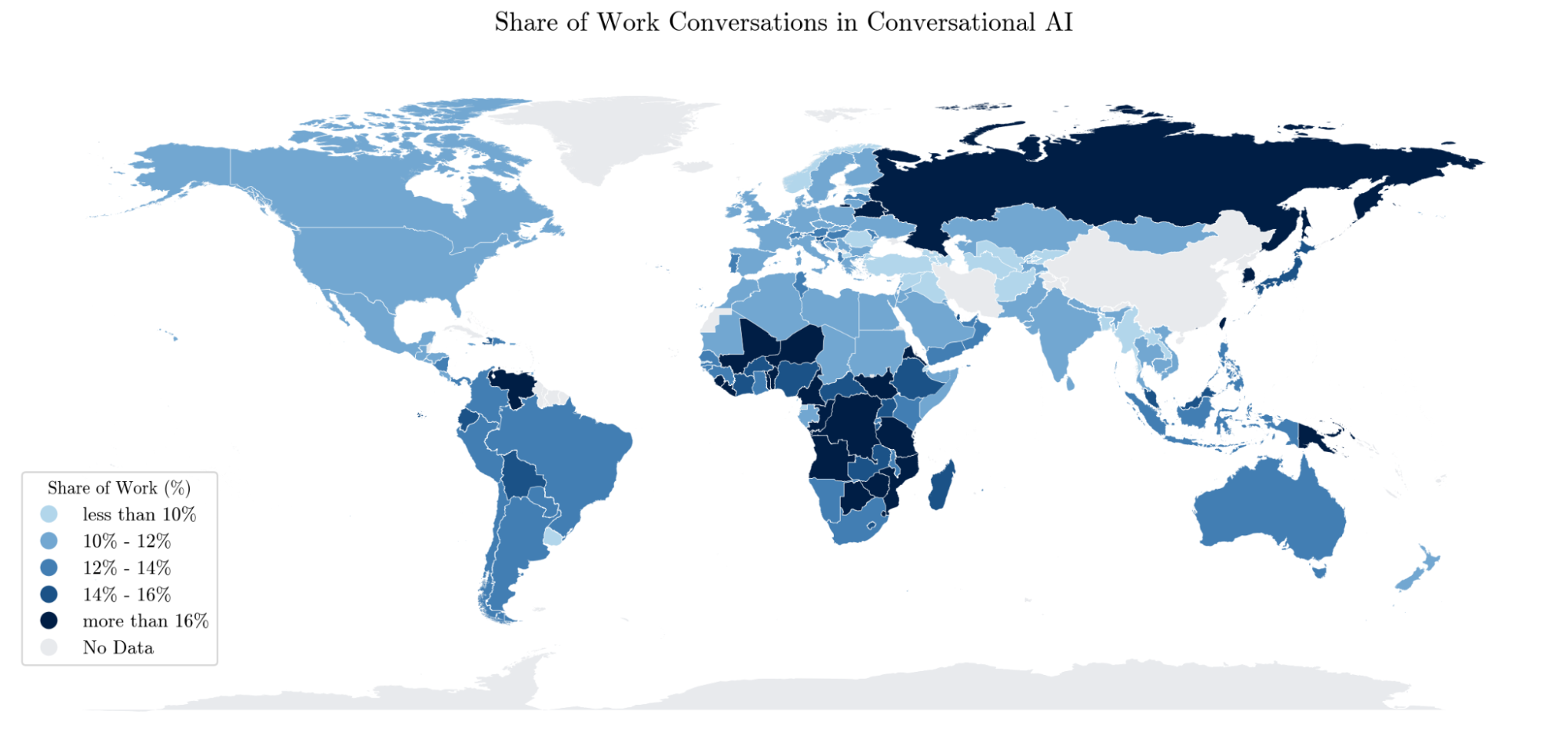}
    \vspace{0.2cm}
    \parbox{\textwidth}{\footnotesize \textit{Notes:} This figure maps the percentage of conversations classified as work-related by country. The work percentage is calculated as the ratio of work conversations to total conversations. Classification of work status is determined by our automated work/non-work classifier applied to conversation content. The analysis excludes China and countries/regions with populations under 1,000,000.}
\end{figure}

However, a surprising inversion occurs when measuring work-related AI conversations not as an absolute per-capita volume, but as a percentage of a country/region’s total AI conversations, as featured on Figure~\ref{fig:figure_4_8}. Under this lens, the global ranking changes considerably: highly connected, high-income regions like the United States and the European Union fall into the lower quintiles of work-related use. Conversely, developing regions, most notably across Africa, surge into the top quintile of work-biased usage, while South America maintains its stronghold in the upper tiers. This striking shift may suggest that professionals in developing economies could be leveraging it more intensely as a productivity multiplier to bypass traditional business constraints while users in developed nations treat AI more as a general-purpose utility. However, this pattern could also reflect a lower baseline of casual, non-professional AI adoption in these regions due to access or cost constraints, artificially concentrating their usage metrics within the workplace.

This inversion is likely driven by differing economic constraints around digital engagement. In lower-income regions, where metered mobile data and limited device access impose higher marginal costs, digital usage may tend to be more goal-directed.\footnote{Lower income countries may also see a tendency to use traditional phone calls and SMS messages to access AI services \cite{bjorkegren2025aileapfrogwebevidence, baroplevine2026afrigpt}.} Users in these markets may leverage conversational AI primarily as a tool to generate economic value, minimizing casual or recreational use. Conversely, in higher-income nations where ubiquitous connectivity reduces marginal costs to near zero, AI functions as both a professional utility and a leisure luxury. The resulting volume of curiosity-driven or entertainment queries likely dilutes the professional share in wealthier markets, lowering their relative rankings.

Alternatively, this pattern could be an artifact of a dataset limitation---specifically, the exclusion of Gemini Apps enterprise subscription data. If corporate sponsorship of enterprise plans is significantly more prevalent in regions like North America and Europe than in emerging markets, omitting this commercial traffic would under-represent professional usage in high-income economies, contributing to the observed downward skew.

How are people engaging with AI? While mapping the global distribution reveals where AI adoption is concentrated, examining some of the characteristics of these interactions and their modalities provides additional insights.

Gemini Apps and AI Mode are predominantly utilized for non-work tasks, which account for  87\% of their respective traffic. In contrast, Gemini API usage is overwhelmingly work-related (98.8\%). Work-related conversations tend to be notably longer, suggesting that professional use cases involve more complex workflows requiring iterative prompting and refinement.

\begin{figure}[htbp]
    \centering
    \caption{Middle East and East Asian Countries and Work Conversations Tend to Have Higher Turn Counts With Conversational AI}
    \label{fig:figure_4_9}
    \includegraphics[width=\textwidth]{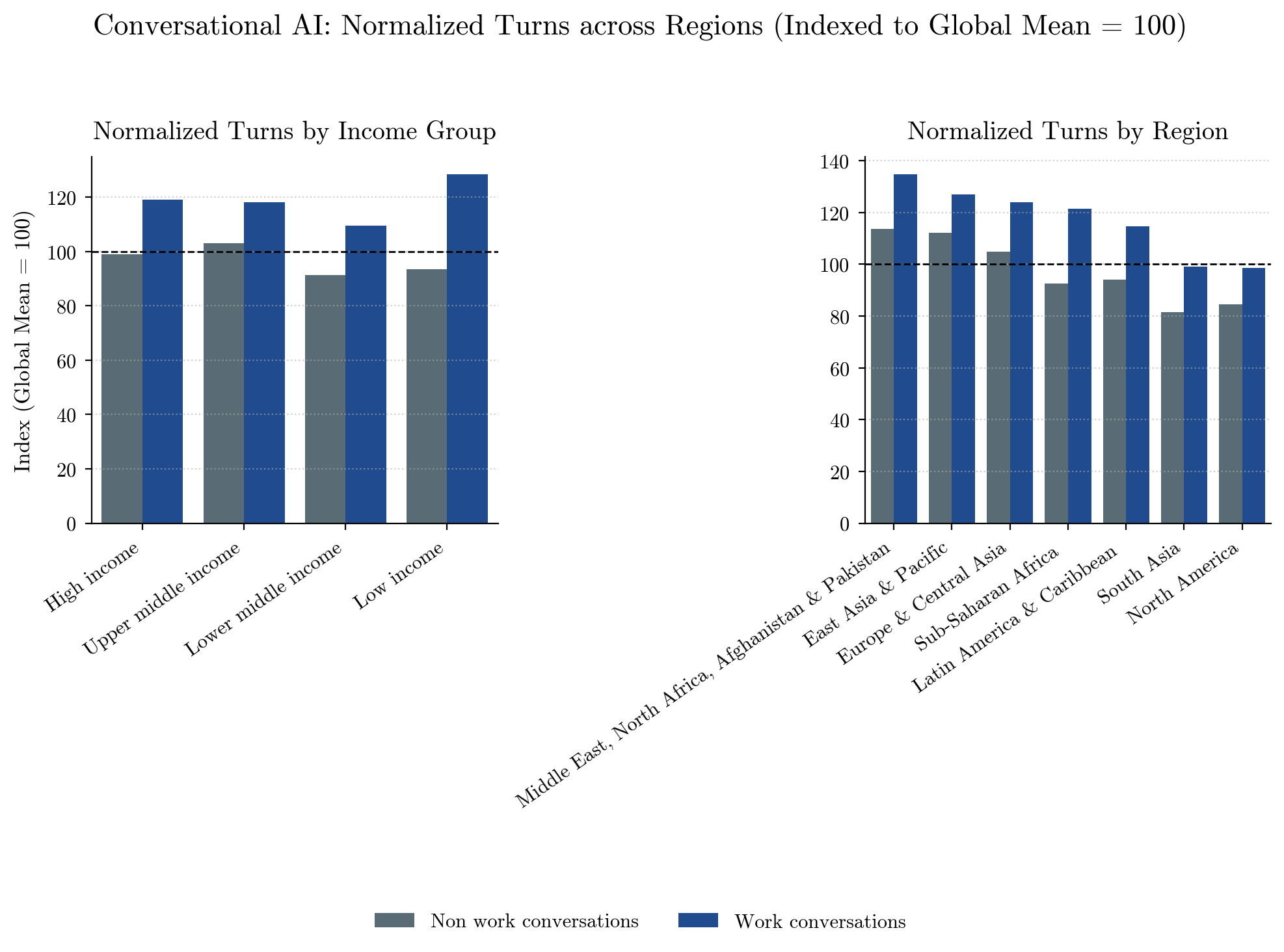}
    \vspace{0.2cm}
    \parbox{\textwidth}{\footnotesize \textit{Notes:} This figure displays normalized average turn counts per conversation, split by income group and region, and categorized by work status. Turn counts are weighted, and averages are indexed to the global mean (Global Mean = 100). The underlying data combines Gemini Apps and AI Mode. The metric does not control for language-specific communication styles, which may influence dialogue length.}
\end{figure}

While adoption volume remains concentrated in high-income nations, analyzing the depth of engagement---measured by the average number of turns per conversation---suggests a distinct behavioral divide between transactional queries and intensive problem-solving. Across every region and income group, professional contexts consistently yield a higher average number of turns than non-work interactions. This uniform trend likely reflects the iterative prompting and ongoing contextual refinement often required by workflows such as coding and data analysis.

Geographically, these behavioral patterns indicate a notable regional divergence in how AI is utilized. The Middle East, North Africa, Afghanistan \& Pakistan region records the deepest engagement (with a number of turns of $\approx 1.35$ times the average), followed closely by East Asia \& Pacific ($\approx 1.27$ times the global average). Conversely, North America and South Asia anchor the bottom of the rankings, just under the global average. For a mature, high-volume market like North America, a lower turn count could indicate greater transactional efficiency, where users might rely more heavily on quick-answer interfaces and rapid execution rather than extended, exploratory dialogue.

\subsection{Languages of AI Usage}

The widespread international diffusion seen in the previous section is enabled by the capacity of AI to converse in different languages. In economic terms, having this option lowers the costs or frictions associated with human-computer \citep{norman1988psychology, card1983psychology} or human-AI interaction in particular. While there is still a gap in the responses model may give in English versus in other languages \citep{ahuja-etal-2023-mega, gupta-etal-2025-found, sirdeshmukh2025multichallengerealisticmultiturnconversation} recent models are improving.\footnote{This is especially true of Gemini models who show great performance in multilingual settings \citep[e.g.,][]{artificialanalysis2026global}.}

This section documents the usage of different languages in our data, explores how language choice interacts with task complexity and multimodal usage and documents the choices of non-primary languages communication. The analysis in this section is based on user interactions on our conversational surfaces (the Gemini App and Google AI Mode). Throughout this section, we refer to these as ``conversational AI'' interactions, to distinguish them from developer interactions via the Gemini API (which are excluded from the main language diversity analysis).

\begin{figure}[htbp]
    \centering
    \caption{Conversational AI Usage is Multilingual, With Non-English Languages Accounting for the Majority of Global Interactions}
    \label{fig:figure_4_10}
    \includegraphics[width=\textwidth]{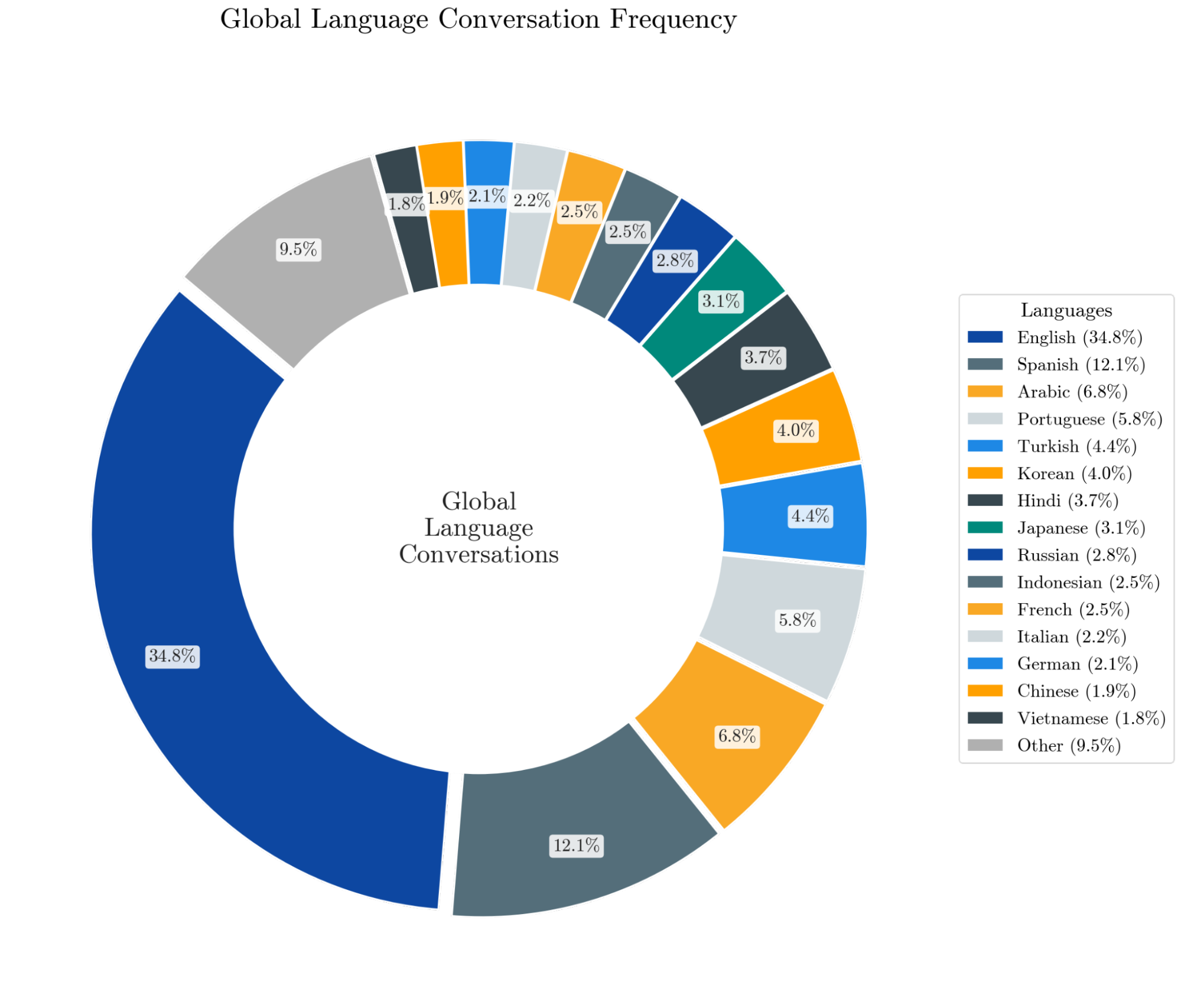}
    \vspace{0.2cm}
    \parbox{\textwidth}{\footnotesize \textit{Notes:} This figure displays the global distribution of primary interaction languages across conversational AI surfaces (Gemini App and Google AI Mode), showing the top 15 individual languages by global request share with remaining languages aggregated under ``Other.'' Regional variations of a base language (e.g., Brazilian and European Portuguese) are consolidated into single language categories. Across 143 privacy-cleared languages, non-English interactions represent the majority of total global conversation volume, underscoring the widespread international adoption of conversational AI beyond English-speaking countries.}
\end{figure}

We find 143 distinct languages that occur in sufficient volume to pass our privacy-preserving thresholds. Figure~\ref{fig:figure_4_10} shows the complete distribution of these conversations by language volume. English serves as the primary language of interaction in our dataset, with just above a third of the total conversation counts. Spanish is the second most frequent at 12\%. Arabic (just below 7\%) and Portuguese (just below 6\%) also have high-volume usage. Beyond the top four, usage frequencies decline smoothly across major global languages. The remainder of the top 15 ranges from Turkish (just above 4\%) and Korean (4\%) down to Vietnamese (at just below 2\%). The ``Other'' category captures nearly a tenth of all conversations. However, because our classification methodology imposes minimum cluster size thresholds to filter out personally identifiable information, this observed diversity is an underestimate. Even with these filters in place, ATLAS covers languages spoken by over 93\% of first-language speakers among the 200 most spoken languages globally \citep{ethnologue2026}.

One hypothesis suggests that, if users in non-English primary countries face a significant AI response quality penalty, we would expect the propensity to use a non-primary language to vary systematically by task. Specifically, users would be incentivized to shift economically important conversations---such as advanced reasoning, professional workflows, and high-value deliverables---into English to maximize model performance. Conversely, for low-stakes conversations like those for non-work personal life, household management, or leisure, we would expect users to default to their primary language.

\begin{figure}[htbp]
    \centering
    \caption{Rates of Non-Primary Language Use are Remarkably Symmetric Across Work and Personal Domains, Disproving High-Stakes Code-Switching}
    \label{fig:figure_4_11}
    \includegraphics[width=\textwidth]{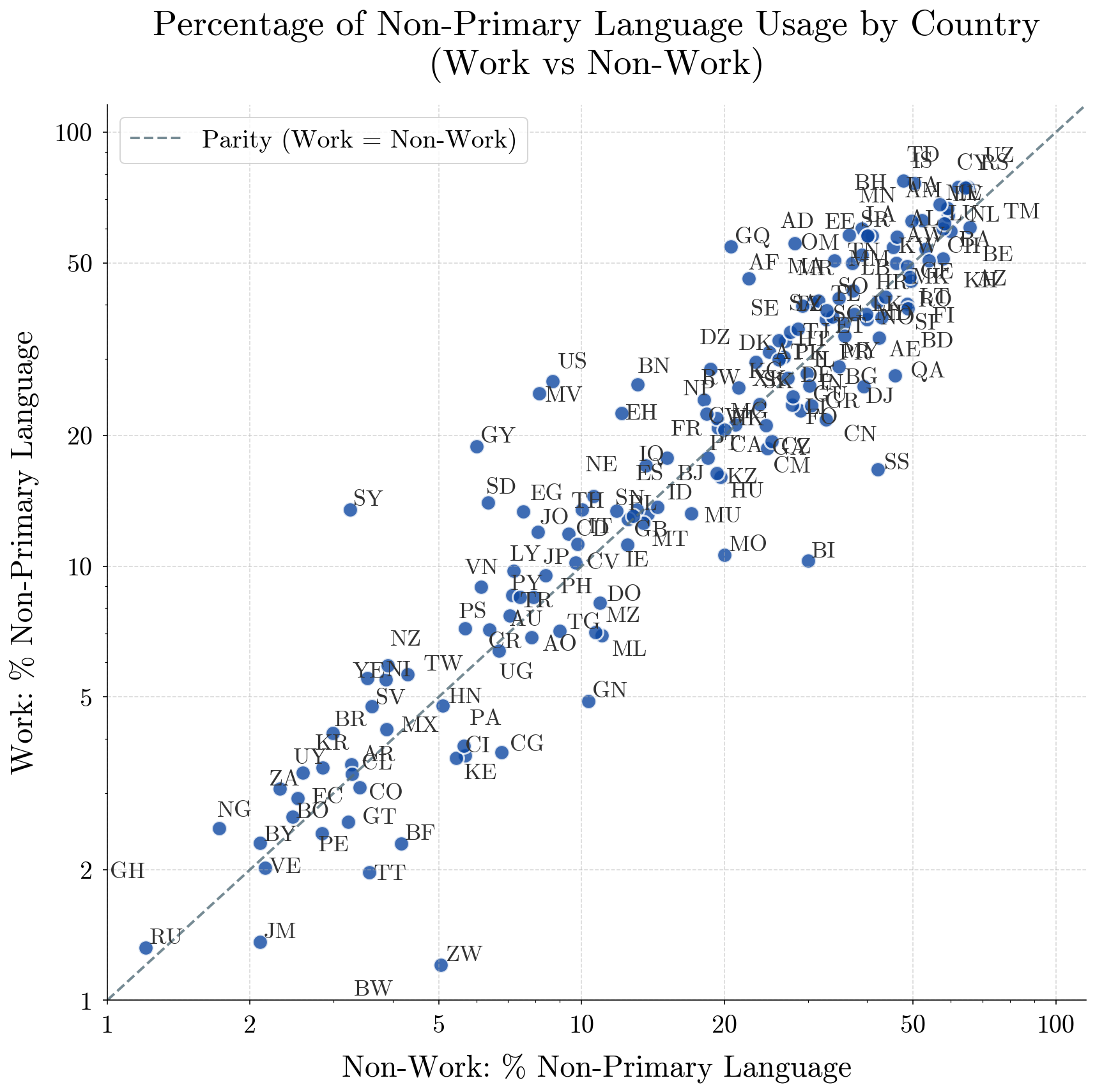}
    \vspace{0.2cm}
    \parbox{\textwidth}{\footnotesize \textit{Notes:} This log-log scatter plot compares the share (\%) of AI conversations conducted in a language other than a country's primary language across work activities (y-axis) versus non-work activities (x-axis) for countries with over 1,000 total conversations. The figure uses for defining work and work related usage conversational ATUS AI conversations combined with the work/non-work classifier bucket. Each point represents an individual country. The dashed 45-degree line indicates parity where non-primary language (defined by the not most spoken language for AI queries inside of a country) propensity is identical across work and personal life. Contrary to the hypothesis that users selectively switch to English or major global languages primarily for high-stakes professional tasks, non-primary language usage exhibits strong symmetry across domains (averaging 26\% for work vs. 24\% for non-work), indicating that language selection is primarily driven by stable country-level demographic characteristics.  }
\end{figure}

Our findings seem to largely challenge this task-sorting hypothesis. Figure~\ref{fig:figure_4_11} plots the share of non-primary language usage for work versus non-work activities across our dataset. Contrary to the hypothesis that professional demands might compel users to switch to a dominant global language like English, the data reveals a symmetry across domains. Specifically, the global average share of non-primary language conversation stands at 26\% for work activities and just below 24\% for non-work activities. Visually, this consistency is reflected in a tight clustering of countries along the 45-degree parity line. These numbers are robust to analyzing English usage in particular. For non-primary English language countries, the percentages of conversations in English are 14\% for work conversations, and 13\% for non-work.

The aggregate symmetry documented in Figure~\ref{fig:figure_4_11} suggests that language choice between work and non-work activities is mostly driven by stable, country-level characteristics. However, this high-level view may obscure important variations across different types of non-work activities. So does the propensity to code-switch into a non-primary language remain uniform across all non-work tasks, or do certain tasks exhibit different linguistic choices?

\begin{figure}[htbp]
    \centering
    \caption{Language-Switching Does Not Seem To Be Systematically Correlated with Economic Value}
    \label{fig:figure_4_12}
    \includegraphics[width=\textwidth]{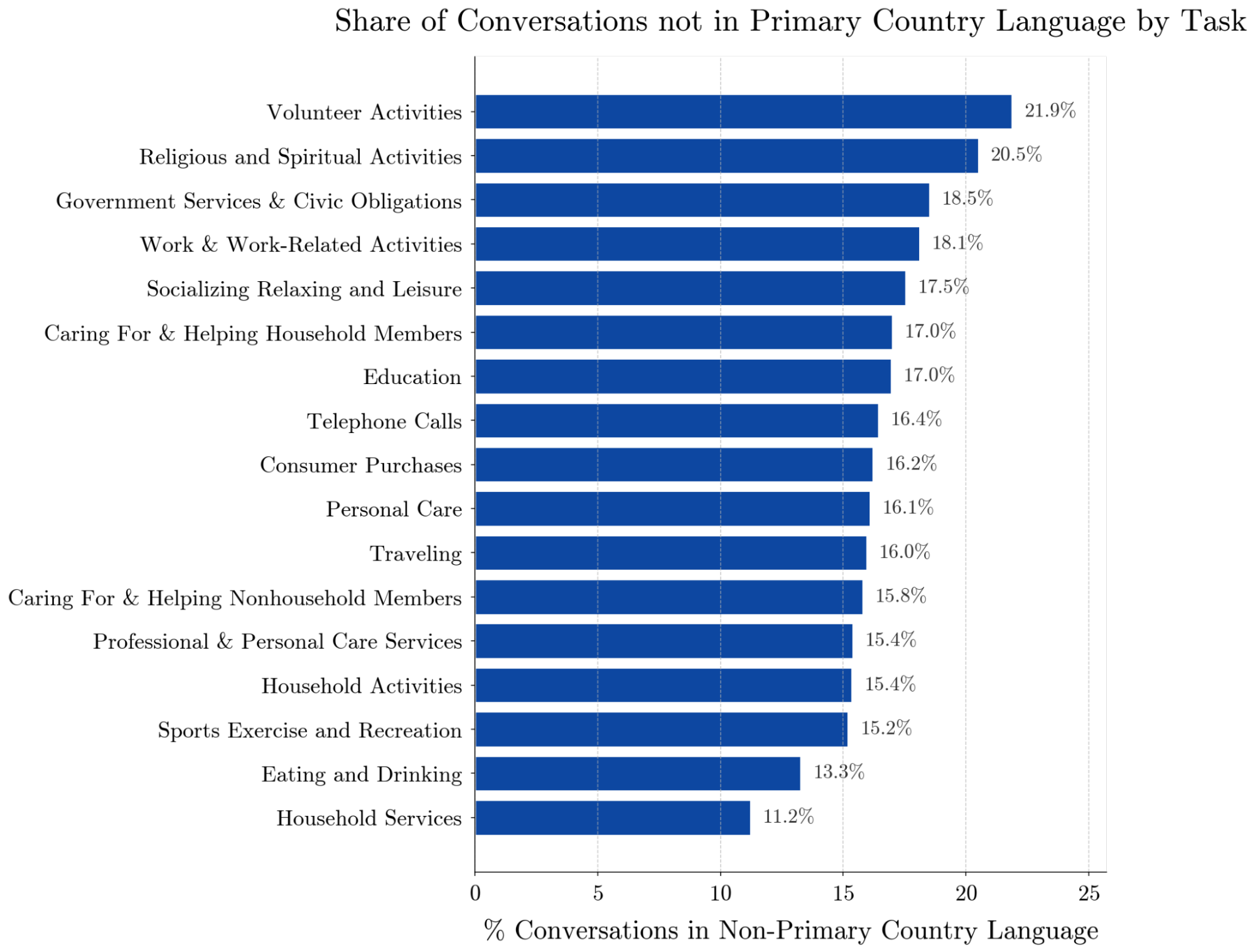}
    \vspace{0.2cm}
    \parbox{\textwidth}{\footnotesize \textit{Notes:} This figure shows the percentage of AI conversations conducted in a non-primary country language across major American Time Use Survey (ATUS) activity domains. Activity categories are sorted by overall non-primary language share (requiring at least 100 conversations per activity-country pair). Non-primary language usage is highest in community, civic, and personal identity domains—led by Volunteer Activities (21.9\%), Religious \& Spiritual Activities (20.5\%), and Government Services \& Civic Obligations (18.5\%), whereas routine domestic household services (11.2\%) and dining (13.3\%) show the lowest rates. Crucially, high-value economic tasks like comparison shopping, education, and professional services sit near the average, suggesting that users do not systematically shift economic tasks away from native languages.}
\end{figure}

To answer this, we examine how language choices vary across the major activities in the American Time Use Survey (ATUS) taxonomy.\footnote{See Section \ref{sec:home_usage} for more information on ATUS and its taxonomy.} The data plotted in Figure~\ref{fig:figure_4_12} shows that the propensity to switch to a non-primary language varies across different tasks. The absolute propensity to use a non-primary language is most highly concentrated in community, spiritual, and professional activities, led by Volunteer Activities (21.9\%), Religious and Spiritual Activities (20.5\%), and Government Services \& Civic Obligations (18.5\%). Conversely, more common daily routines, such as Household Services (11.2\%) and Eating and Drinking (13.3\%), exhibit the lowest overall rates of non-primary language usage.\footnote{Ultimately, this granular breakdown does not offer very strong evidence of the task-sorting hypothesis. If users were code-switching primarily to bypass AI quality penalties on high-stakes reasoning or complex tasks, we would expect non-primary language usage to spike heavily in technical or economically dense domains. Instead, the prominence of non-primary language usage in volunteer, religious, and civic activities implies a different mechanism. It suggests that language choice in human-AI interaction is less about optimizing for model performance, and is instead structurally driven by the sociolinguistic context of the activity itself—such as translating spiritual texts or perhaps interacting with global communities.} Nonetheless, this breakdown does not strongly suggest that more economically important activities occur outside of primary languages, with activities like education, purchases or professional services being around the average non-primary language usage.

Finally, our analysis highlights that there might be efficiency costs associated with AI conversations outside primary languages. We find that interacting in a non-primary country language, such as English, may increase session length and verbosity. Even after isolating the language effect from task sorting via granular ATUS activity and language fixed effects\footnote{We note a related NLP literature on ``tokenization bias'' \citep{Lundin2026,ahia2023languagescostsametokenization}. Because standard large language model tokenizers are often optimized for Western, Latin-script languages, non-English scripts may require more tokens to represent the same text.} non-primary English conversations drive 9–12\% higher turn counts and 18–20\% more total tokens, alongside a slight increase in technical complexity. These results suggest that utilizing AI in a non-primary language may be associated with a higher interaction ``cost,'' or that users systematically sort more complex queries (within a task category) into English. Ultimately, this underscores that investments in high-quality multilingual AI capabilities \citep[e.g.,][]{bjorkegren2025aileapfrogwebevidence, baroplevine2026afrigpt} could yield long-term dividends not only for user welfare and access, but also for underlying human-AI interaction efficiency.\footnote{For the full empirical specification—including our progression of ATUS fixed effects—and detailed regression results, please see Appendix \ref{app:geography}.}

\section{Discussion and Open Questions}\label{sec:discussion}

One defining characteristic of a General Purpose Technology \citep{Lipsey2005, NBERw4148}, from the steam engine to the computer, is its pervasiveness. To drive aggregate productivity upward, a technology must first diffuse widely across sectors, geographies, and the routines of daily life. Our analysis of 15 million AI interactions covering more than 800 occupations, 4000 tasks, 300 household activities, 150 countries, and 140 languages suggests that generative AI is rapidly clearing this hurdle.

However, the extent to which AI will drive positive economic and societal impacts remains subject to debate. Those optimistic about the effects of AI focus on the evidence of labor augmentation \citep[based on evidence like][]{DellAcqua2026, Noy2023}, skill-leveling \citep[based on evidence like][]{Brynjolfsson2025, Wilmers2024Generative}, and foresee a surge in total factor productivity \citep[as explored in e.g.,][]{brynjolfsson2023macroeconomic, arnon2025projected}. Others take a more cautionary view, warning of severe chances of labor market dislocation \citep[e.g.,][]{amodei2026adolescence}, a decline in entry-level hiring \citep[based on evidence like][]{brynjolfsson2025canaries, HosseiniMaasoum2025, KleinTeeselink2025}, and a widening digital divide \citep[e.g.,][]{Gmyrek2024, gmyrek2026disruption}.\footnote{Economists (e.g. \cite{Nguyen2026}) already estimate the dividends from the AI economy to be in the hundreds of billions. It is largely an open question how high these numbers will be, depending on not just future capabilities, but also the ability of firms and households to translate them to valuable market and non-market tasks. Nonetheless, the data on the macroeconomic impact, both from a labor market and outside of the labor market perspective is still mixed and preliminary (e.g. \cite{kolko2026ai, tedeschi2026decline, NBERw33777})} By mapping real-world usage logs to standardized economic frameworks, ATLAS provides an early empirical basis for evaluating some of these competing narratives, acknowledging that significantly more study is needed to validate and explore these findings.

First, we find robust initial evidence of widespread but ``shallow'' diffusion at work. AI has reached 68\% of occupations globally yet it covers only a median of 21\% of the constituent tasks within those occupations. This pattern is largely consistent with the labor-augmenting view that technology may help automate specific tasks rather than entire jobs. We also find that Gemini usage is most concentrated in non-routine cognitive tasks, and to some extent in routine cognitive tasks, whereas manual and interpersonal tasks are less affected. Moreover, a majority of the usage seems to be collaborative in nature rather than seeking out task automation (e.g., less than 10\% for non-routine cognitive tasks). Nonetheless, reassurance in the present should not preclude vigilance in the future, especially as the capabilities of AI models are improving fast \citep[e.g.][]{metr-2026-time-horizons}.

Second, while a lot of productivity debates are focused in the workplace, we observe most conversational AI usage happens at home: over 86\% of conversations occur outside formal work. In the household, AI usage is highly correlated with human daily time use, meaning that individuals use AI for a variety of physical and non-physical household tasks. We also see a striking over-representation for high-friction services, such as seeking financial and legal help or navigating government and civic services. Many of these interactions are happening outside of standard working hours, allowing users to get advice on nights and weekends. Putting all of this together, under conservative assumptions, this household production translates to tens of billions in annual economic value just in the US— gains likely omitted from traditional GDP metrics.

Third, the data offers cause for concern about a digital divide: we observe a steep positive gradient between AI usage and occupational earnings, where a 1\% wage increase is associated with a more than 2.5\% increase in AI usage intensity. In line with the previous literature, this is further mirrored at the macro level, where national AI adoption scales nearly proportionally with GDP per capita. Indeed, countries in the lowest-usage quintile of AI usage represent 17\% of the global population generate just 2\% of usage, while the top quintile represents only 11\% of the population yet drive 30\% of total AI conversations. This suggests the need for further research on the causes of this stubbornly high disparity and what may be done to alleviate it.

Our findings also highlight some important open questions that require further investigation. First, because ATLAS measures de-identified interactions rather than productivity outcomes, we cannot yet answer how (and if) these micro-level task savings translate into macro-level total factor productivity increases. Connecting these signals to realized performance remains a question for future research. Second, because our dataset has yet to include enterprise-contracted logs, analysis of corporate-level integrations and the emerging impact of agentic workflows will come in the future. Third, while our data is rich in descriptive correlations, it remains fundamentally observational, leaving the causal impacts of generative AI on topics like labor substitution, wage distribution, and skill formation an open question for future (quasi-)experimental research.

Ultimately, we view the ATLAS study as the foundation for a broader research agenda for our team. We look forward to both extracting deeper insights from our current dataset while expanding our focus to investigate these emerging frontiers of AI adoption and its current and future economic impact.

\newpage

\bibliography{references.bib}

\newpage
\appendix
\textbf{\Huge Appendix}

\section{Data and Methodology}
\label{app:app1}

This section provides further detail about the data sources for ATLAS, the methodological steps in preparing the dataset and privacy preservation measures we took at every step.

In brief, the fully automated pipeline for data processing proceeds as follows. We sample a large set of conversations from Gemini App, Gemini AI Mode and Gemini API, redact any personal information, and use Gemini to classify these redacted conversations into work and non-work usage of AI (see Appendix \ref{app:app_1_1}). This distinction is important to establish early on as work and non-work activities require different summary facets and map to different statistical taxonomies, so they need to be processed in different data pipelines. After the triage into separate work and non-work pipelines, conversations are summarized, and those summaries grouped into semantically similar clusters using Observation Clustering and Taxonomy Organisation (OCTO), a bespoke tool developed for this purpose by Google DeepMind. The clusters are then mapped by OCTO to official statistical taxonomies (American Time Use Survey for non-work and O*NET-SOC categories for work) (see Appendix \ref{app:app_1_2}). The final processing stage involves adding additional metadata annotations, applying weights to our sample to make it representative of the population of AI use cases on the three sources we use (see Appendix \ref{app:app_1_3}). At every stage, the pipeline involves extensive privacy preserving features: multiple lines of defense of personal data redaction, conversation and cluster summaries being restricted to topics required for downstream classification, filtering out of any clusters or segments which do not meet our k-anonymization bar of containing 10 distinct users, and strict access restrictions to the final processed dataset.

The rest of this section provides additional detail on all these aspects of our methodology.

\subsection{Data Ingestion and Pre-processing}
\label{app:app_1_1}

\subsubsection{Data Sources}

AI is increasingly powering features across many technology products, so measuring the scope of economically relevant AI interactions on Google platforms is at least to some degree a matter of judgement. Our choices for the first version of ATLAS dataset were informed by (i) availability of log data, and (ii) scope for multi-turn, user-driven interactions which are likely to contain more context on user intent to enable clustering algorithms to group similar activities together, and (iii) overall usage. This led us to focus on Gemini App and Google AI Mode (AIM) among potential conversational AI tools and Gemini API for its high relevance for automation and developer-focused use cases. Table~\ref{tab:table_app1_1} provides further detail on each of these sources, and the key exclusions from the data.

\begin{table}[htbp]
    \centering
    \caption{ATLAS Data Sources and Scope}
    \label{tab:table_app1_1}
\small {\renewcommand{\arraystretch}{1.5} 
\begin{tabular}{
    >{\raggedright\arraybackslash}p{2cm}
    >{\raggedright\arraybackslash}p{3.8cm}
    >{\raggedright\arraybackslash}p{3.8cm}
    >{\raggedright\arraybackslash}p{3.8cm}
}
        \toprule
        \textbf{Surface} & \textbf{Scope limitations} & \textbf{Data Unit} & \textbf{Geographic Coverage} \\
        \midrule
        Gemini App & Interactive Gemini App conversations, on the Web and Mobile apps (Android/iOS) & Turn-based conversation sequence (excluding file attachments) & Available globally (250 countries and territories) \\
        Google AI Mode (AIM) & All interactive AI Mode conversations. & Turn-based conversation sequence (excluding file attachments and non-AI mode search queries) & Adheres to standard official AI mode availability \\
        Gemini API & Direct API calls and Google AI Studio (excluding Vertex AI/GCP and Gemini LM) & Individual API request and responses (no persistent conversation ID) & Only request counts (not content) are available for the EEA, the UK, Switzerland and all paid services \\
        \bottomrule
    \end{tabular}}
\end{table}

\subsubsection{Sampling}
The sample for the current version of ATLAS is drawn from interaction logs between April 6, 2026 and April 19, 2026. On each of the surfaces noted above, we sample approximately 5 million interactions---full conversations (for Gemini App and AIM) and request-response pairs (for Gemini API)---which satisfy basic data cleaning and materiality filters.\footnote{The actual number of sampled observations exceeds 5m per surface to partly offset the sample attrition from k-anonymization filtering after clustering, which drops clusters with fewer than 10 unique users. Due to request and response text not being logged for paid Gemini API usage, the sample for the Gemini API component of the main ATLAS pipeline was drawn from free API requests only.} These samples are then re-weighted by the overall interaction shares of each source for analyses featured in this report. Queries that are below 10 tokens, originate from internal Google accounts, or fail to generate a model response (for example due to the prompt violating Google usage policies) are excluded from this sampling process.

\subsubsection{PII Removal}
A core tenet of the origination pipeline is the isolation and stripping of Personally Identifiable Information (PII) before semantic clustering begins. Via Google’s internal privacy-preserving tools, the pipeline uses a series of extensive public Data Loss Prevention (DLP) filters to redact PII from conversations including personal contact details, financial information, security credentials, government-issued identifiers and health information, as well as specific hardware indicators. In addition to that, all internal identifiers in the logged data are replaced with mathematically unlinked Universally Unique Identifiers (UUIDs) specific to the ATLAS pipeline, which fully obfuscates user identities and prevents data from being joined back to product logs.

\subsubsection{Work/Non-Work Classification and Summaries}
It is important to separate work and non-work observations early on because the downstream economic analysis of these two domains is distinct. For instance, different statistical taxonomies would be relevant for these two domains, meaning different summary information is relevant for identifying the appropriate taxonomy labels.

For this reason, the first step of our analytical pipeline is to triage the redacted observations into two separate processing sequences using a classifier to determine whether the conversation involves a task performed in the course of active employment (henceforth ``the work/non-work classifier''). Due to our interest in aligning work activities with occupation definitions used by statistical agencies, we define ``active employment'' to exclude activities such as job seeking, interview preparation, clear academic study or general non-business-related upskilling. As this is the only classification step that uses the text of the original conversation, the classifier also provides an estimate of whether the way in which the query was expressed is indicative of high, medium or low ``domain knowledge'' relating to the subject of conversation.\footnote{``Domain knowledge'' about the activity is distinct from the inherent expertise of the activity. The former is defined with respect to the complexity and domain specificity of language and conceptual frameworks used in asking the query, whereas the latter is about the inherent complexity of the objective the query aims to achieve.}

After conversations are routed into work and non-work pipelines, each is summarized using facets that focus specifically on the objectives of downstream classification into professional or household time use taxonomies. For instance, work conversation summary facets include a 1-2 sentence description of the core professional task to which the request likely contributes and a list of most pertinent ``Specialized Knowledge and Tools'' involved in the request. To maintain privacy throughout the data processing, both work and non-work summarization prompts reiterate explicit instructions to avoid including any PII, and the summarization facets are closely aligned with downstream classification category definitions to avoid retaining information that is not directly critical for the objective of analysis.

\subsection{Cluster Creation and Analysis}
\label{app:app_1_2}

\subsubsection{Clustering Methodology}
Text data analysis, and large language model use specifically, are increasingly enabling economic research in academia \citep{10.1257/jel.20181020, 10.1257/jel.20231736, NBERw33941} and industry \citep{tamkin2024clioprivacypreservinginsightsrealworld, Eloundou2024, tomlinson2025workingaimeasuringapplicability,verma2026how,NBERw34255}. To support research, as well as other internal use cases, Google DeepMind has developed Observation Clustering and Taxonomy Organisation (OCTO), a tool for clustering and hierarchical taxonomy assignment of text logs at Google scale (this is similar in purpose to Anthropic's \citet{tamkin2024clioprivacypreservinginsightsrealworld}).

OCTO performs the clustering step as follows. First, text data inputs (e.g. conversation summaries) are embedded into high-dimensional vector space using the Gemini text embedding model \citep{https://doi.org/10.48550/arxiv.2503.07891}. Next, a weighted, undirected graph of these representations is constructed, with nodes representing embeddings of the individual observations and edge weights corresponding to the similarity between the embedding vectors. For scalability, edges are formed only between embeddings that share the same Locality Sensitive Hash \citep[SimHash,][]{Charikar2002} to avoid evaluating probabilistically distant pairs. To maximize recall, this process is repeated across multiple independent projection vectors to yield an approximate nearest neighbor graph. Finally, the resulting graph acts as the input to affinity clustering, a large-scale hierarchical agglomerative clustering approach \citep{46700}. The algorithm iteratively merges nodes with their most similar neighbours to form distinct clusters based on the underlying similarity of the summarised conversations. The number of clusters is selected automatically at each level of the hierarchy based on user-defined parameters, such as the target cluster size and the minimum edge weight threshold required for a merge.

Following clustering, a Gemini model is used to generate a title and summary for each granular cluster. Similar to \citet{tamkin2024clioprivacypreservinginsightsrealworld}, this step uses sampled conversation summaries from the focal cluster as well as contrastive samples from neighbouring but similar clusters to further refine the labels.

In ATLAS, we process our data in separate work and non-work subsets with OCTO to aggregate the de-identified conversation summaries into semantically similar clusters. We use Gemini 3.1 Flash Lite to label and summarize these clusters and apply prompts customized for work and non-work pipelines, which focus on extracting summary facets that are most relevant for downstream taxonomy assignment.

All clusters representing activity of fewer than 10 unique users are automatically discarded and are not subject to any further classification or analysis. After the full processing pipeline described in the rest of this section, the final ATLAS dataset contains 550,407 conversation clusters representing 14,653,926 interactions, which are split approximately equally between the three surfaces.\footnote{The number of interactions represented in the final sample is lower than the numbers sampled at the outset and no longer exactly equal across surfaces, due to clusters containing fewer than 10 unique users being eliminated from the data immediately at the k-anonymization part of our clustering pipeline.}

\subsubsection{Primary Statistical Taxonomies}
From the earliest stages of ATLAS development, we set the goal of mapping both work and non-work data to existing statistical taxonomies. Official taxonomies enable connections to external datasets and comparative analyses with other studies that organize the data in the same way, both of which are critical for meaningful economic analysis. No individual way of taxonomizing work or human life is perfect, so our choice of the specific taxonomies was guided by maximizing (i) granularity and (ii) breadth of compatible economic research and data sources.

We map our work clusters to US Bureau of Labor Statistics (BLS) 2018 Standard Occupational Classification (SOC) \citep{bls_soc_2018}, further broken down into occupational titles and tasks from Occupational Information Network (O*NET) Database v30.2 \citep{onet_30_2}.\footnote{All analysis in this report used O*NET Database v30.2, unless stated otherwise.} This combined taxonomy for economic analysis contains 23 major groups, 95 minor groups,  1,016 occupation titles and 18,797 tasks. Our synthetic data testing revealed that naive taxonomy tree traversal is sub-optimal for work classification, due to the large number of available occupations, and the number of taxonomy levels, each presenting a cumulative misclassification risk. Instead, for the actual process of taxonomy assignment in ATLAS we ``compress'' the SOC taxonomy by grouping the occupation titles (accompanied by granular Gemini-generated descriptions based on metadata from BLS and O*NET) into 34 custom labelled clusters using OCTO.\footnote{The result is a three-level occupational taxonomy, with 34 bespoke categories replace the top three layers of the SOC, directly nesting detailed occupation titles, which in turn are mapped to their constituent O*NET tasks.}  We select the clustering parameters by performing hyperparameter tuning, which optimizes a loss function based on cluster density penalized for deviation from the target number of clusters.

Non-work clusters are mapped to BLS 2024 American Time Use Survey (ATUS) Activity Lexicon \citep{bls_atus_lexicon_2024} categories, consisting of 17 Major categories, 108 Tier 2 categories and 456 Tier 3 categories.

Mapping global AI interactions to taxonomies from US statistical agencies creates a higher risk of misclassification in regions with significant occupations, tasks or activities that have no US equivalents yet attract sufficient AI usage to pass privacy thresholds in our data. One mitigating factor, however, is that both O*NET tasks and ATUS categories are fairly general and wide ranging. In circumstances where differences between countries arise primarily from different emphasis on generic constituent parts (e.g. different relative importance of specific tasks within a job, or share of total employment for a particular occupation), our description-based classifiers should still assign relevant labels. Nonetheless, our current approach makes it impossible to assess whether an occupation or activity is over-represented in a non-US country’s AI use data relative to relevant employment shares or population time use, as there is little reason to assume the US distribution of jobs or activities within the population would apply to other countries. 

\subsubsection{Taxonomy Assignment Methodology}
In both work and non-work pipelines, labeled clusters are mapped to their respective hierarchical taxonomies using OCTO. Starting at the top of the target taxonomy (e.g. American Time Use Survey Major Category), we use Gemini 3.1 Flash Lite to map clusters to one of the top-level categories.\footnote{Gemini 3.1 Flash Lite is used in this pipeline for scale, but we have found its classifications to be robust to using more compute-heavy models, e.g. 3.1 Flash, 3.1 Pro.} Each inference call uses constrained decoding \citep{https://doi.org/10.48550/arxiv.2407.08103} to ensure classification outputs conform to one of the provided taxa only. This approach is then applied recursively to traverse the taxonomy, assigning clusters at each level to the most suitable category within the set of descendants from the category selected at a higher level. Due to the evidence of systematic bias in LLM classification arising from the order in which options are presented \citep{https://doi.org/10.48550/arxiv.2308.11483, https://doi.org/10.48550/arxiv.2506.14092}, the order of options is fully randomised for each classification call.

To provide greater context for classification, taxonomy categories at each level were accompanied by LLM-generated descriptions, which summarize additional available metadata from statistical agencies in ways that align with the cluster summary facets. To generate the summaries, we traverse the taxonomy from its most detailed level to higher order ones, recursively aggregating and summarizing the metadata at each node. The sources of metadata used in this exercise include e.g. Occupation Descriptions and Technology Skills datasets from Occupational Information Network (O*NET) Database for work taxonomies and `Example’ fields from 2024 ATUS Activity Lexicon or Occupation Descriptions from SOC. This process is fully automated; at no point does a human view the content of the conversations being clustered. 

\subsection{Post-processing for Analysis}
\label{app:app_1_3}

We perform additional post-processing of the taxonomized work and non-work clusters described above. First, we augment the cluster annotations dataset with additional bespoke classifiers unrelated to official statistical taxonomies. In some cases, for instance the work task intent assessment,  the classification uses cluster summaries (and/or additional labels) as inputs.  In other cases, we perform classification at the level of taxonomy category like O*NET tasks or ATUS Tier 3 activities, and then map the annotation labels to clusters based on their assigned taxonomy category. As the scope of these classifiers is narrow, their methodologies are explained in the sections of the report where they are used. 

Second, due to some of the metadata (e.g. country, language, flags for multimodal use) in the ATLAS dataset cutting across clusters, we enforce a strict constraint that no data segment (i.e. a desired combination of variables, such as country x occupation) containing fewer than 10 unique users can be materialized or used for analysis at any stage.

Finally, when calculating the number of conversations associated with any combination of variables that cuts across multiple data sources, we apply statistical weights to rebalance our sample based on the underlying global population of conversations in each country-surface cell. This ensures that drawing equal samples across differently sized surfaces does not give smaller surfaces a disproportionately large weight in calculating aggregate trends. 

\subsection{Data Governance}
The production of the ATLAS dataset is overseen by internal product counsel and dedicated Privacy Working Groups (PWG) to ensure strict compliance with internal privacy and data handling policies.  The processing of the underlying data is fully automated, and the content of conversations or even individual conversation summaries are not retained in the final ATLAS dataset. Access to the dataset is heavily restricted to a small team of researchers. All artifacts are bound by stringent automated wipeout policies and managed via robust data governance tags that prevent inadvertent cross-system leakage.

\pagebreak

\section{Classifier Validation}
\label{app:classifier_validation}

This Appendix summarizes the challenges of assessing performance of classifiers that map AI use logs to existing statistical taxonomies, and our developing thinking about the solutions to these issues. We also report results of three kinds of classifier validation: accuracy testing with synthetic validation samples, inter-rater agreement, and human approval of AI labels.\footnote{More broadly, economics, political science and other social sciences showed an increasing focus on using text data and automated classifier; see e.g., \citet{Grimmer2022} or \citet{10.1257/jel.20181020}.} We view transparent reporting of classification accuracy as an important contribution to the broader economic work on AI usage measurement. We describe some of our innovations for understanding the strengths and limitations of statistical taxonomy assignment to text data on AI interactions.

The methodology and classifier performance reported here is a starting point for ATLAS, not the steady-state.  We consider the improvement of both classification performance and of researchers’ understanding of inaccuracies and bias (if any) in these classifiers a critical aspect of research of economic impact on AI in the long term. 

\subsection{Measurement Challenges}
\subsubsection{Fundamental Challenges of Taxonomy Assignment}
Classifying open-ended AI conversations into standardized economic taxonomies is intrinsically difficult. Consider the scale of the category space alone. At granular levels, the ATUS taxonomy contains 456 distinct activity codes, and O*NET-SOC contains 1,016 occupational titles. A classifier choosing at random would hit expected accuracies of just 0.2\% and 0.1\%, respectively. Even at the coarse SOC 2-digit level (23 categories), random assignment is correct less than 4.4\% of the time.

But the difficulty extends beyond the sheer number of categories. Many conversations are ambiguous in their occupational or activity content. A user asking an AI assistant to "help me understand the tax implications of this contract" could plausibly be classified as a lawyer, an accountant or a business operations specialist, all of which are distinct occupational titles in O*NET-SOC. Even at the highest level of aggregation of the SOC taxonomy, major groups can be hard to distinguish. A user asking about "managing a departmental budget and preparing quarterly financial forecasts" could work in a Management Occupation (SOC 11) or Business and Financial Operations Occupation (SOC 13). The same issues apply to time use categories: an adult asking AI for instructions on building a nest box with a 10-year old could do so for their own child (`Arts and crafts with household children') or their visiting nephew (`Arts and crafts with non-household children'). And yet these two activities fall under distinct categories even at the most aggregated level of the ATUS taxonomy.

While the challenges are considerable, they are not necessarily unique to applying statistical taxonomies to AI use. The ambiguity we describe is largely inherent in the categories themselves due to their semantic similarity and the degree of overlap. For instance, `helping household adults', `looking after household adults' and `organization and planning for household adults', are all distinct Tier 3 ATUS categories. Any differences between them appear highly subjective. For occupations, it is not self-evident whether `managing departmental budgets' falls within the scope of responsibilities of a worker in `Management' or `Business and Financial Operations' even if there is no AI involved, and the definitions of those job roles can vary across companies.

In fact, material measurement errors arising from inconsistencies in humans’ applications of statistical taxonomy categories are well documented. \cite{mellow1983} found that 17\% of CPS respondents disagreed with the information provided by their employers about their major (1 digit) occupations (comprising only 11 categories at the time), rising to 42\% disagreeing about the detailed (3 digit) occupation. \cite{10.1086/269327} finds even greater employee-employer disagreement on occupation coding at 24\% and 48\% for 1 digit and 3 digit occupations, respectively.

Given this context, significant uncertainty in LLM-based mapping of text data to granular statistical taxonomies is unavoidable. Understanding the degree and nature of these inaccuracies is a critical research priority for us, as these methodologies form the backbone of economic research on AI adoption. What is understood can be managed. We have already used accuracy measurement to improve our classifiers in ATLAS, and understanding noise patterns has guided our focus toward more robust task aggregations. Further research can help determine how traditional econometric knowledge regarding measurement error \citep[e.g.,][]{Bound2001} can best improve economic research on AI use.

\subsubsection{Classifier Validation Under Information Constraints}
Assessing classifier performance in contexts like ATLAS is inherently difficult due to lack of  ground truth data. Even if unrestricted access to raw logs were possible, the correct category for any given interaction would still remain ambiguous unless users explicitly stated their occupation and O*NET task they were performing. In practice, of course, the summaries of clusters of summaries of user interactions contain even less context for definitive category annotation for a labelled training set. 

To solve this, we used Gemini 3.1 Flash Lite to generate a synthetic ground-truth dataset comprising 16,050 work-related and 2,751 non-work interactions. We seeded the each generation request with randomly chosen target categories from the most granular tiers of the O*NET-SOC and ATUS taxonomies, instructing the model to produce authentic prompts a worker might realistically submit. Because no personal information was involved, we could extensively validate the output to ensure diversity and realism.

To ensure these synthetic validation sample reflects the complexities and imperfections of production logs, we injected three specific empirical challenges:
\begin{itemize}
    \item \textbf{Short Fragments}: For a material subset of observations, chosen at random, we instructed the model to produce brief and underspecified single-sentence queries lacking full context. 
    \item \textbf{Long and Complex Requests}: For another randomly chosen substantial subset of observations, we instructed the model to produce a long and detailed prompt paired with realistic, and frequently messy, attachments (e.g., raw pasted CSVs, messy meeting transcripts, or synthetic code blocks).
    \item \textbf{Style variation and informal text}: Prompts contained explicit instructions to use natural human language, including typos, abbreviations or fragmented thoughts. Style variation was also injected.
\end{itemize}

While rigorously validated, we recognize that synthetic data remains an imperfect benchmark that risks introducing two opposing biases. First, classification may be artificially easier if the model inadvertently over-shares context compared to real users or if, in contrast to real human language, Gemini-generated descriptions contain subtle language cues especially recoverable by a Gemini-based classifier. This is a bias we cannot fully measure without inspecting private logs at scale, which is clearly not a viable option. Conversely, classification might also be artificially harder. Given the inherent ambiguity of taxonomy categories, a generated prompt might legitimately match a different occupation or task better than its seeded ground truth. In such cases, an accurate classifier could be unjustly penalized for picking the best semantic fit over the imperfect transmission of the original seed.

\subsection{Assessing Accuracy with Synthetic Conversation Data}
\subsubsection{Result Overview}
We pass synthetic conversations through the processing pipeline (summarization, clustering, cluster summarization, and hierarchical classification).\footnote{While the clustering algorithm was the same, to produce meaningful clustering results for downstream analysis we had to change the number of compression rounds and merge similarity thresholds, reflecting a much smaller input dataset in this exercise compared to the original pipeline.} Finally, we compare the classifier's outputs against the known true labels.

The first and perhaps most consequential classification decision is whether a conversation is work-related or not. This binary determination gates all subsequent analysis: work-classified conversations are routed to the SOC/O*NET pipeline, while non-work conversations are routed to the ATUS pipeline. An error at this stage propagates through all downstream results.

This is a difficult problem. For example, a user asking for help understanding a calculus concept might be a student (non-work) or a tutor/instructor preparing materials (work). Similarly, conversations involving legal research, financial planning, or creative writing frequently straddle the work/non-work boundary, since these domains feature prominently in both professional and personal contexts. Nonetheless, our work/non-work classifier achieves an accuracy of 93.7\% on a balanced synthetic validation set that pools work and non-work observations.\footnote{Because the full validation dataset is skewed toward work, we report classification accuracy as a mean across 100 independent bootstrapping simulations. In each iteration, accuracy was measured in a balanced sample containing random samples of 2,750 conversations, each from work and non-work synthetic validation samples.}

Moving to the official statistical taxonomies, the accuracy of the work classifier varies systematically across levels of taxonomic granularity, as one would expect. Coarser classifications are substantially more accurate than fine-grained ones. Table~\ref{tab:table_app2_1} reports cluster-level accuracy---meaning the rate at which the classifier assigns a label to a cluster which matches the modal (or tied modal) category in the cluster---at each level of the SOC/O*NET hierarchy alongside baseline random chance.

\begin{table}[htbp]
    \centering
    \caption{Automated Classifier Performance Significantly Outperforms Random Chance, Especially in Fine-Grained Taxonomies}
    \label{tab:table_app2_1}
 \begin{tabular}{lccccc}
        \toprule
        \textbf{Taxonomy} & \textbf{Granularity Level} & \textbf{N} & \textbf{Random \% } & \textbf{Acc. \%} & \textbf{Ratio} \\
        \midrule
        ATUS (Non-Work) & Tier 1 (Major) & 17 & 5.88\% & 72.84\% & 12.4x \\
        ATUS (Non-Work) & Tier 2 (Intermed.) & 108 & 0.93\% & 42.91\% & 46.3x \\
        ATUS (Non-Work) & Tier 3 (Detailed) & 456 & 0.22\% & 23.70\% & 108.1x \\
        SOC (Work) & Major Group & 23 & 4.35\% & 71.57\% & 16.5x \\
        SOC (Work) & Minor Group & 98 & 1.02\% & 58.44\% & 57.3x \\
        SOC (Work) & Occupation Title & 1,016 & 0.10\% & 42.47\% & 432x \\
        O*NET (Work) & Specific Tasks & 18,797 & 0.005\% & 22.58\% & 4,244x \\
        O*NET (Work) & Autor-Thompson  & 5 & 20\% & 70.44\% & 3.5x \\
        \bottomrule
    \end{tabular}
    \vspace{0.2cm}
    \parbox{\textwidth}{\footnotesize \textit{Notes:} This table evaluates the classification accuracy of the automated ATLAS pipeline across hierarchical levels of official non-work (ATUS) and work (SOC / O*NET-SOC) economic taxonomies. Baseline Random Chance represents uniform random assignment across all candidate categories in a given tier. Accuracy measures the percentage of sample conversation clusters assigned to the correct modal ground-truth category. The Informational Lift Ratio measures accuracy relative to random chance. While absolute accuracy naturally declines as candidate categories expand (e.g., from 71.6\% for 23 SOC Major Groups down to 22.6\% for 18,797 specific O*NET tasks), the model maintains or even improves the discriminative lift (4,244x over chance at the task level). Grouping tasks by \cite{NBERw33941} functional task types yields 70.4\% agreement.}
\end{table}

Three features of these results deserve emphasis. First, despite the taxonomy's inherent ambiguity, accuracy at aggregated levels remains remarkably high. The classifier achieves nearly 72\% accuracy at the SOC major group (2-digit) level, approaching the employee-employer agreement levels observed by \cite{10.1086/269327}, even though the taxonomy at the time contained 11 categories instead of 23 currently. Furthermore, exact accuracy stays above 57\% even at the narrower SOC minor group.

Second, while absolute accuracy naturally declines as the level of granularity of taxonomy level increases and the number of options explodes, the model retains substantial discriminative power. At the specific O*NET task level, for example, the classifier is over 4,200 times more likely to select the correct task than random assignment.

Finally, although accuracy in matching the ground truth for exact O*NET task statements is 22.6\%, the classifier assigns a task with the same \cite{NBERw33941} task type as the ground truth for 70.4\% of clusters. This highlights the robustness benefit of grouping tasks in economically meaningful ways when individual assignments may be noisy. These validation findings have motivated our focus on broad categories of tasks in ATLAS analysis, as well as choice of metrics that are more robust to individual misclassifications (fact of an individual task occurring as compared to its exact frequency).

We also note that accuracy rates for ATUS categories in part reflect the taxonomy’s large number of granular activities which are not meaningfully different in terms of their economic implications (e.g. travel or waiting split by purpose) and are defined by context unlikely to be provided in AI conversations. For instance, a user asking for help building a lego set rarely specifies whether the activity is with a household child (their own) or a non-household child (e.g., a nephew). Similarly, a user asking for travel directions from 'point A to point B' has no economic reason to disclose the purpose of the trip (e.g., distinguishing travel related to household management from travel for consumer purchases), yet the ATUS taxonomy splits these into separate categories. Under alternative aggregations that pool these more ambiguous categories together (rolling up travel categories and merging household and non-household pairs), Tier 2 accuracy improves from 42.9\% to 51.7\% (+8.8 pp) and Tier 3 accuracy improves from 23.7\% to 32.4\% (+8.7 pp).

\subsubsection{Patterns of Misclassification}
Analyzing the empirical distribution of classifier errors provides insight into whether misclassifications are random or systematic. While systematic errors may seem more daunting as they reflect an underlying cause (e.g. conceptual ambiguity) that confounds the classifier, they can also provide a constructive way forward for better models. If established rigorously, systematic patterns of errors could motivate error-prone category merges (where the distinction not economically meaningful), classifier prompt/cluster summary clarification context necessary to draw out differences between problematic pairs, or even ex-post weights to ``reassign'' a share of observations labelled with a particular category to an overlooked but often correct alternative. We present some of the emerging confusion lift matrix analysis below for both transparency. As our synthetic data and models improve we expect these analyses to inform the next iteration of classifiers.

\begin{figure}[htbp]
    \centering
    \caption{Occupational Classifier Misclassifications Concentrate Most Often Within Closely Related White-Collar Domains}
    \label{fig:appendix_figure_2_1}
    \includegraphics[width=\textwidth]{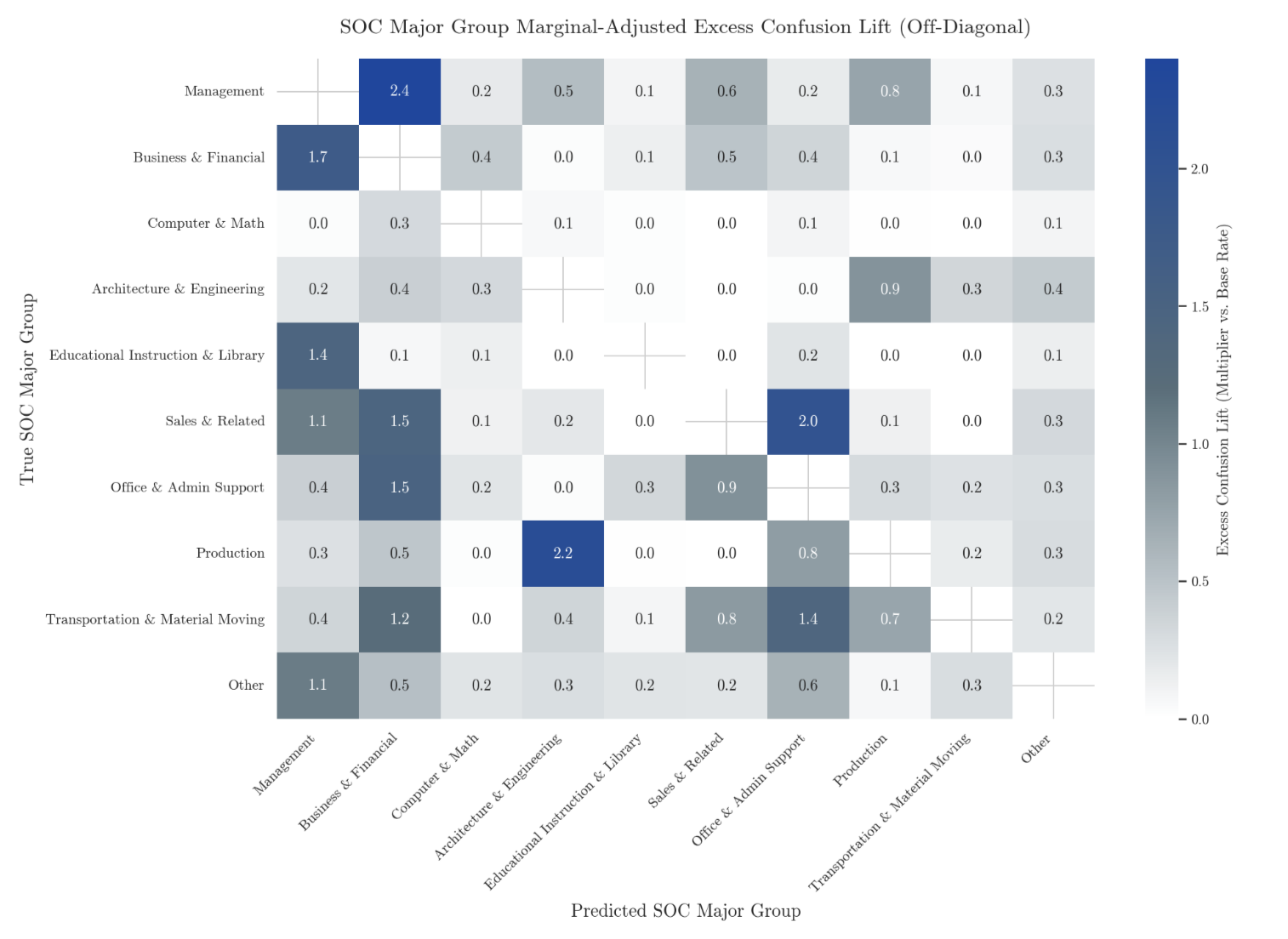}
    \vspace{0.2cm}
    \parbox{\textwidth}{\footnotesize \textit{Notes:} This heatmap displays off-diagonal Excess Confusion Lift multipliers for SOC Major Groups (SOC 2-digit occupations), isolating conceptual classification overlap from overall category sample size. A multiplier of 1.0x indicates that misclassifications occur at the exact rate expected by random chance based on category size. Values above 1.0x indicate systematic conceptual overlap (e.g., 2.0x–3.0x confusion lift between Management and Business/Financial Operations, or between Sales and Administrative Support). Diagonal true-positive matches are omitted to highlight error distributions. The concentration of errors within semantically adjacent fields suggests that pipeline misclassifications generate only small distortions, between economically hard to distinguish activities.}
\end{figure}

\begin{figure}[htbp]
    \centering
    \caption{Time-Use Misclassifications Are Heavily Confined to Context-Silent Tasks}
    \label{fig:appendix_figure_2_2}
    \includegraphics[width=\textwidth]{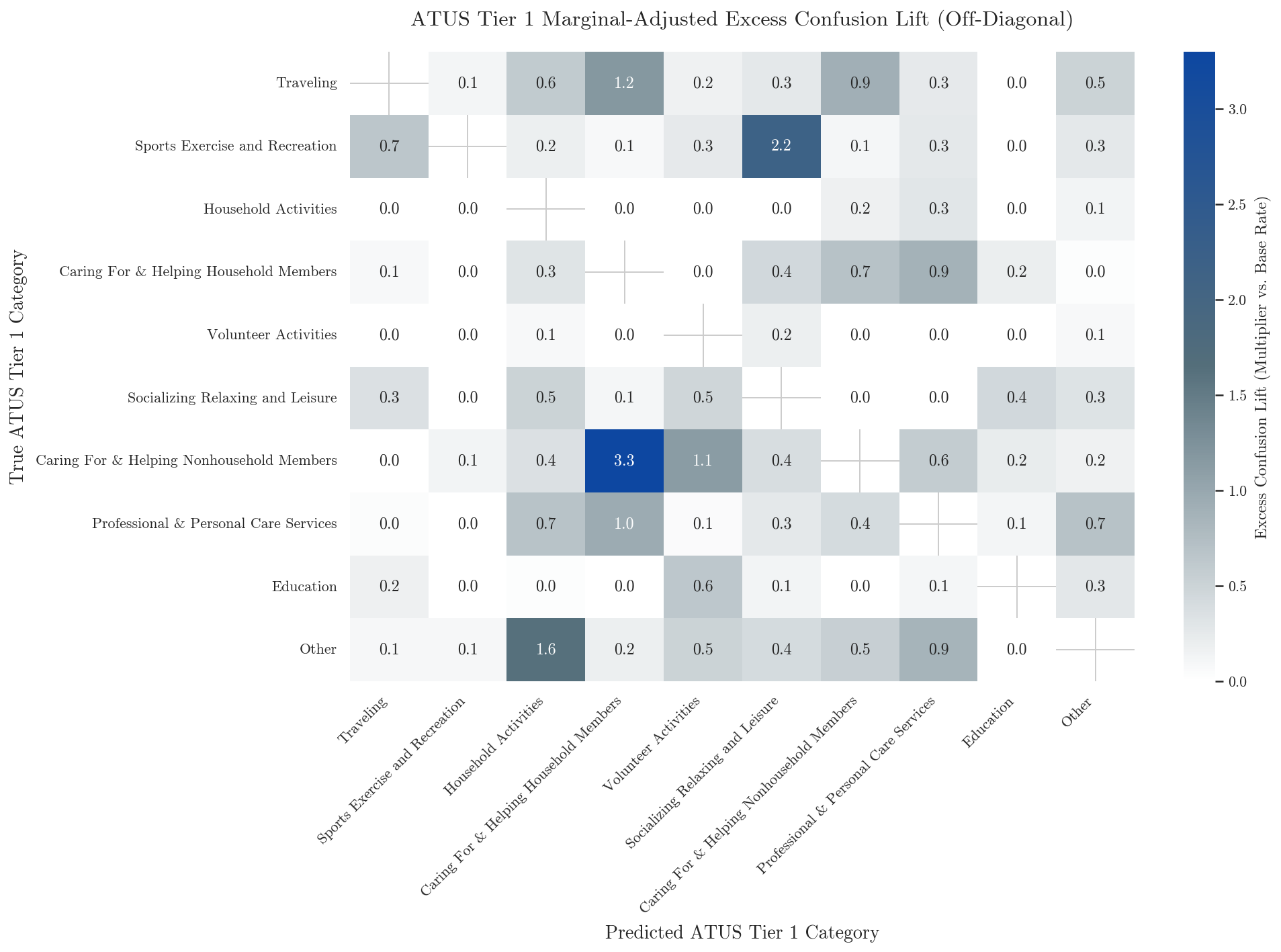}
    \vspace{0.2cm}
    \parbox{\textwidth}{\footnotesize \textit{Notes:} This heatmap displays off-diagonal Excess Confusion Lift multipliers for ATUS Tier 1 major activity categories, measuring systematic misclassifications patterns beyond base-rate category frequency. A value of 1.0x represents base-rate chance error. Systematic confusion is tightly localized to categories where conversational summaries lack context, such as distinguishing whether a caregiving or assembly activity is performed for a household member versus a non-household member. Modest overall confusion lift values across non-work domains confirm that classifier errors do not significantly distort non-market time allocation estimates.}
\end{figure}

Figure~\ref{fig:appendix_figure_2_1} presents the empirical error distribution for SOC Major Groups, while Figure~\ref{fig:appendix_figure_2_2} presents the corresponding distribution for ATUS Tier 1 activities. To draw out the systematic pairs of categories that are confounded particularly often, we normalize each conditional misclassification probability by the category's overall share in the validation sample. To read the numbers in the figures below, think of them simply as multipliers compared to random chance. A value of 1.0 means that when the classifier makes an error, it lands in that target category at the exact normal rate you would expect based purely on how big that category is in the overall share of the validation sample. A value of 2.0 or 3.0 means those two fields are confused two or three times more often than normal. Conversely, a value near 0.0 means those fields are almost never confused. 

Two overarching insights emerge from these Figures. First, the errors make intuitive economic sense. Rather than being scattered randomly across disparate fields (for example, confusing Computer \& Mathematical roles with Production or Transportation), misclassifications concentrate within similar domains. In the occupational space (Figure~\ref{fig:appendix_figure_2_1}), the primary set of confusions occurs between Management and Business \& Financial Operations. Similarly, Sales roles exhibit overlap with Office \& Administrative Support. In the non-work domain (Figure~\ref{fig:appendix_figure_2_2}), errors concentrate heavily where conversational context is silent. For example, caregiving errors occur often between helping household versus non-household members 

Second, across the vast majority of off-diagonal category pairs, the excess confusion multipliers remain modest and smaller than the random-chance baseline. Even in the most overlapping work categories, the lift rarely exceeds a few multiples of the base rate. Because the errors are localized to near-synonymous occupations and modest in magnitude, they are unlikely to generate severe distortions in our aggregate labor market or time-use measurement.

\subsection{Comparing Model Performance to Human Annotators}
As discussed above, accuracy measures could be significantly affected by category ambiguity and by some synthetic conversations resembling other similar categories even more than their ‘ground truth’ as a result. We therefore also validate our automated classifiers against human judgment on real datasets  as an alternative lens on performance. We start by exploring inter-rater agreement metrics for work/not work classification as well as the higher levels of our core statistical taxonomies.  However, these measures are not meaningfully applicable for granular taxonomy levels where the number of possible options vastly exceeds the number of observations that can be feasibly rated by humans.  We therefore also report metrics of human assessment of ‘reasonableness’ of labels assigned at those granular levels. 

For both inter-rater agreement and reasonableness, we used ratings by three in-house annotators. Annotators were instructed to spend a maximum of 4 minutes per cluster to avoid diminishing (and possibly negative) returns of over-deliberation.

For the ATUS and O*NET-SOC classification validation, annotators were provided with samples of 120 summaries of clusters from the relevant work or non-work subset of the final ATLAS dataset. Each of the three surfaces contributed 40 clusters for annotation. As the work/non-work classifier operates at the level of the original conversation, we relied  on publicly available WildChat dataset \citep{zhao2024wildchat1mchatgptinteraction} for examples of classified conversations for human raters to review to protect the privacy of our users.

\subsubsection{Inter-rater Agreement}
We start the measurement of inter-rater agreement by  identifying the plurality consensus among the three human  annotators, defined as the category chosen by at least two out of the three human raters. Next we evaluate the model's performance directly against the human plurality consensus using Cohen's Kappa, applying conservative bounds defined by maximum and minimum possible agreement for observations that were missing plurality. 

Next, we evaluate inter-rater reliability across both human raters and all raters (inclusive of the model), and calculate the delta arising from the LLM being added to the set of raters. For all classifiers, we calculate the Mean Pairwise Cohen's Kappa (the arithmetic mean of all individual annotator pairs' Cohen's Kappas \citep{Light1971}) and  Fleiss's Kappa (the statistical measure for assessing the reliability of agreement between multiple raters classifying items into mutually exclusive categories \citep{Fleiss1971}).

Table \ref{rateragreement} summarizes the results of this analysis. We find high LLM agreement with human plurality across tasks, falling into ‘substantial’ or ‘almost perfect’ agreement categories on the widely cited \citep{Landis1977} kappa evaluation scale. Mean pairwise Cohen’s and Fleiss’ kappas show ‘moderate’ or ‘substantial’ agreement between the LLM classifiers and human annotators. For example, the SOC major groups classification achieved a plurality agreement kappa of 0.83, and Fleiss’ kappa of 0.69, suggesting a high degree of agreement between the model and humans on occupational signals. The extent to which the exact values of kappa are interpretable and comparable has been questioned due to the sensitivity of the metric to the underlying distributions in the data \citep{Sim2005, Warrens2010}. Nonetheless, within the same dataset, the metrics should be  more comparable. In this context, introducing the LLM as an additional rater either significantly improved (p < 0.05 for Work and SOC Minor Group) or did not significantly shift (p > 0.1 for all ATUS, SOC Major Group) the inter-agreement metrics for all tasks and metrics. This establishes that  the model performs comparably to  (or may be even better) than a human across some of the range of metrics studied. 

There are two likely contributing factors to the observed agreement levels on the work classifier being lower than the rest of the observations. First, despite the Wildchat dataset being filtered to conversations in English before sampling, it still contained a small number of observations which contained text in foreign languages (with the most common case being computer code accompanied by instructions in a language other than English). Second, the LLM received clear instructions that interactions related to job-seeking and education should not be considered work, as well as guidance on how to identify education-related queries. To avoid biasing the human raters by replicating the LLM prompt, they did not receive the same level of detail on category definitions. As a result, humans may have found the task more ambiguous in these marginal cases. Overall, due to work/non-work sample coming from a non-ATLAS dataset, the applicability of the exact agreement values is less clear than for other variables in which raters had access to labeled ATLAS clusters. However, this is the best signal on inter-rater agreement for this metric we can currently obtain within the current technical constraints.    

We do not report inter-rater agreement statistics for finer-grained taxonomy levels below ATUS Tiers 1/2 and SOC Major and Minor Groups as those are highly unreliable in circumstances where the number of potential categories far exceeds the number of observations. First, kappas are based on observed choice probabilities within the sample. As a result, hundreds (or for O*NET thousands) of alternative categories that do not appear in the 120 rated clusters do not contribute to the probability of chance agreement, leading it to be (somewhat) overestimated and kappa underestimated.  Second, and much more importantly,  if the probability of agreement varies across different categories, a sample containing a small proportion of the potential categories is highly unlikely to be meaningfully informative about agreement in the population. The calculated kappas will be primarily driven by the characteristics of the categories that happened to be sampled, with no way to establish whether they are representative.   

\begin{table}[htbp]
    \centering
    \caption{Inter-Rater Agreement With Human and LLM Raters}
    \label{rateragreement}
    \resizebox{\textwidth}{!}{%
\begin{tabular}{l *{9}{c}}
\toprule
 & \multicolumn{2}{c}{Base Metrics} & \multicolumn{2}{c}{Human Only} & \multicolumn{2}{c}{With LLM} & \multicolumn{2}{c}{\begin{tabular}{@{}c@{}}Delta \\ (with - without)\end{tabular}} & \\
\cmidrule(lr){2-3} \cmidrule(lr){4-5} \cmidrule(lr){6-7} \cmidrule(lr){8-9}
Classifier & \begin{tabular}{@{}c@{}}Plurality \\ Agreement\end{tabular} & Kappa & \begin{tabular}{@{}c@{}}PW \\ Kappa\end{tabular} & Fleiss & \begin{tabular}{@{}c@{}}PW \\ Kappa\end{tabular} & Fleiss & \begin{tabular}{@{}c@{}}PW \\ Kappa\end{tabular} & Fleiss & N \\
\midrule
Work & \begin{tabular}{@{}c@{}}0.85 \\ {\scriptsize [0.85 - 0.85]}\end{tabular} & \begin{tabular}{@{}c@{}}0.71 \\ {\scriptsize [0.71 - 0.71]}\end{tabular} & \begin{tabular}{@{}c@{}}0.48 \\ {\scriptsize (0.37 - 0.58)}\end{tabular} & \begin{tabular}{@{}c@{}}0.48 \\ {\scriptsize (0.35 - 0.59)}\end{tabular} & \begin{tabular}{@{}c@{}}0.52 \\ {\scriptsize (0.43 - 0.61)}\end{tabular} & \begin{tabular}{@{}c@{}}0.53 \\ {\scriptsize (0.43 - 0.63)}\end{tabular} & \begin{tabular}{@{}c@{}}0.05 \\ {\scriptsize (0.032)}\end{tabular} & \begin{tabular}{@{}c@{}}0.06 \\ {\scriptsize (0.020)}\end{tabular} & 110 \\
ATUS (Major) & \begin{tabular}{@{}c@{}}0.83 \\ {\scriptsize [0.78 - 0.84]}\end{tabular} & \begin{tabular}{@{}c@{}}0.77 \\ {\scriptsize [0.72 - 0.79]}\end{tabular} & \begin{tabular}{@{}c@{}}0.74 \\ {\scriptsize (0.67 - 0.80)}\end{tabular} & \begin{tabular}{@{}c@{}}0.77 \\ {\scriptsize (0.70 - 0.83)}\end{tabular} & \begin{tabular}{@{}c@{}}0.74 \\ {\scriptsize (0.67 - 0.80)}\end{tabular} & \begin{tabular}{@{}c@{}}0.75 \\ {\scriptsize (0.69 - 0.81)}\end{tabular} & \begin{tabular}{@{}c@{}}-0.01 \\ {\scriptsize (0.523)}\end{tabular} & \begin{tabular}{@{}c@{}}-0.02 \\ {\scriptsize (0.124)}\end{tabular} & 116 \\
ATUS (2nd tier) & \begin{tabular}{@{}c@{}}0.82 \\ {\scriptsize [0.67 - 0.85]}\end{tabular} & \begin{tabular}{@{}c@{}}0.78 \\ {\scriptsize [0.62 - 0.82]}\end{tabular} & \begin{tabular}{@{}c@{}}0.58 \\ {\scriptsize (0.50 - 0.64)}\end{tabular} & \begin{tabular}{@{}c@{}}0.60 \\ {\scriptsize (0.52 - 0.67)}\end{tabular} & \begin{tabular}{@{}c@{}}0.58 \\ {\scriptsize (0.51 - 0.65)}\end{tabular} & \begin{tabular}{@{}c@{}}0.59 \\ {\scriptsize (0.52 - 0.66)}\end{tabular} & \begin{tabular}{@{}c@{}}0.00 \\ {\scriptsize (0.668)}\end{tabular} & \begin{tabular}{@{}c@{}}-0.00 \\ {\scriptsize (0.800)}\end{tabular} & 116 \\
SOC (Major) & \begin{tabular}{@{}c@{}}0.85 \\ {\scriptsize [0.78 - 0.87]}\end{tabular} & \begin{tabular}{@{}c@{}}0.83 \\ {\scriptsize [0.75 - 0.85]}\end{tabular} & \begin{tabular}{@{}c@{}}0.66 \\ {\scriptsize (0.58 - 0.73)}\end{tabular} & \begin{tabular}{@{}c@{}}0.68 \\ {\scriptsize (0.60 - 0.75)}\end{tabular} & \begin{tabular}{@{}c@{}}0.68 \\ {\scriptsize (0.60 - 0.73)}\end{tabular} & \begin{tabular}{@{}c@{}}0.69 \\ {\scriptsize (0.63 - 0.75)}\end{tabular} & \begin{tabular}{@{}c@{}}0.01 \\ {\scriptsize (0.299)}\end{tabular} & \begin{tabular}{@{}c@{}}0.01 \\ {\scriptsize (0.576)}\end{tabular} & 120 \\
SOC (Minor) & \begin{tabular}{@{}c@{}}0.83 \\ {\scriptsize [0.71 - 0.85]}\end{tabular} & \begin{tabular}{@{}c@{}}0.81 \\ {\scriptsize [0.69 - 0.84]}\end{tabular} & \begin{tabular}{@{}c@{}}0.55 \\ {\scriptsize (0.48 - 0.61)}\end{tabular} & \begin{tabular}{@{}c@{}}0.57 \\ {\scriptsize (0.49 - 0.64)}\end{tabular} & \begin{tabular}{@{}c@{}}0.58 \\ {\scriptsize (0.52 - 0.64)}\end{tabular} & \begin{tabular}{@{}c@{}}0.61 \\ {\scriptsize (0.54 - 0.67)}\end{tabular} & \begin{tabular}{@{}c@{}}0.03 \\ {\scriptsize (0.000)}\end{tabular} & \begin{tabular}{@{}c@{}}0.04 \\ {\scriptsize (0.002)}\end{tabular} & 120 \\
\bottomrule
\end{tabular}%
}

    \vspace{0.2cm}
    
    \parbox{\textwidth}{\footnotesize \textit{Notes:}  This table reports inter-rater reliability measures across various classification schemes. Base Metrics compare the human labels against the LLM prediction. Plurality Agreement is the proportion of cases where the LLM prediction exactly matches the human plurality label. The bounds for Base Metrics Plurality Agreement and Cohen's Kappa represent Manski bounds constructed using forced agreement (indicated with square brackets): the upper bound assumes the LLM prediction matches any human ties, while the lower bound assumes complete disagreement on ties. Human Only reports standard reliability metrics (Pairwise Cohen's Kappa and Fleiss's Kappa) calculated exclusively among human annotators, with bootstrapped 95\% confidence intervals (2000 resamples) in parentheses. With LLM recalculates these metrics treating the LLM as an additional rater. Delta (with - without) is the difference between the With LLM and Human Only point estimates, with two-tailed p-values in parentheses (calculated using the standard error of differences from the bootstrap resamples).}
\end{table}

\subsubsection{Human Approval of AI Labels}
To shed additional light on the performance of our classifiers at fine-grained taxonomy levels, and explore effects of the ambiguity of the underlying categories on human annotation, we performed a human-AI agreement assessment of model classification. Three human annotators evaluated new samples of 120 work cluster summaries alongside the SOC major group, ONET-SOC occupation title and O*NET task statements assigned to these clusters by our pipeline, and 120 non-work cluster summaries  accompanied by their respective ATUS Tier 1 to ATUS Tier 3 activity labels. For each of the clusters, the raters answered a single Yes/No questions for each of the taxonomy levels based on whether the following statement applied: 

``\textit{Based on available information, most of the activities in the cluster could either fit the assigned label, or are close enough that this minor mismatch would not significantly distort the economic interpretation of the activity.}''

The purpose of the exercise was to detect the frequency of significant classification disagreement, as distinguished from cases where the LLM making a defensible assignment given the ambiguity of the cluster summary (e.g. multiple activities mentioned) or ambiguity and/or duplication of the statistical categories themselves. Table~\ref{tab:table_app2_2} summarizes the simple unweighted average across raters of the share of rated clusters where the assigned label was rated as ``reasonable''.

\begin{table}[htbp]
    \centering
    \caption{Human Expert Review Validates Automated Taxonomic Classifications as ``Reasonable''}
    \label{tab:table_app2_2}
    \begin{tabular}{lc|lc}
        \toprule
        \textbf{Work Category} & \textbf{\% Reasonable} & \textbf{Non-work Category} & \textbf{\% Reasonable} \\
        \midrule
        SOC Major Group & 96.1\% & ATUS Tier 1 (Major) & 98.1\% \\
        Occupation Title & 92.8\% & ATUS Tier 2 (Intermediary) & 96.1\% \\
        Task Statement & 85.8\% & ATUS Tier 3 (Detailed) & 92.2\% \\
        \bottomrule
    \end{tabular}
    \vspace{0.2cm}
    \parbox{\textwidth}{\footnotesize \textit{Notes:} This table reports agreement rates from a human validation study where three independent domain expert raters evaluated 120 work conversation clusters and 120 non-work conversation clusters randomly sampled across Gemini App, Google AI Mode, and Gemini API. Experts evaluated whether the label assigned by the automated LLM pipeline at each taxonomic depth was "reasonable" (defined as fitting the activity or being close enough that any minor mismatch would not distort economic interpretation). Approval rates averages range from 96\% (work) and 98\% (non-work) at major group levels to 86\% at the most granular O*NET task statement level.}
\end{table}

Three things stand out about these results. First, raters rarely disagreed with the assigned labels, and at the highest taxonomy levels for both work and non-work, rater approval was almost universal. Even when evaluating the most granular task statements, raters found 86\% of the Gemini-selected labels to be either a strong fit for the cluster or close enough that the economic substance remained intact. While a label might accurately describe a broader cluster without perfectly matching every single conversation inside it, these high approval ratings provide strong reassurance that the pipeline is generating economically meaningful classifications.

The second noteworthy fact is the difference between the rate of agreement between the plurality of human raters and the LLM ratings in the previous section, and the 11 and 15pp higher rate of raters not disagreeing with the model for comparable work and non-work top-level taxonomy categories. Even at the highest level of aggregation, the statistical taxonomies are sufficiently ambiguous for it to be possible for raters to pick different options unprompted but still agree with an alternative if one is presented.  This problem is not new, but the methodological challenge of producing accurate and economically meaningful usage statistics in circumstances where multiple options can often be a reasonable fit merits further research. 

Finally, there is a large gulf between high human rater acceptance and the pipeline's lower accuracy rates on synthetic data, particularly at highly granular levels. The drivers of this divergence, and its implications for improving classifier validation require further investigation. But even with the current evidence, approval and accuracy provide useful bounds on the degree of confidence about the highly granular taxonomy assignments like O*NET tasks and  ATUS Tier 3 activities. The high human approval rates suggest these granular labels hold real economic meaning. On the other hand, however, the lower accuracy in assigning exact options at the most detailed taxonomy levels  serves as a caution against over-relying on hyper-specific task analysis. This dynamic informs the approach of the report. By focusing on broader distributions, using metrics more robust to measurement error (e.g. presence vs exact counts), and aggregating these specific labels into  less granular economic categories (e.g, \cite{NBERw33941} task types) it is possible to significantly reduce the impact of individual classification noise without sacrificing the economic value of the analysis.

\pagebreak

\section{Additional Tables and Figures, Occupation Section}
\label{app:app_3}

\begin{figure}[htbp]
    \centering
    \caption{Gemini Detailed Occupation Task Saturation Shares With O*NET Importance Weighting}
    \label{fig:appendix_figure_1a_1}
    \includegraphics[width=\textwidth]{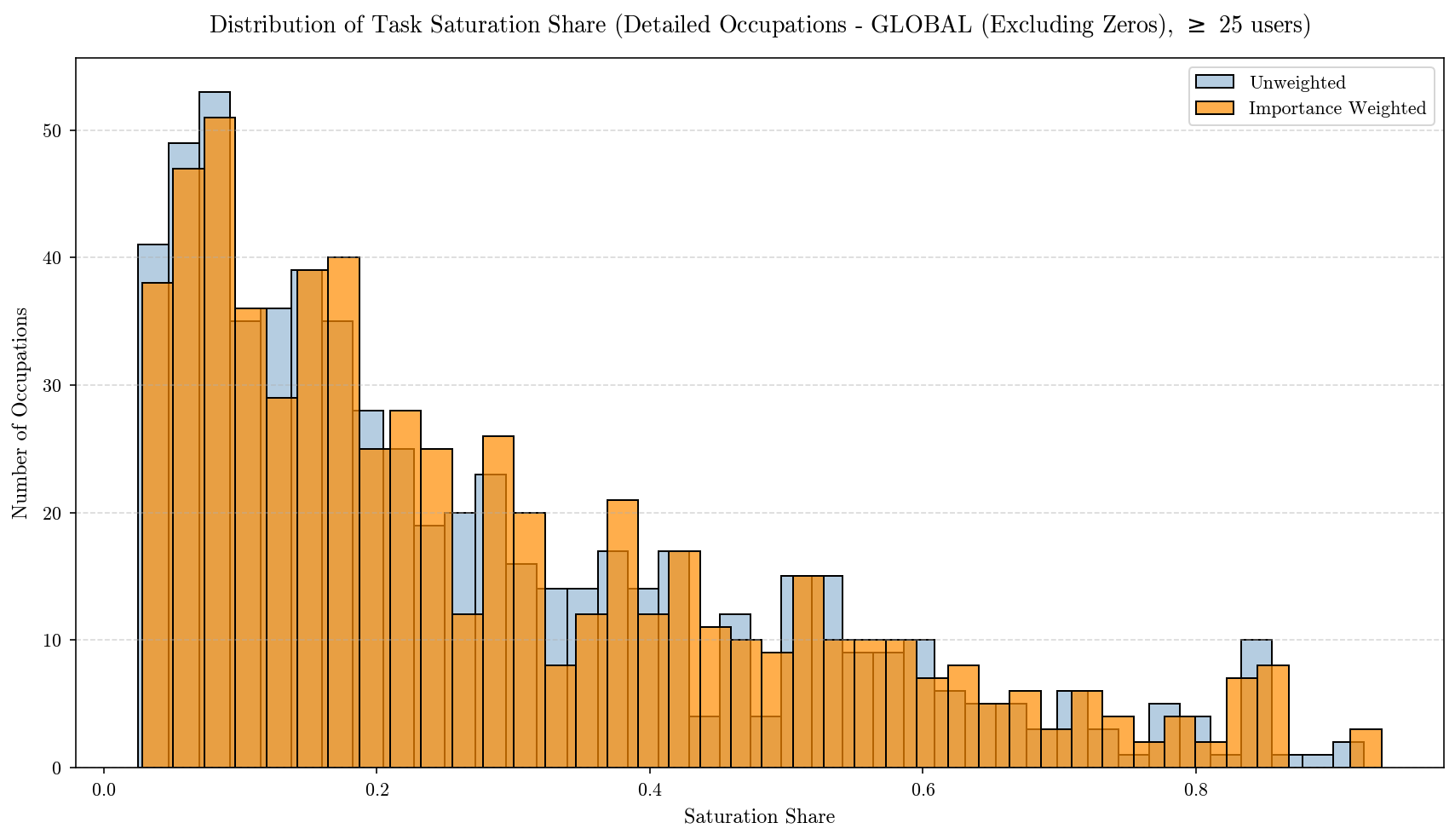}
    \vspace{0.2cm}
    \parbox{\textwidth}{\footnotesize \textit{Notes:} Similar to Figure \ref{fig:figure_2_4}. Task importance ratings (scale 1-5) from O*NET are applied to weight the task saturation distribution across occupations. For calculating the importance weighted share, we use (importance - 1) to omit tasks that are not important. We exclude any occupation for which fewer than 50\% of tasks have available importance ratings.}
\end{figure}

\begin{figure}[htbp]
    \centering
    \caption{Distribution of Task Expertise Scores Across the O*NET Task Universe}
    \label{fig:appendix_figure_1a_2}
    \includegraphics[width=\textwidth]{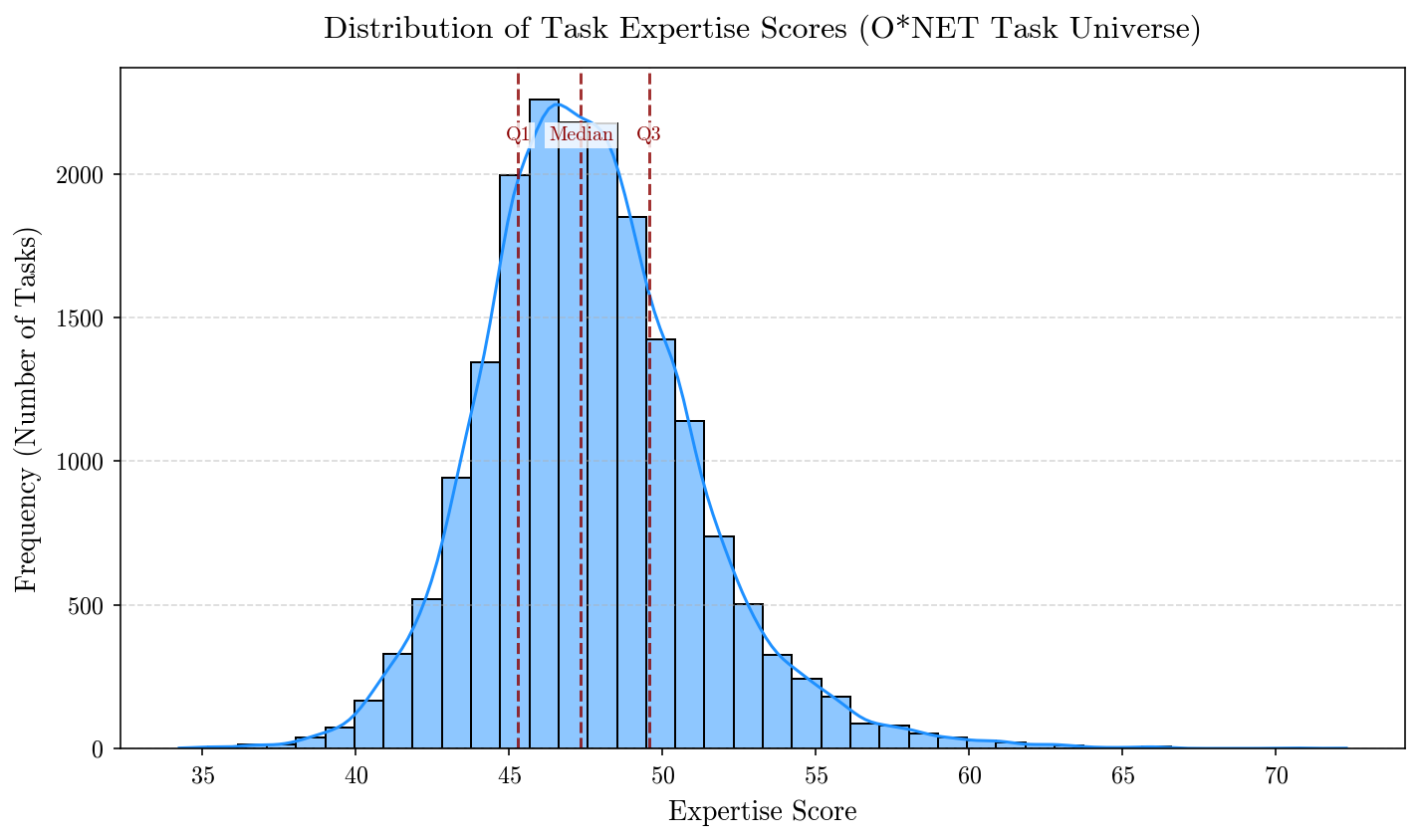}
    \vspace{0.2cm}
    \parbox{\textwidth}{\footnotesize \textit{Notes:} We calculate scores measuring the expertise of tasks in the O*NET universe by replicating the methodology in \cite{NBERw33941}. Expertise for each task is defined as 100 - the average Standard Frequency Index (SFI) of the lemmatized terms in its O*NET task statement, where SFI is an index combining word frequency and word entropy provided for 140,000 English language words in \cite{Zeno1995-sf}. This figure depicts the distribution of resulting values. The dashed lines indicate the quartiles referenced in Figure \ref{fig:figure_2_8}.}
\end{figure}

\newpage
\section{Additional Tables and Figures, Household Section}
\label{app:household_section}

\begin{figure}[htbp]
    \centering
    \caption{Granular Task Comparison Highlights AI Over-Representation in Specialized Knowledge and Under-Representation in Routine Chores}
\label{fig:abs-differences-representation}
    \includegraphics[width=\textwidth]{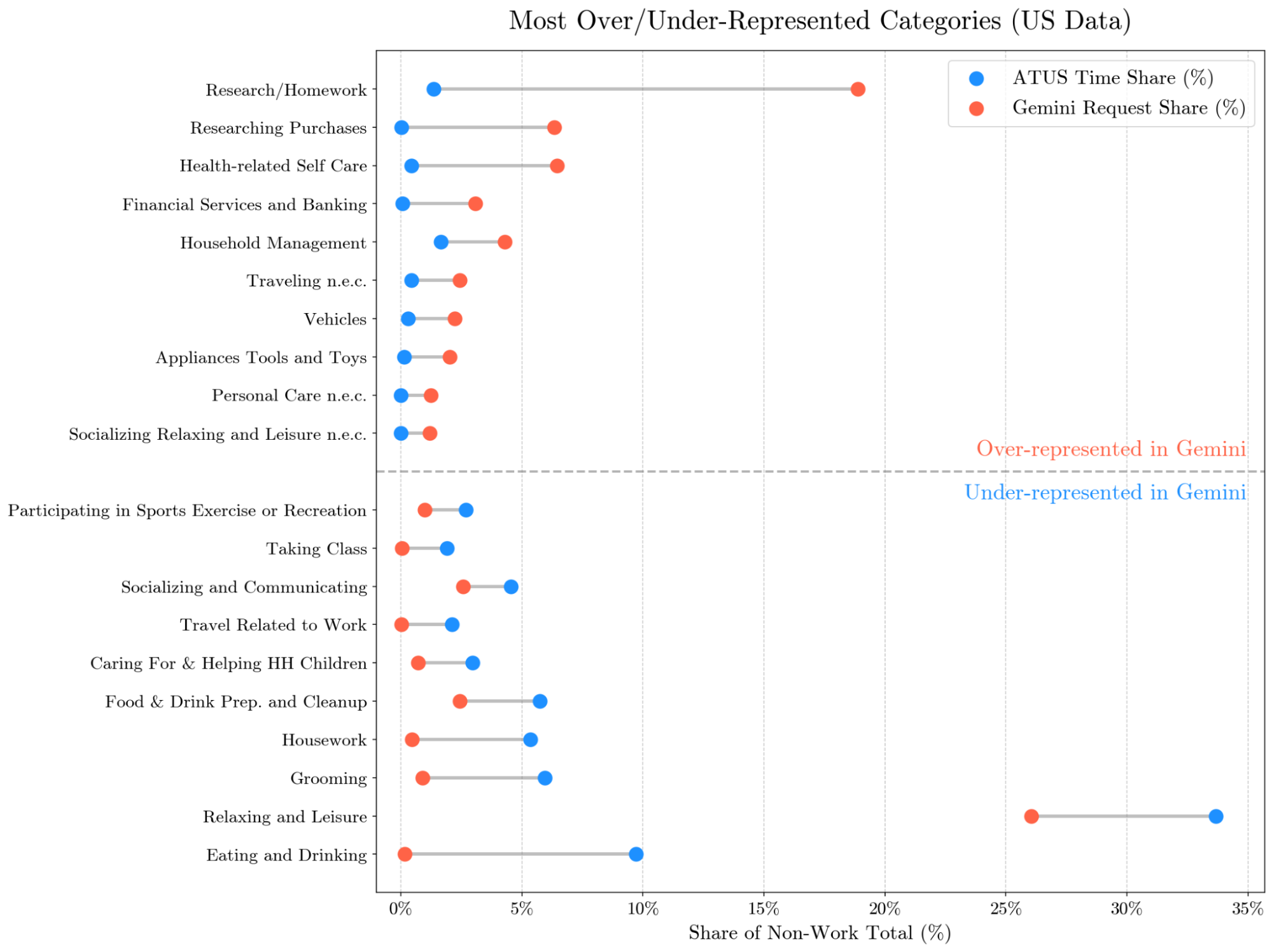}
    \vspace{0.2cm}
    \parbox{\textwidth}{\footnotesize \textit{Notes:} Compares U.S. AI conversation shares against ATUS human time shares across the 10 most over-represented and 10 most under-represented detailed daily tasks (minor categories), ranked by absolute percentage point difference (excluding working category and sleep). The dumbbell plots illustrate that at the most granular level, AI assistance disproportionately concentrates in education work, comparison shopping, financial planning, and government service queries. Conversely, human time heavily outweighs AI usage in physical daily necessities such as television viewing, dining, washing/dressing, interior cleaning, and meal prep.}
\end{figure}

\begin{figure}[htbp]
    \centering
        \caption{Human Time Use Positively Predicts AI Consultation Across Hundreds of Granular Household Activities}
    \label{fig:tier3-correlations}
    \includegraphics[width=\textwidth]{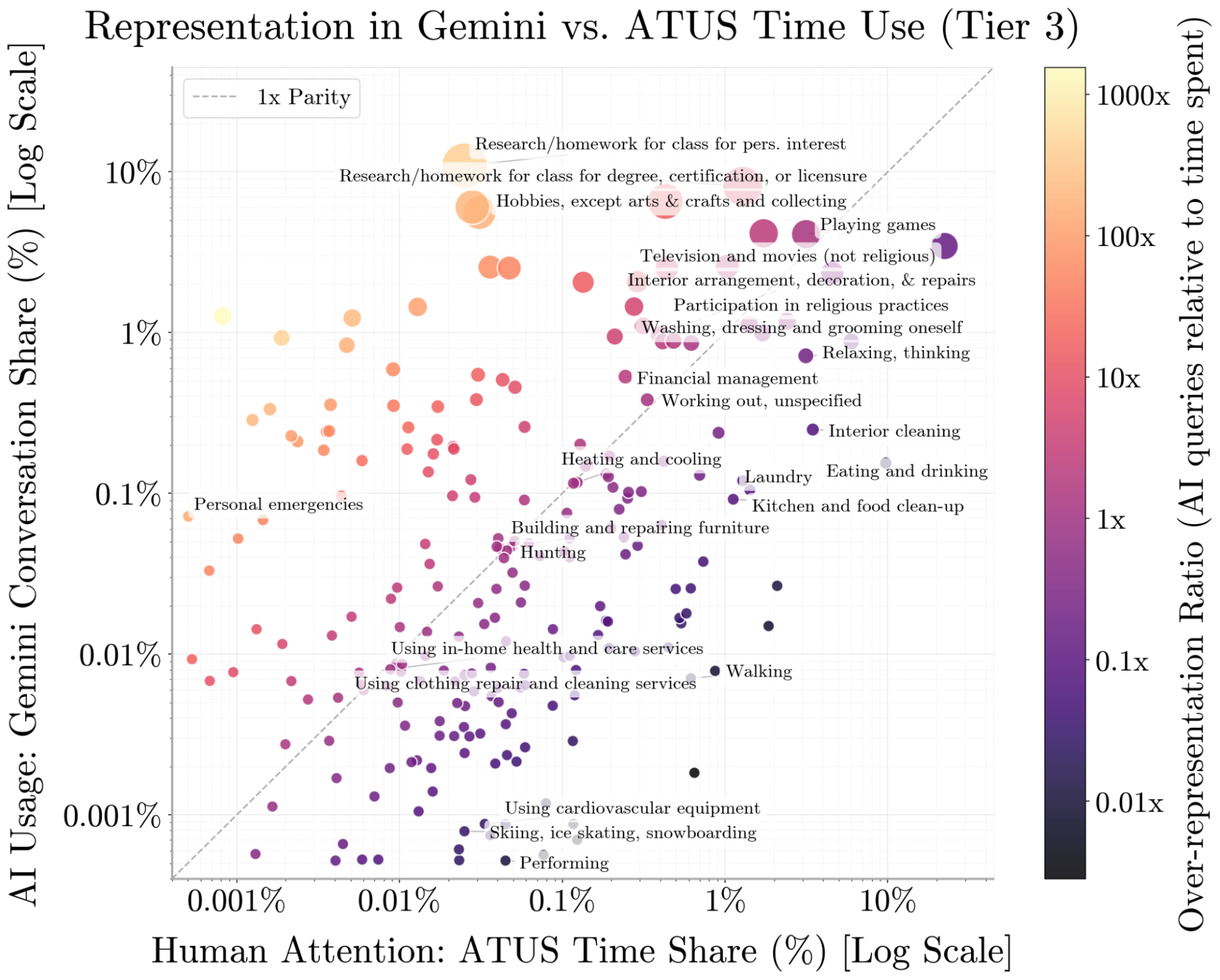}
    \vspace{0.2cm}
    \parbox{\textwidth}{\footnotesize \textit{Notes:} Log-log scatter plot illustrating human time use shares versus AI conversation shares across all individual ATUS Tier 3 minor activity categories (excluding sleeping, working, and uncoded tasks). Each point represents a distinct granular task. The solid 45-degree line indicates parity between human time spent and AI consultation share. Despite wide variation across hundreds of niche activities, the overall positive relationship persists even at this micro-level, confirming that more time-consuming daily activities generally elicit higher volumes of AI conversations.}
\end{figure}

\begin{figure}[htbp]
    \centering
    \caption{Bureaucratic Compliance, Licenses, Taxes, and Civic Obligations Dominate Government-Related AI Conversations}
    \label{fig:gvt-tasks-decomp}
    \includegraphics[width=1\textwidth]{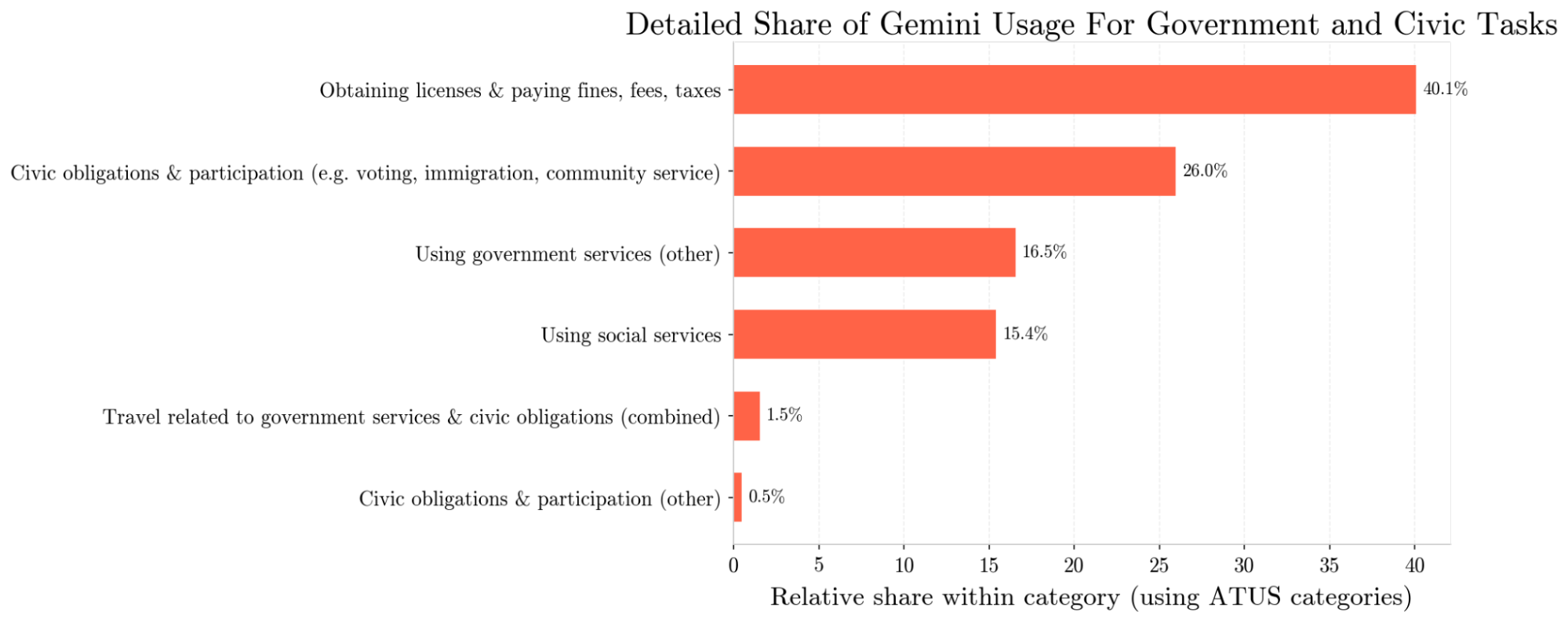}
    \label{fig:appendix_figure_4_3}
    \vspace{0.2cm}
    \parbox{\textwidth}{\footnotesize \textit{Notes:} Relative share of specific tasks within the Government Services \& Civic Duties domain (and associated travel). The x-axis displays the relative share (\%) of each specific task out of all AI conversations classified under government and civic interactions. Obtaining government licenses, paying taxes/fines/fees, and navigating civic obligations (such as voting requirements, local civic procedures, and immigration compliance) represent the largest share of citizen consultations.}
\end{figure}

\begin{figure}[htbp!]
    \centering
    \caption{Privacy-Preserving Conversation Summarizing Clusters Also Reveal High Citizen Demand for Help with Legal Requirements, Procedures, and Compliance}
\label{fig:gvt-tasks-word-map}
    \includegraphics[width=\textwidth]{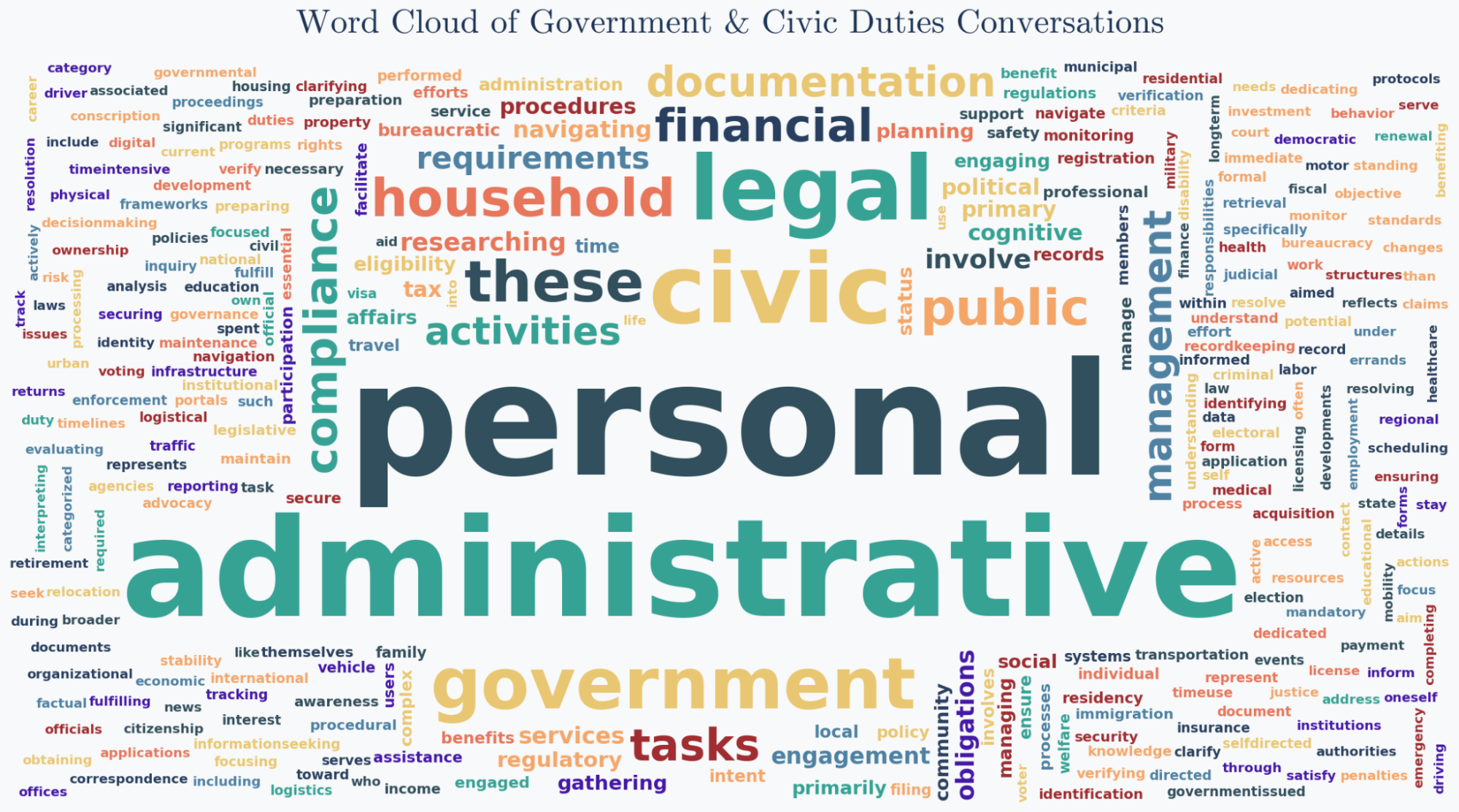}
    \label{fig:appendix_figure_4_4}
    \vspace{0.2cm}
    \parbox{\textwidth}{\footnotesize \textit{Notes:} Word cloud visualizing frequent substantive terms across AI conversation summaries in government and civic tasks (with word size scaled by relative frequency). Some prominent terminology centers are ``administrative'', ``documentation'', ``compliance'', ``requirements'', ``application'', ``procedures'', ``legal'', and to some extent ``bureaucratic'', ``immigration'', ``taxes'', ``driving'', and ``visa''.}
\end{figure}

\begin{figure}[htbp!]
    \centering
    \caption{Nearly Half of All Medical, Legal, Financial, and Government and Civic Services AI Conversations Occur Outside Standard Business Hours}
\label{fig:tasks-working-hours}
    \includegraphics[width=.95\textwidth]{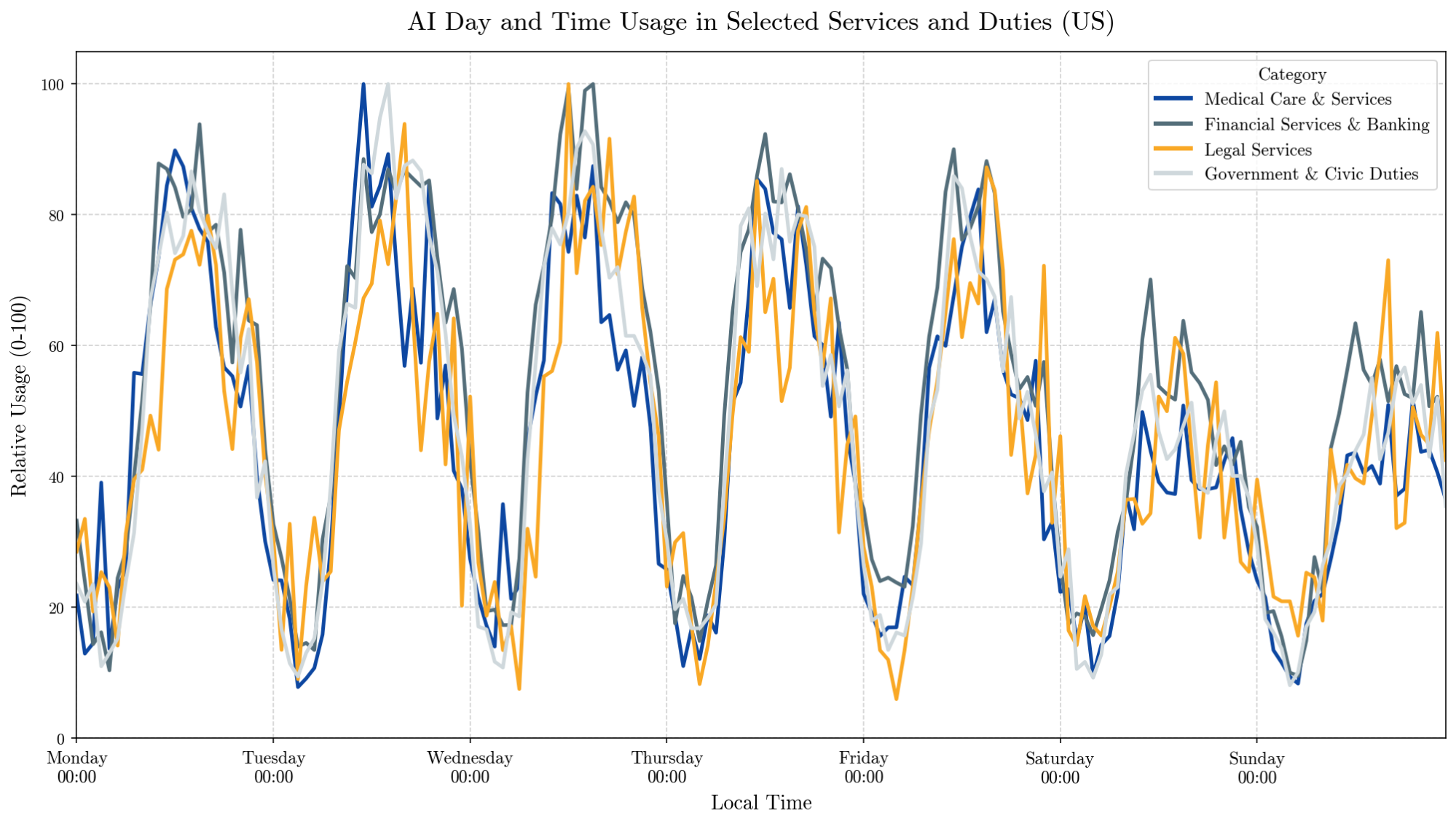}
    \label{fig:appendix_figure_4_5}
    \vspace{0.2cm}
    \parbox{\textwidth}{\footnotesize \textit{Notes:} Tracks hourly AI conversation volume across key professional and civic service domains throughout the week (aggregated continuously from Monday 00:00 to Sunday 23:00 local time and normalized to a 0--100 scale). Notably, just below 50\% of all conversations across these bureaucratic domains occur during evenings (M--F outside 9 AM--5 PM) and weekends.}
\end{figure}

\begin{figure}[htbp!]
    \centering
    \caption{Non-Physical Cognitive Tasks Show Significantly Higher AI Representation Ratios than Physical Activities}
    \label{fig:appendix_figure_4_6}
    \includegraphics[width=.95\textwidth]{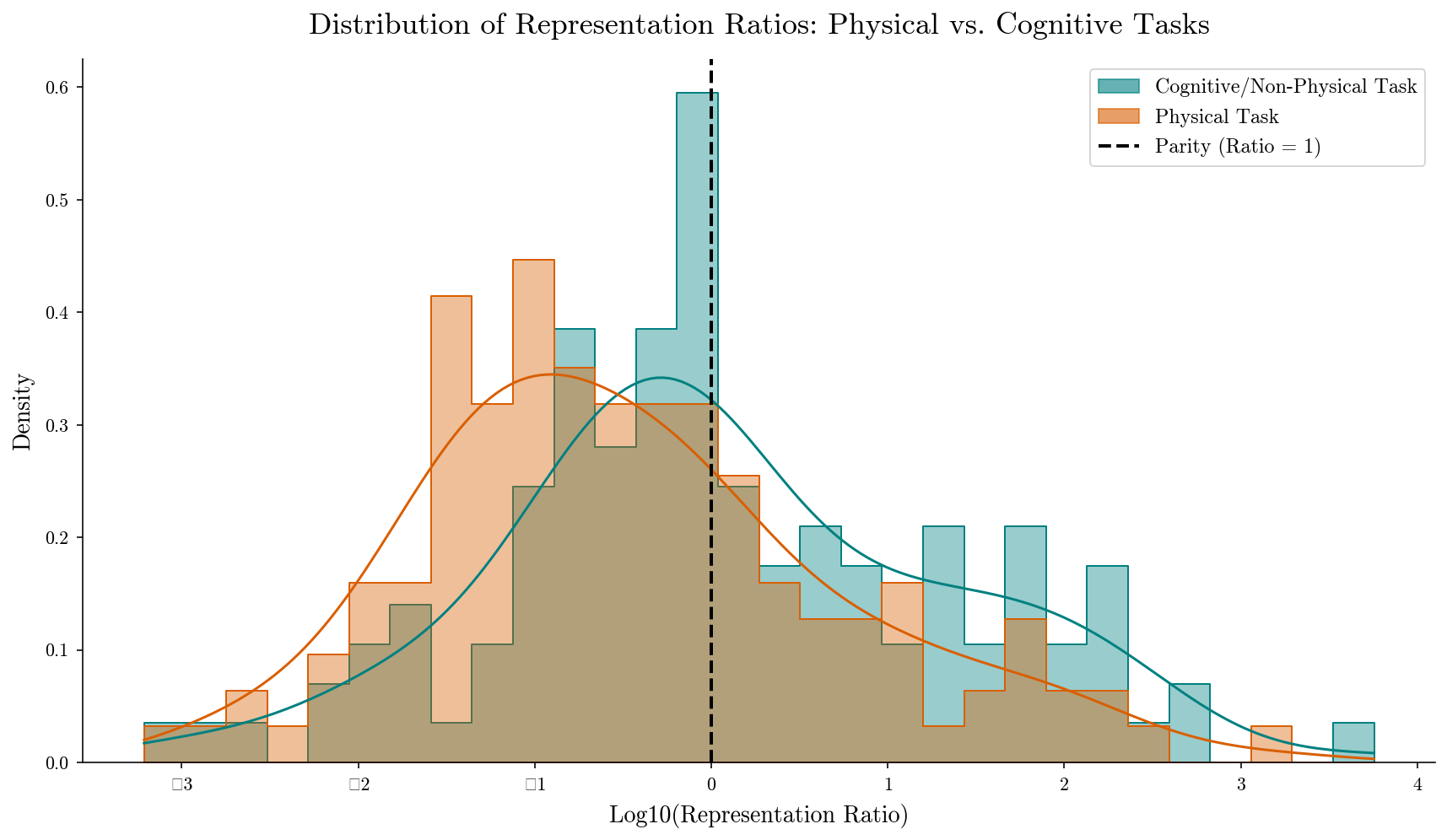}
    \vspace{0.2cm}
    \parbox{\textwidth}{\footnotesize \textit{Notes:} Probability density functions of AI representation ratios for physical vs. cognitive ATUS Tier 3 tasks. We use Gemini to classify a set of 456 activities in the ATUS taxonomy into physical or cognitive tasks. The vertical dashed line marks exact parity between AI share and human time share. The distribution for cognitive/non-physical tasks is shifted substantially to the right, with a mean representation ratio (defined as the AI conversation share divided by the ATUS human time share, plotted on a logarithmic scale) approximately twice as high and a median roughly two thirds higher than that of physical tasks.}
\end{figure}

\newpage

\section{Geography and Language Section Further Results}
\label{app:geography}

\begin{figure}[htbp!]
    \centering
    \caption{Conversational AI Usage Per Capita Shows Higher Adoption in Higher-Income Countries, Even Prior to Gemini Penetration Adjustment}
    \label{fig:appendix_figure_5_1}
    \includegraphics[width=\textwidth]{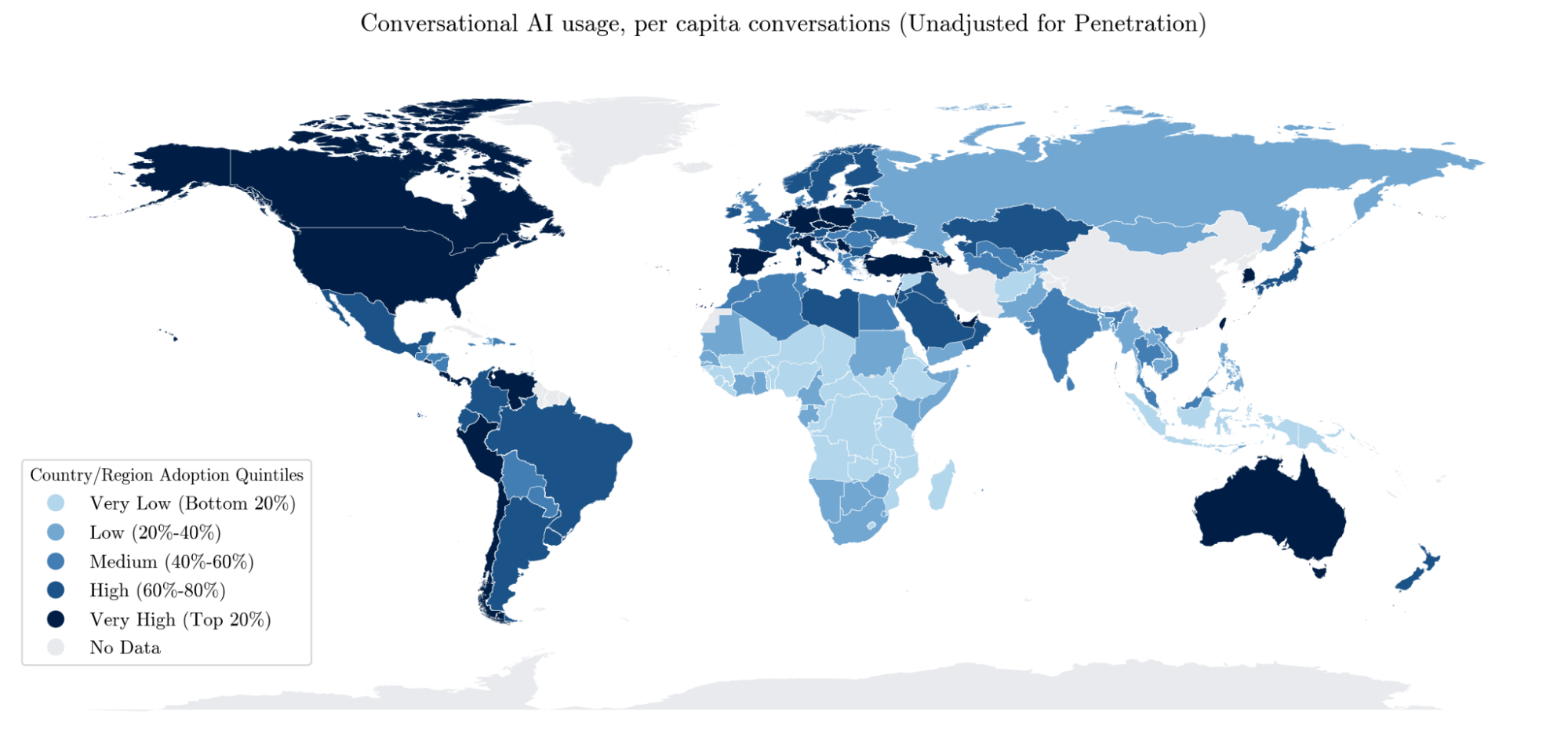}
    \vspace{0.2cm}
    \parbox{\textwidth}{\footnotesize \textit{Notes:} Global distribution of conversational AI (Gemini Apps and AI Mode) per capita usage by country/region quintiles (unadjusted for penetration). Calculations are based on weighted conversation counts across both platforms. Several countries/regions are excluded due to consumer-level access being unavailable, or queries primarily occurring through enterprise Workspace accounts. Per-capita rates are calculated using \cite{worldbank_population_2024} population estimates from the World Development Indicators (WDI) database (or \cite{imf_weo_oct2024} WEO where missing). Only countries/regions with more than 1,000,000 inhabitants are included. The figure utilize the same methodology as in Section \ref{sec:geo_diffusion} (unadjusted for penetration version).}
\end{figure}

\begin{figure}[htbp]
    \centering
    \caption{Unadjusted Per Capita Usage Highlights High Differences in Adoption Between Countries}
    \label{fig:appendix_figure_5_2}
    \includegraphics[width=\textwidth]{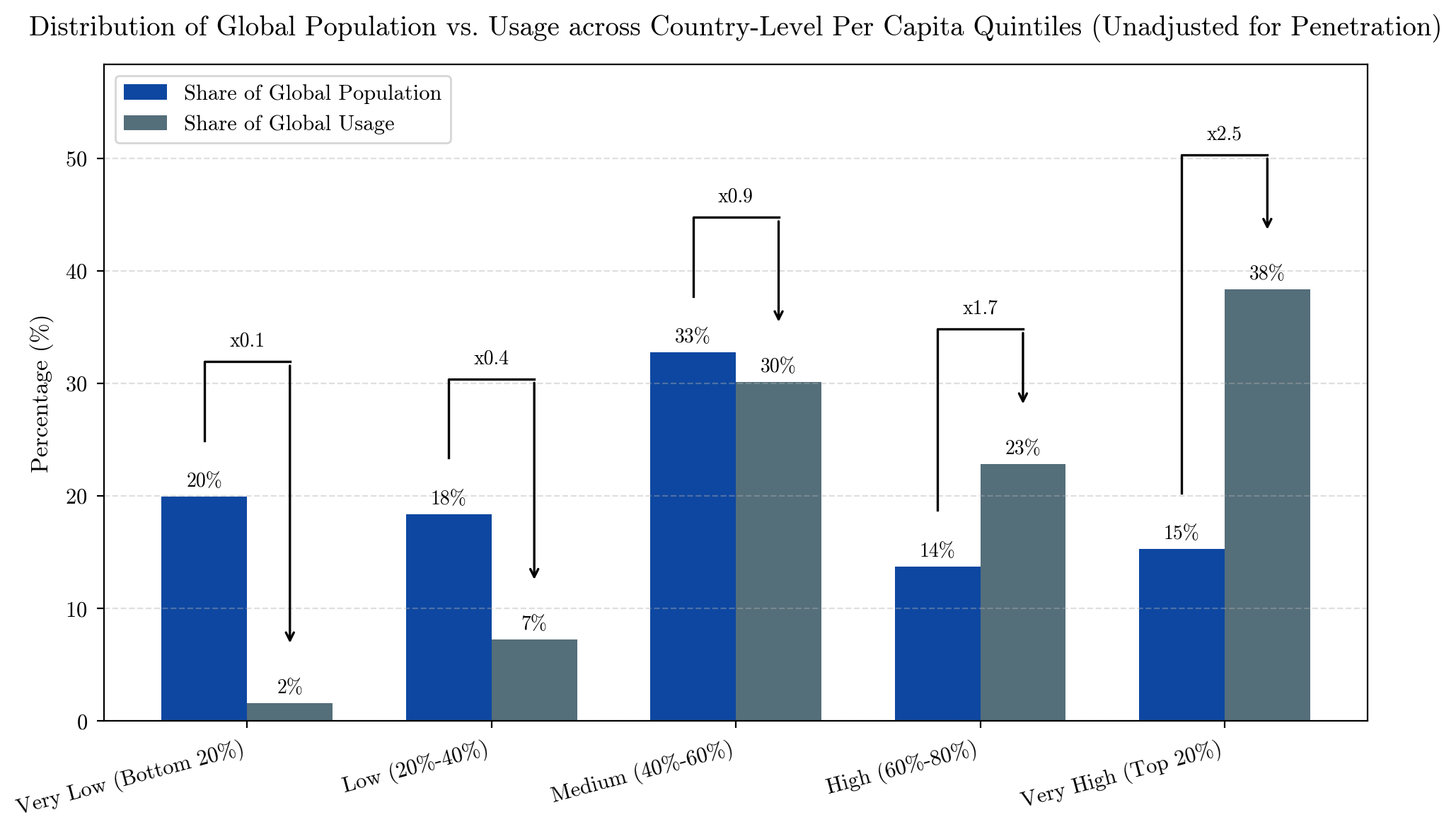}
    \vspace{0.2cm}
    \parbox{\textwidth}{\footnotesize \textit{Notes:} Global population share versus unadjusted AI conversation share across five per-capita usage quintiles at the country/region level. Multipliers denote the ratio of usage share to population share for each quintile. China and regions with populations under 1,000,000 are excluded. The figure utilize the same methodology as in Section \ref{sec:geo_diffusion} (unadjusted for penetration version). Note that the analysis aggregates country-level data, which may mask within-country inequality in adoption and usage.}
\end{figure}

\begin{figure}[htbp]
    \centering
    \caption{Unadjusted Conversational AI Usage Intensity Maintains a Strong Positive Correlation With GDP Per Capita}
    \label{fig:appendix_figure_5_3}
    \includegraphics[width=\textwidth]{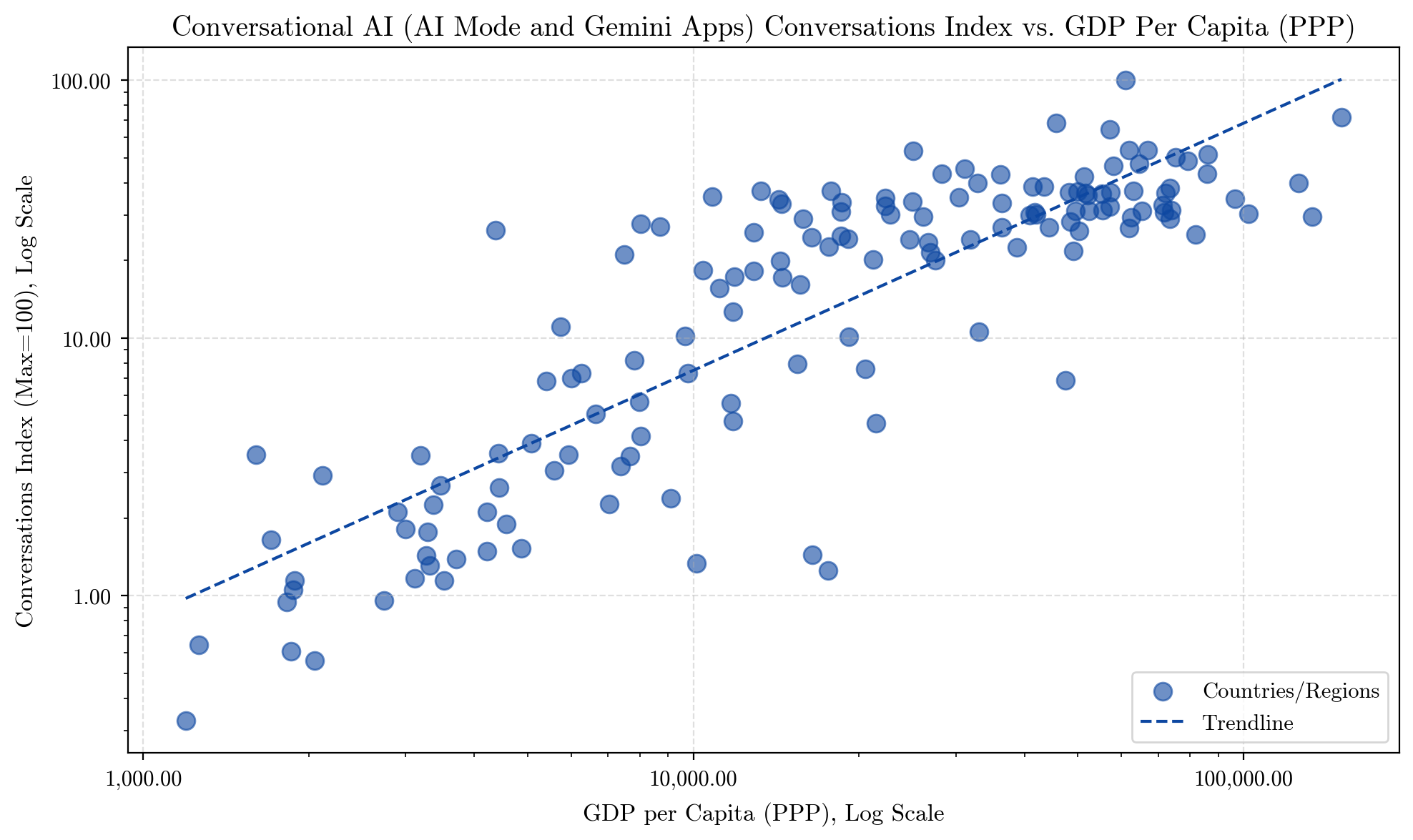}
    \vspace{0.2cm}
    \parbox{\textwidth}{\footnotesize \textit{Notes:} Unadjusted per-capita conversation counts versus PPP GDP per capita on a log-log scale. Per-capita conversations are indexed to a maximum value of 100. GDP per capita data is sourced from the \cite{worldbank_gdppc_2024} WDI database (or \cite{imf_weo_oct2024} WEO where missing). Only countries/regions with more than 1,000,000 inhabitants with official availability of Gemini Apps are featured; Mainland China is excluded. The figure utilize the same methodology as in Section \ref{sec:geo_diffusion} (unadjusted for penetration version).}
\end{figure}

\begin{figure}[htbp]
    \centering
    \caption{Adjusting for Internet Access Elevates Relative Adoption Metrics for Connectivity-Constrained Regions}
    \label{fig:appendix_figure_5_4}
    \includegraphics[width=\textwidth]{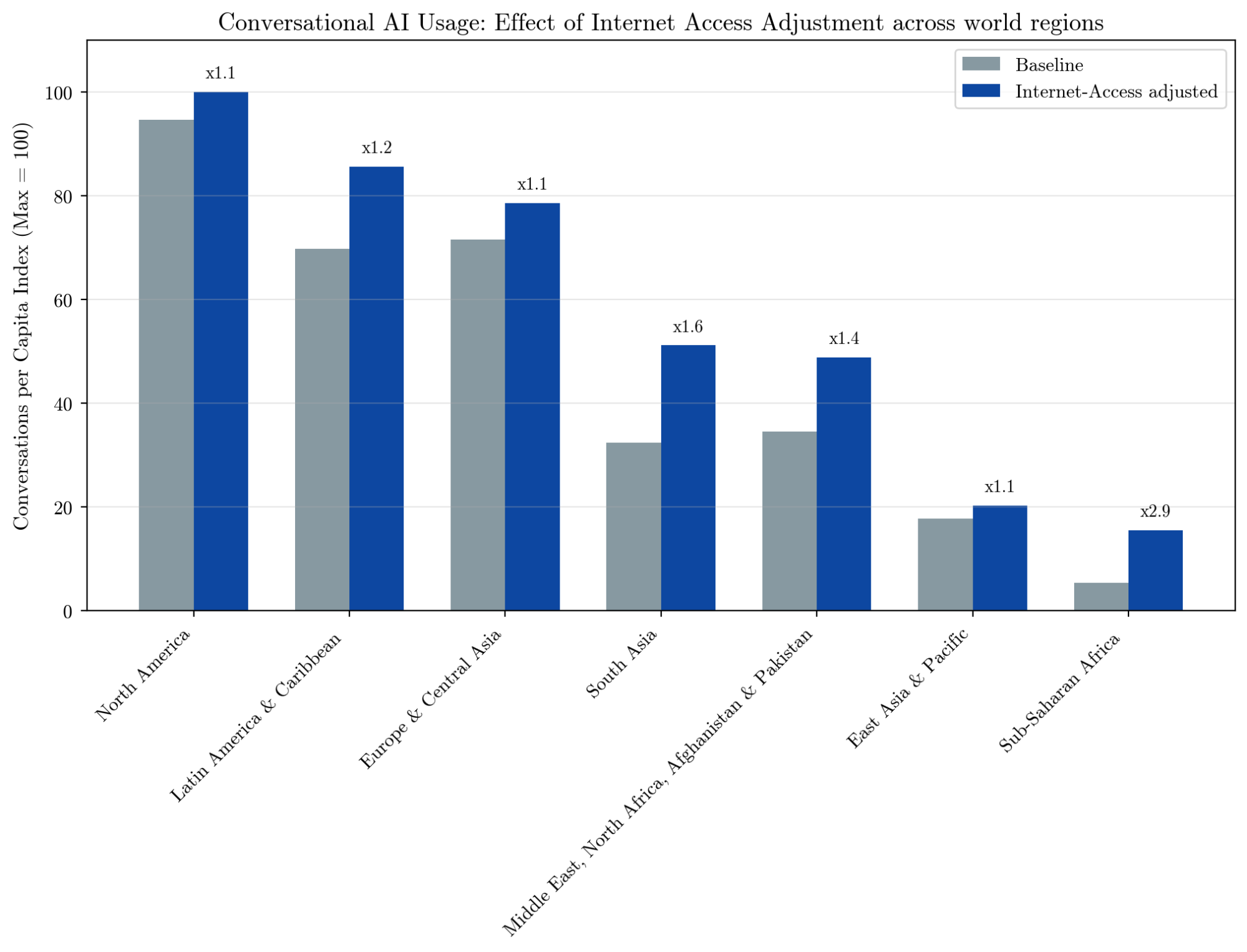}
    \vspace{0.2cm}
    \parbox{\textwidth}{\footnotesize \textit{Notes:} Regional comparison of per-capita vs. per-internet-user adoption (unadjusted for Gemini penetration). The internet-access unadjusted metric utilizes internet users (calculated using \cite{worldbank_internet_2024} internet access percentages from WDI) in the denominator instead of total population. The comparison is aggregated at the regional level, which masks country-specific variation and digital divides within regions. Both metrics are scaled to a maximum regional value of 100.  The figure utilize the same methodology as in Section \ref{sec:geo_diffusion} (unadjusted for penetration version).}
\end{figure}

\begin{figure}[htbp]
    \centering
    \caption{Established Digital Hubs Maintain Per-Capita Leadership in Work-Related AI Conversations Prior to Penetration Adjustment Like Sub-Saharan Africa}
    \label{fig:appendix_figure_5_5}
    \includegraphics[width=\textwidth]{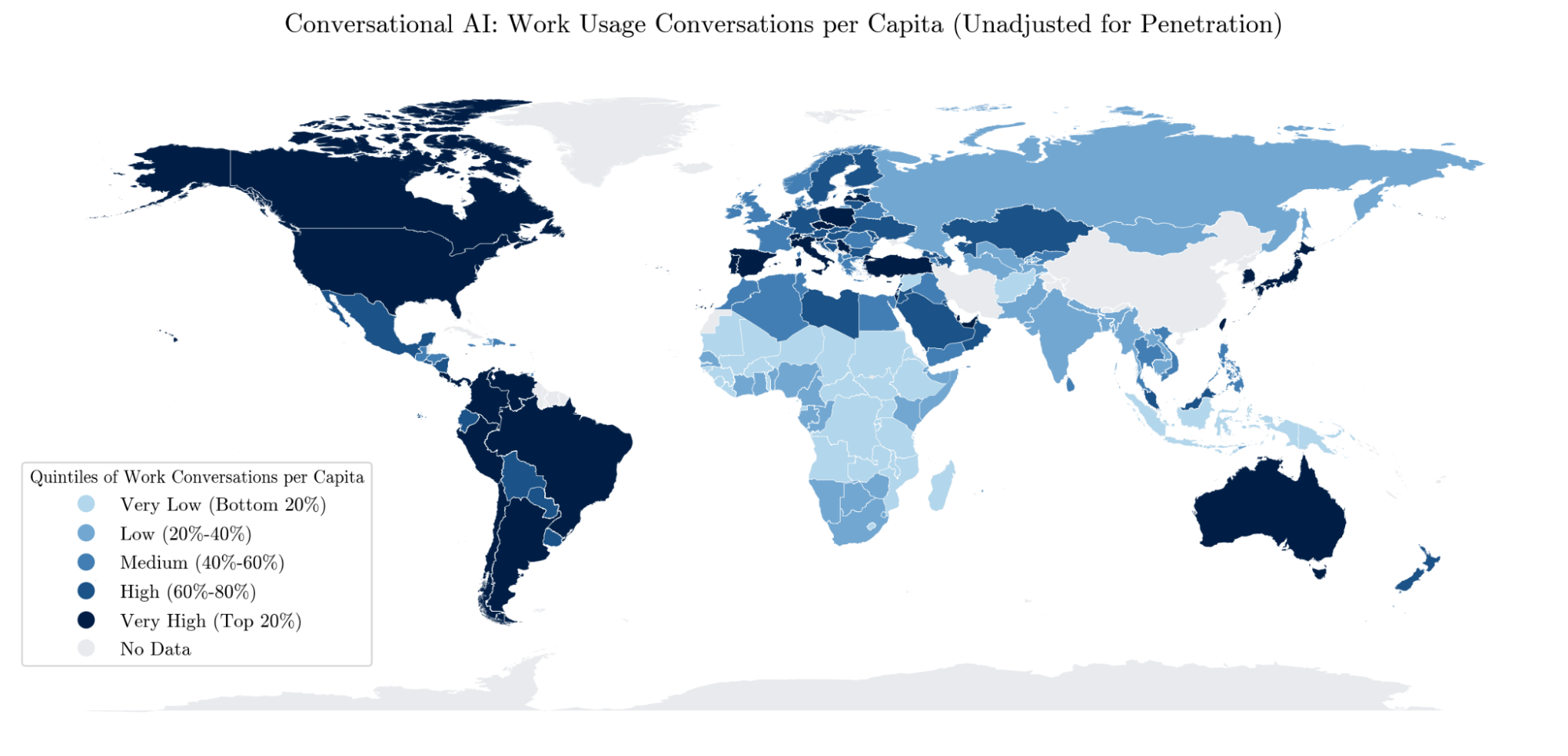}
    \vspace{0.2cm}
    \parbox{\textwidth}{\footnotesize \textit{Notes:} This figure displays the per-capita intensity of work-related conversations, unadjusted for usage share, by country/region quintiles. The metric divides work conversations per capita by the StatCounter Gemini usage share proxy. China and countries with populations under 1,000,000 are excluded. To prevent extreme outliers from small sample sizes, countries with a Gemini usage share of 0.1\% or lower are excluded. .}
\end{figure}

\newpage

\begin{figure}[htbp]
    \centering
    \caption{Interacting in a Non-Primary Language Drives Longer Sessions and Higher Token Verbosity}
    \label{fig:appendix_figure_5_6}
    \includegraphics[width=\textwidth]{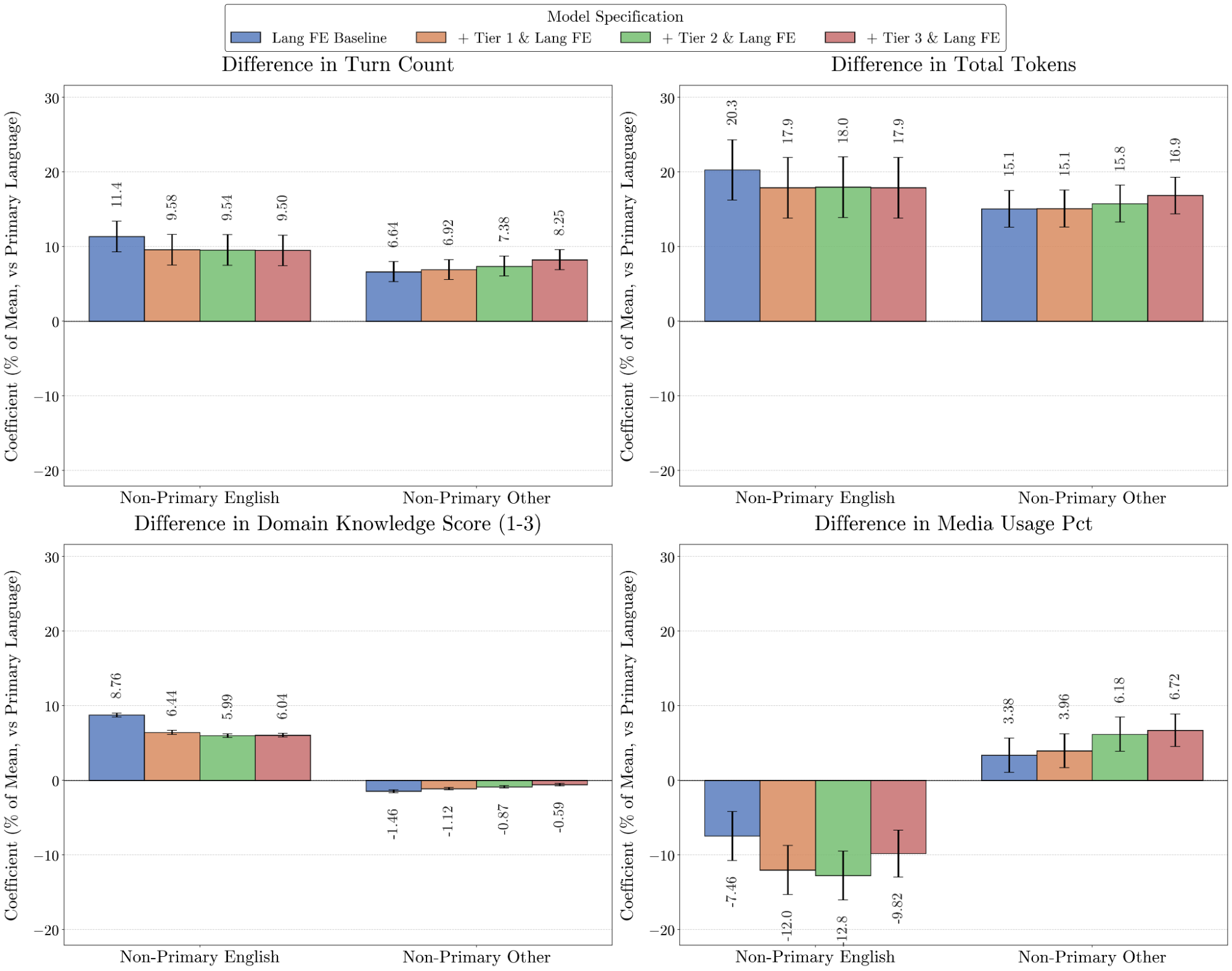}
    \vspace{0.2cm}
    \parbox{\textwidth}{\footnotesize \textit{Notes:} Regression point estimates (with 95\% confidence intervals) evaluating how interacting in a non-primary language affects session length (turn count), verbosity (total tokens), domain technical depth score, and media usage relative to primary-language interactions. The Domain Knowledge Score maps qualitative ratings (Low, Medium, High) to a 1--3 scale. Four progressive model specifications are shown, incorporating language fixed effects and increasingly granular ATUS activity category fixed effects (Tiers 1 through 3)  so we can get closer to testing whether observed variations in session length, verbosity, technical depth, and modality are a direct function of the language of interaction, or are rather a byproduct of task sorting (e.g., shifting more complex tasks into English). The results regarding token costs and turn counts are similar for non-primary other languages. This suggests it is perhaps more ``expensive'' to use AI in your non-primary language, or that users may bring slightly more complex queries to English.}

\end{figure}

\end{document}